\documentclass[reprint,prc,twocolumn,
amsmath,
amssymb,
amsart,
aps,
superscriptaddress,
longbibliography,
floatfix]{revtex4-2} 
\usepackage{graphicx}
\usepackage{lipsum}
\usepackage{tabularx}
\usepackage{float}
\usepackage[font= small, labelfont=bf,justification=justified,format=plain]{caption}
\usepackage{url}
\usepackage[labelformat=simple]{subcaption}
\usepackage[switch]{lineno}
\usepackage[labelformat=simple]{subcaption}

\usepackage[dvipsnames]{xcolor}
\usepackage{ragged2e}
\renewcommand{\justify}{\leftskip=0pt \rightskip=0pt plus 0cm}
\newcommand{\justifyy}{\leftskip=0pt \rightskip=0pt plus 1em}
\usepackage{enumitem}
\usepackage[english]{babel}
\begin{document}
\preprint{APS/123-QED}
\title{Exclusive $\pi^{-}$ electroproduction off the neutron in deuterium in the resonance region}

\newcommand*{\SCAROLINA}{University of South Carolina, Columbia, South Carolina 29208, USA}
\newcommand*{\SCAROLINAindex}{1}
\affiliation{\SCAROLINA}
\newcommand*{\SU}{Syracuse University, Syracuse, New York 13244, USA}
\newcommand*{\SUindex}{2}
\affiliation{\SU}
\newcommand*{\JLAB}{Thomas Jefferson National Accelerator Facility, Newport News, Virginia 23606, USA}
\newcommand*{\JLABindex}{3}
\affiliation{\JLAB}
\newcommand*{\ODU}{Old Dominion University, Norfolk, Virginia 23529, USA}
\newcommand*{\ODUindex}{4}
\affiliation{\ODU}
\newcommand*{\ANL}{Argonne National Laboratory, Argonne, Illinois 60439, USA}
\newcommand*{\ANLindex}{5}
\affiliation{\ANL}
\newcommand*{\TEMPLE}{Temple University,  Philadelphia, Pennsylvania 19122, USA}
\newcommand*{\TEMPLEindex}{6}
\affiliation{\TEMPLE}
\newcommand*{\INFNFE}{INFN, Sezione di Ferrara, 44100 Ferrara, Italy}
\newcommand*{\INFNFEindex}{7}
\affiliation{\INFNFE}
\newcommand*{\INFNGE}{INFN, Sezione di Genova, 16146 Genova, Italy}
\newcommand*{\INFNGEindex}{8}
\affiliation{\INFNGE}
\newcommand*{\ITEP}{National Research Centre Kurchatov Institute - ITEP, Moscow, 117259, Russia}
\newcommand*{\ITEPindex}{9}
\affiliation{\ITEP}
\newcommand*{\UTFSM}{Universidad T\'{e}cnica Federico Santa Mar\'{i}a, Casilla 110-V Valpara\'{i}so, Chile}
\newcommand*{\UTFSMindex}{10}
\affiliation{\UTFSM}
\newcommand*{\DUQUESNE}{Duquesne University, 600 Forbes Avenue, Pittsburgh, Pennsylvania 15282, USA}
\newcommand*{\DUQUESNEindex}{11}
\affiliation{\DUQUESNE}
\newcommand*{\BRESCIA}{Universit\`{a} degli Studi di Brescia, 25123 Brescia, Italy}
\newcommand*{\BRESCIAindex}{12}
\affiliation{\BRESCIA}
\newcommand*{\INFNPAV}{INFN, Sezione di Pavia, 27100 Pavia, Italy}
\newcommand*{\INFNPAVindex}{13}
\affiliation{\INFNPAV}
\newcommand*{\INFNCAT}{INFN, Sezione di Catania, 95123 Catania, Italy}
\newcommand*{\INFNCATindex}{14}
\affiliation{\INFNCAT}
\newcommand*{\MESSU}{Universit\`{a} degli Studi di Messina, 98166 Messina, Italy}
\newcommand*{\MESSUindex}{15}
\affiliation{\MESSU}
\newcommand*{\FAIR}{Fairfield Universiy, 1073 North Benson Rd, Fairfield, Connecticut 06824, USA}
\newcommand*{\FAIRindex}{16}
\affiliation{\FAIR}
\newcommand*{\SACLAY}{IRFU, CEA, Universit\'{e} Paris-Saclay, F-91191 Gif-sur-Yvette, France}
\newcommand*{\SACLAYindex}{17}
\affiliation{\SACLAY}
\newcommand*{\INFNRO}{INFN, Sezione di Roma Tor Vergata, 00133 Rome, Italy}
\newcommand*{\INFNROindex}{18}
\affiliation{\INFNRO}
\newcommand*{\ROMAII}{Universit\`{a} di Roma Tor Vergata, 00133 Rome, Italy}
\newcommand*{\ROMAIIindex}{19}
\affiliation{\ROMAII}
\newcommand*{\JLU}{II Physikalisches Institut der Universit\"{a}t Gie\ss{}en, 35392 Gie\ss{}en, Germany}
\newcommand*{\JLUindex}{20}
\affiliation{\JLU}
\newcommand*{\HFHF}{Helmholtz Forschungsakademie Hessen f\"ur FAIR (HFHF) GSI Helmholtzzentrum f\"ur Schwerionenforschung Campus Gie\ss{}en, D-35392 Gie\ss{}en, Germany}
\newcommand*{\HFHFIindex}{21}
\affiliation{\HFHF}
\newcommand*{\GWUI}{The George Washington University, Washington, DC 20052, USA}
\newcommand*{\GWUIindex}{22}
\affiliation{\GWUI}
\newcommand*{\UCONN}{University of Connecticut, Storrs, Connecticut 06269, USA}
\newcommand*{\UCONNindex}{23}
\affiliation{\UCONN}
\newcommand*{\FIU}{Florida International University, Miami, Florida 33199, USA}
\newcommand*{\FIUindex}{24}
\affiliation{\FIU}
\newcommand*{\MSU}{Skobeltsyn Institute of Nuclear Physics and Physics Department, Lomonosov Moscow State University, 119234 Moscow, Russia}
\newcommand*{\MSUindex}{25}
\affiliation{\MSU}
\newcommand*{\MISS}{Mississippi State University, Mississippi State, MS 39762-5167, USA}
\newcommand*{\MISSindex}{26}
\affiliation{\MISS}
\newcommand*{\FERRARAU}{Universit\`{a} di Ferrara, 44121 Ferrara, Italy}
\newcommand*{\FERRARAUindex}{27}
\affiliation{\FERRARAU}
\newcommand*{\YORK}{University of York, York YO10 5DD, United Kingdom}
\newcommand*{\YORKindex}{28}
\affiliation{\YORK}
\newcommand*{\LAMAR}{Lamar University, 4400 MLK Blvd, P.O. Box 10046, Beaumont, Texas 77710, USA}
\newcommand*{\LAMARindex}{29}
\affiliation{\LAMAR}
\newcommand*{\YEREVAN}{Yerevan Physics Institute, 375036 Yerevan, Armenia}
\newcommand*{\YEREVANindex}{30}
\affiliation{\YEREVAN}
\newcommand*{\OHIOU}{Ohio University, Athens, Ohio  45701, USA}
\newcommand*{\OHIOUindex}{31}
\affiliation{\OHIOU}
\newcommand*{\ORSAY}{Universit\'{e} Paris-Saclay, CNRS/IN2P3, IJCLab, 91405 Orsay, France}
\newcommand*{\ORSAYindex}{32}
\affiliation{\ORSAY}
\newcommand*{\FSU}{Florida State University, Tallahassee, Florida 32306, USA}
\newcommand*{\FSUindex}{33}
\affiliation{\FSU}
\newcommand*{\INFNTUR}{INFN, Sezione di Torino, 10125 Torino, Italy}
\newcommand*{\INFNTURindex}{34}
\affiliation{\INFNTUR}
\newcommand*{\UNH}{University of New Hampshire, Durham, New Hampshire 03824-3568, USA}
\newcommand*{\UNHindex}{35}
\affiliation{\UNH}
\newcommand*{\URICH}{University of Richmond, Richmond, Virginia 23173, USA}
\newcommand*{\URICHindex}{36}
\affiliation{\URICH}
\newcommand*{\WM}{College of William and Mary, Williamsburg, Virginia 23187-8795, USA}
\newcommand*{\WMindex}{37}
\affiliation{\WM}
\newcommand*{\GLASGOW}{University of Glasgow, Glasgow G12 8QQ, United Kingdom}
\newcommand*{\GLASGOWindex}{38}
\affiliation{\GLASGOW}
\newcommand*{\VT}{Virginia Tech, Blacksburg, Virginia   24061-0435, USA}
\newcommand*{\VTindex}{39}
\affiliation{\VT}
\newcommand*{\KNU}{Kyungpook National University, Daegu 41566, Republic of Korea}
\newcommand*{\KNUindex}{40}
\affiliation{\KNU}
\newcommand*{\VIRGINIA}{University of Virginia, Charlottesville, Virginia 22901, USA}
\newcommand*{\VIRGINIAindex}{41}
\affiliation{\VIRGINIA}
\newcommand*{\NSU}{Norfolk State University, Norfolk, Virginia 23504, USA}
\newcommand*{\NSUindex}{42}
\affiliation{\NSU}
\newcommand*{\CUA}{Catholic University of America, Washington, DC 20064, USA}
\newcommand*{\CUAindex}{43}
\affiliation{\CUA}
\newcommand*{\RPI}{Rensselaer Polytechnic Institute, Troy, New York 12180-3590, USA}
\newcommand*{\RPIindex}{44}
\affiliation{\RPI}
\newcommand*{\INFNFR}{INFN, Laboratori Nazionali di Frascati, 00044 Frascati, Italy}
\newcommand*{\INFNFRindex}{45}
\affiliation{\INFNFR}
\newcommand*{\JMU}{James Madison University, Harrisonburg, Virginia 22807, USA}
\newcommand*{\JMUindex}{46}
\affiliation{\JMU}
\newcommand*{\NMSU}{New Mexico State University, P.O. Box 30001, Las Cruces, New Mexico 88003, USA}
\newcommand*{\NMSUindex}{47}
\affiliation{\NMSU}
\newcommand*{\UCR}{University of California Riverside, 900 University Avenue, Riverside, California 92521, USA}
\newcommand*{\UCRindex}{48}
\affiliation{\UCR}
\newcommand*{\CNU}{Christopher Newport University, Newport News, Virginia 23606, USA}
\newcommand*{\CNUindex}{49}
\affiliation{\CNU}
\newcommand*{\CSUDH}{California State University, Dominguez Hills, Carson, California 90747, USA}
\newcommand*{\CSUDHindex}{50}
\affiliation{\CSUDH}
\newcommand*{\GSI}{GSI Helmholtzzentrum f\"ur Schwerionenforschung GmbH, D-64291 Darmstadt, Germany}
\newcommand*{\GSIIindex}{51}
\affiliation{\GSI}
\newcommand*{\CMU}{Carnegie Mellon University, Pittsburgh, Pennsylvania 15213, USA}
\newcommand*{\CMUindex}{52}
\affiliation{\CMU}
\newcommand*{\CANISIUS}{Canisius College, Buffalo, New York 14208, USA}
\newcommand*{\CANISIUSindex}{53}
\affiliation{\CANISIUS}


\newcommand*{\NOWISU}{Idaho State University, Pocatello, Idaho 83209, USA.}
\newcommand*{\NOWJLAB}{Thomas Jefferson National Accelerator Facility, Newport News, Virginia 23606, USA.}

\author{Y.~Tian }
\thanks{Corresponding author: yetian@syr.edu}
\affiliation{\SCAROLINA}
\affiliation{\SU}
\author {R.W.~Gothe} 
\affiliation{\SCAROLINA}
\author{V.I.~Mokeev}
\affiliation{\JLAB}
\author{G.~Hollis}
\affiliation{\SCAROLINA}
\author {M.J.~Amaryan} 
\affiliation{\ODU}
\author {W.R.~Armstrong} 
\affiliation{\ANL}
\author {H.~Atac} 
\affiliation{\TEMPLE}
\author {H.~Avakian} 
\affiliation{\JLAB}
\author {L. Barion} 
\affiliation{\INFNFE}
\author {M.~Battaglieri} 
\affiliation{\INFNGE}
% ...

\collaboration{The CLAS Collaboration}


\begin{abstract}
New results for the exclusive and quasifree cross sections off neutrons bound in deuterium ${\gamma}_vn(p) \rightarrow p{\pi}^{-} (p)$ are presented over a wide final state hadron angle range with a kinematic coverage of the invariant mass ($W$) up to 1.825~GeV and the 
four-momentum transfer squared ($Q^{2}$) from 0.4 to 1.0~GeV$^2$. The exclusive structure functions were extracted and their Legendre moments were obtained. Final-state-interaction contributions have been kinematically separated from the extracted quasifree cross sections off bound neutrons solely based on the analysis of the experimental data. These new results will serve as long-awaited input for phenomenological analyses to extract the $Q^{2}$ evolution of previously unavailable 
$n \to N^{*}$ electroexcitation amplitudes and to improve state-of-the-art models of neutrino scattering off nuclei by augmenting the already available results from free protons.
\end{abstract}

\maketitle


\section{Introduction}
The studies on nucleon resonance electroexcitation amplitudes (also referred to as $\gamma_vpN^*$ electrocouplings or transition form factors) from the data on exclusive meson electroproduction off protons have been proven to be an effective tool in the exploration of the nucleon resonance ($N^*$) structure \cite{Aznauryan:2011qj,Burkert:2019kxy,Burkert:2017djo,Mokeev:2020vab}. These studies have provided unique information on many facets of the strong interaction dynamics in the region where the QCD running coupling and the emergence of hadron mass are largest. This so-called strong QCD (sQCD) regime defines the manifestation of all nucleon excited states with various quantum numbers and distinctively different structures \cite{Carman:2020qmb}. It makes the exploration of nucleon resonance electroexcitations an important direction in contemporary hadron physics that focuses on gaining insights into sQCD from the experimental results for the spectrum of the ground and excited hadron states and their characteristic structures \cite{Brodsky:2020vco,Barabanov:2020jvn}.

The CLAS detector at Jefferson Lab \cite{Mecking:2003zu} has provided the dominant part of all available experimental results on  differential cross sections and polarization asymmetries for exclusive meson electroproduction off protons in the resonance region at invariant masses $W\leqslant$ 2.01~GeV and photon virtualities $Q^2\leqslant$ 6.0~GeV$^2$ \cite{Aznauryan:2011qj,Carman:2020qmb}. The numerical data on the measured observables are stored in the CLAS Physics Database~\cite{CLAS:DB}. The wealth of the experimental data from CLAS enabled us in this kinematic regime to determine the $\gamma_vpN^*$ electrocouplings of most nucleon resonance states based on independent studies of the exclusive $\pi^+n$, $\pi^0p$ \cite{Ungaro:2006df,Egiyan:2006ks,Aznauryan:2009mx,Park:2014yea}, $\eta p$ \cite{Denizli:2007tq}, and $\pi^+\pi^-p$ \cite{Mokeev:2012vsa,Mokeev:2015lda,Burkert:2019opk,Mokeev:2020hhu} electroproduction channels.
Consistent $\gamma_vpN^*$ electrocoupling results obtained in these independent studies make it possible to establish systematic uncertainties for the extraction of these quantities imposed by the reaction models.


The CLAS results on the $\gamma_vpN^*$ electrocouplings have had a considerable impact on the exploration of the excited nucleon state structure. It was found that all nucleon resonance structures studied so far are consistent with an interplay between the inner core of three dressed quarks and an external meson-baryon cloud. This conclusion is based on independent studies of the $Q^2$ evolution of $\gamma_vpN^*$ electrocouplings within quark models \cite{Aznauryan:2018okk,Obukhovsky:2019xrs,Lyubovitskij:2020gjz,Giannini:2015zia,Ramalho:2018wal} and the advanced coupled-channels approach developed by the Argonne-Osaka group \cite{Kamano:2011ut,Suzuki:2010yn}.

 Coupled-channels approaches in general are making progress towards the extraction of the $\gamma_vpN^*$ electrocouplings from combined analyses of meson photo-, electro-, and hadroproduction data. Recently, the $\pi N$ and $\eta p$ electroproduction multipoles, which are directly related to the $\gamma_vpN^*$ electrocouplings, were determined from CLAS data within a multichannel analysis \cite{Mai:2021vsw,Mai:2021aui}.

A successful description of the $\Delta(1232)3/2^+$ and $N(1440)1/2^+$ $\gamma_v pN^*$ electrocouplings  has been achieved at $Q^2>$ 0.8~GeV$^2$ and $Q^2>$ 2.0~GeV$^2$, respectively, by a continuum QCD approach with a traceable connection to the QCD Lagrangian \cite{Segovia:2014aza, Segovia:2015hra}. The $\gamma_vpN^*$ electrocouplings of these resonances are well reproduced by employing the same QCD-inferred momentum-dependent dressed quark mass function  \cite{Roberts:2021nhw} that was also used for the successful description of the pion and nucleon elastic form factors \cite{Segovia:2014aza,Horn:2016rip}. This success demonstrates the ability to gain insights into the dynamical hadron mass generation from combined studies of the pion, nucleon elastic, and $N \to N^*$ transition form factors. Therefore, further studies on various nucleon resonance electroexcitations are of particular importance in order to address the key open problem of the Standard Model on the emergence of hadron mass \cite{Brodsky:2020vco,Roberts:2020hiw}.

Currently available data on exclusive meson electroproduction in the resonance region are limited to the results off hydrogen targets, as the results on exclusive meson production off bound neutrons are mostly limited to photoproduction data only \cite{Ireland:2019uwn}. As a consequence, only photocouplings for resonance excitations off bound neutrons are currently available \cite{Kamano:2016bgm, Anisovich:2017afs}. The experimental results on the $\gamma_vnN^*$ electrocouplings of bound neutrons are of particular importance for the isospin decomposition of the electromagnetic $N \rightarrow N^*$ transition currents, addressing important open problems in the exploration of the $N^*$ structure and sQCD dynamics that underline $N^*$ generation from quarks and gluons.

Analyses of the $\gamma_vpN^*$ electrocouplings demonstrated that while the relative contributions from the meson-baryon cloud to the $N^*$ electrocouplings decreases with $Q^2$ towards quark core dominance at high $Q^2$, the interplay between the meson-baryon cloud and quark core  depends substantially on the resonance spin, parity, and isospin projection. For instance, the meson-baryon cloud contribution to the $A_{1/2}$ $\gamma_vpN^*$ electrocoupling of the  $N(1440)1/2^+$ resonance changes from being substantial at $Q^2<$ 1.0~GeV$^2$ to being negligible at $Q^2>$ 2.0~GeV$^2$. In contrast, contributions from the meson-baryon cloud to the $A_{1/2}$ electroexcitation amplitude of the $N(1520)3/2^-$ resonance remain modest over the entire $Q^2$ range covered by the measurements \cite{Mokeev:2015lda,Burkert:2019opk}. The electroexcitation of the $N(1675)5/2^-$ resonance is expected to demonstrate a pronounced dependence on the isospin projection \cite{Aznauryan:2014xea}. While the  $A_{1/2}$ electroexcitation amplitude of the $N(1675)5/2^-$ off protons is dominated by the meson-baryon cloud, the corresponding $A_{1/2}$ amplitude off neutrons is expected to be determined by a more complex interplay between the inner core of three dressed quarks and the external meson-baryon cloud. This shows that the combined studies of $N^*$ electroexcitation off both free protons and bound neutrons are of particular importance in order to explore the emergence of the meson-baryon cloud in the sQCD regime and other isospin breaking effects. 

First predictions of the $Q^2$ evolution for the $\gamma_vnN^*$ electrocouplings and a light-quark flavor separation have become available in the continuum QCD approach \cite{Rodriguez-Quintero:2019yec,Segovia:2019jdk}. A successful description of the measured $\gamma_vNN^*$ electrocouplings off free protons and quasifree neutrons with the same dressed quark mass function will further validate the credible insight into the hadron mass generation dynamics.

The studies of $\pi^-p$ photo- and electroproduction off bound neutrons play an important role in addressing these open problems in $N^*$ physics. Differential $\pi^-p$ photoproduction cross sections off deuterons in the resonance region have been measured with CLAS \cite{CLAS:2017dco} over a wide range of final state pion emission angles in the center-of-mass (CM) frame. Substantial progress has been achieved in reaction models accounting for the $\pi N$ final state interaction (FSI) within deuterons \cite{Tarasov:2015sta,Tarasov:2011ec,Nakamura:2018cst,Kamano:2016bgm,Anisovich:2017afs,briscoe2021photoproduction}. Previously published results on $\pi^-p$ electroproduction off bound neutrons at photon virtualities covered by our measurements of $Q^2<$ 1.0~GeV$^2$ \cite{Morris:1979mj, Wright:1980ff, Gaskell:2001fn} are scarce and have very limited pion azimuthal angle coverage in the CM frame, making it virtually impossible to determine exclusive structure functions from these measurements.

In Sec.~\ref{results} we present differential cross sections and virtual photon polarization dependent structure functions for the exclusive $\pi^-p$ electroproduction off bound neutrons in the reaction,
\begin{equation}
{\gamma}_{v}+D\rightarrow \pi^{-}+p+p_{s}\; ,
\label{eq:pdpim}
\end{equation}
where $p_s$ is the spectator proton in the deuteron. This process has been measured with the CLAS detector at Jefferson Lab during the ``e1e" run period within the kinematically accessible region of $W<$ 1.825~GeV and photon virtualities 0.4 $<Q^2<$ 1.0~GeV$^2$. The experiment conditions and the data analysis procedures are described in Secs.~\ref{sec:level3}-\ref{sec:cros}. The Legendre moments of the exclusive structure functions have also been extracted by analyzing their polar angle distributions; see Sec.~\ref{sec:legendre}. The results on pion electroproduction off both protons \cite{PhysRevC.101.015208} and deuterons \cite{YeTian-thesis,Gary-thesis} have now become available under the same experimental conditions. The latter offers additional opportunities to investigate pion electroproduction off bound protons and bound neutrons in detail and to minimize the impact of the initial and final state interactions within deuterons on the measured observables. The obtained results presented here provide experimental input for the phenomenological extraction of the nucleon resonance electroexcitation amplitudes off bound  neutrons (see Sec.~\ref{sec:sum}). 


For the kinematics of the scattering process off a bound moving neutron in a deuteron, we have to consider the influence of Fermi motion, off-shell effects, and the final state interactions on the measured cross sections. These effects are introduced next. 

\subsection{Fermi motion}
In the process of Eq.~\eqref{eq:pdpim}, the initial state neutron is moving around in the deuteron rest frame or in the laboratory frame. 
Due to energy and momentum conservation, the sums of the four-momenta of the initial virtual photon $q^\mu$ and deuteron-taget $D^{\mu}$ and of the final hadrons should be equal,   
\begin{equation}
\begin{split}
q^{\mu}+D^{\mu}&=(\pi^{-})^{\mu}+p^{\mu}+p^{\mu}_{s} \; \text{or}\\
q^{\mu}+p^{\mu}_{i}+n^{\mu}&=(\pi^{-})^{\mu}+p^{\mu}+p^{\mu}_{s}\;,
\end{split}
\label{eq:Pdu}
\end{equation}
where $q^{\mu}$
is the four-momentum of the virtual photon and $D^{\mu}$=($m_{D},\vec{0}$) is the 
four-momentum of the deuteron in the laboratory frame, while $n^{\mu}$ and $p^{\mu}_{i}$ correspond to the four-momenta of the initial state neutron and proton, respectively, which are moving and loosely bound in the deuteron. $p^{\mu}$ stands for the four-momentum of the final state proton in the $\pi^-p$ electroproduction channel off bound neutrons.
The four-momentum of the outgoing proton from the deuteron breakup, $p^{\mu}_{s}$, was not directly measured in CLAS due to its acceptance limitation for protons to momenta larger than 0.25 GeV, but it can be fully reconstructed 
by
\begin{equation}
p^{\mu}_{s}=q^{\mu}+D^{\mu}-(\pi^{-})^{\mu}-p^{\mu}, 
\label{eq:Ps4p}
\end{equation}
and hence the three-momentum of this proton in the laboratory frame (i.e., the deuteron rest frame) is determined by
\begin{equation}
\vec{p}_{s}= \vec{q}-\vec{\pi}^{-}-\vec{p}. 
\label{eq:PsP}
\end{equation}
For the quasifree process of the reaction in Eq.~\eqref{eq:pdpim},
where the initial state proton is treated as a ``spectator'' that is totally unaffected by the interaction; 
it follows that $\vec{p}_{i}=\vec{p}_{s}$ in the laboratory frame, and 
ignoring the off-mass-shell effects, 
we can rewrite Eq.~\eqref{eq:Pdu} as 
\begin{equation}
q^{\mu}+n^{\mu}=(\pi^{-})^{\mu}+p^{\mu}\;,
\label{eq:Pnu}
\end{equation}  
and the initial state neutron momentum is reconstructed by
\begin{equation}
\vec{n}= \vec{\pi}^{-}+\vec{p}-\vec{q}. 
\label{eq:Pnmu}
\end{equation}
For the quasifree process, by comparing Eq.~\eqref{eq:PsP} with Eq.~\eqref{eq:Pnmu}, we get 
\begin{equation}
\vec{p}_{s}=\vec{p}_{i}= -\vec{n}. 
\label{eq:Psn}
\end{equation}
The Fermi motion causes changes in the kinematics compared to scattering off a neutron at rest frame. 

\subsection{Off-shell effects}
\begin{figure}[hbt!]
\centering
\includegraphics[width=0.5\textwidth,height=0.4\textwidth]{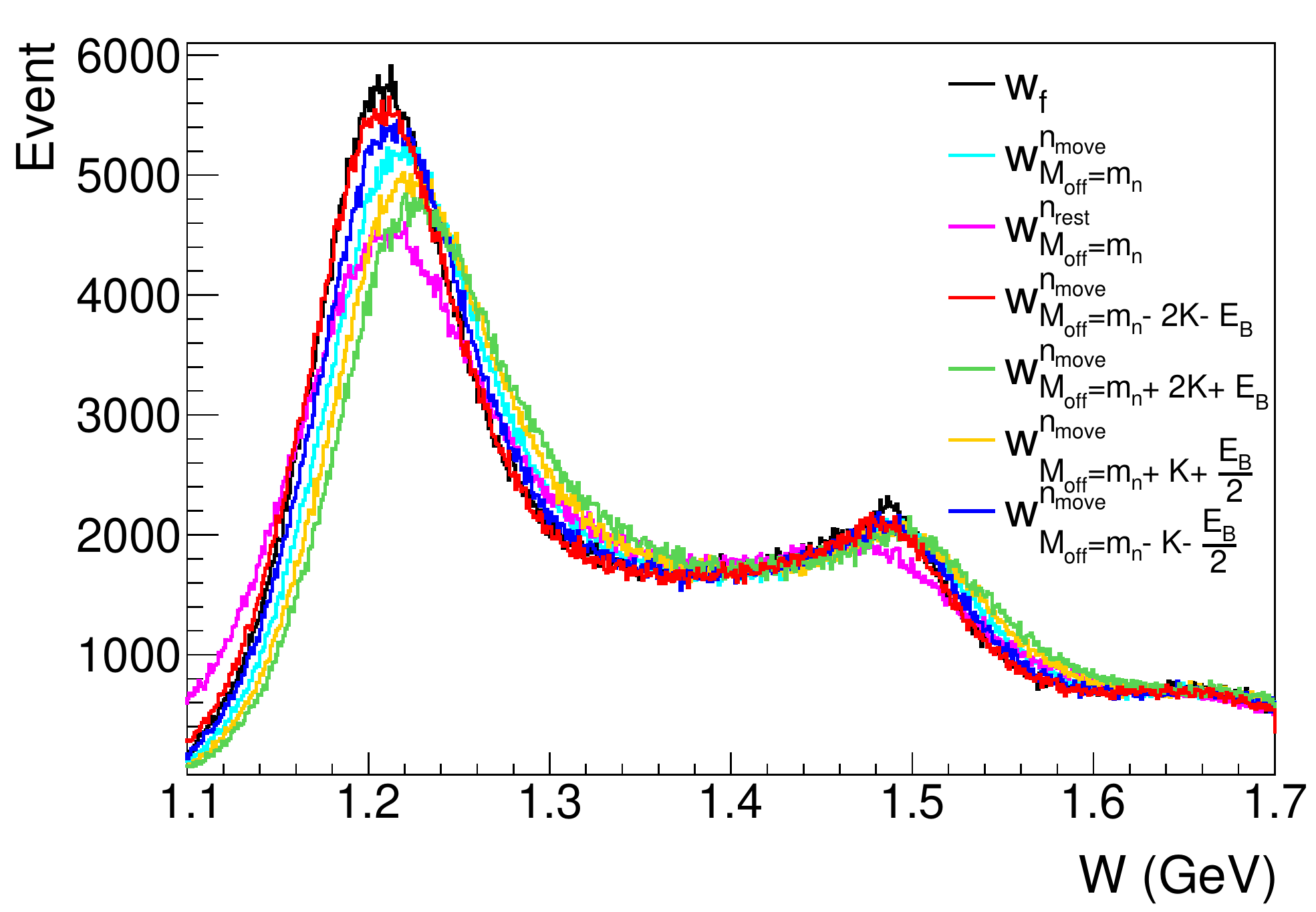}
\caption{\justify{The comparison of $W$ distributions for  $\pi^-p$ electroproduction events in quasifree kinematics within the entire $Q^2$ range covered by this measurement from 0.4 to 1.0 GeV$^2$, where the kinematic energy K$=\frac{p^{2}_{s}}{2m_{n}}$. The black line presents $W_{f}$ calculated from the measured four-momenta of the final $\pi^-$ and proton. The red line shows $W_{i}$ calculated by setting $n^{\mu}=(E_{n},\vec{n})$ with $E_{n}=\sqrt{(-\vec{p}_{s})^{2}+(M_{off})^{2}}$ and $M_{off}$ according to Eq.~\eqref{eq:En} achieving best agreement with $W_{f}$.}} 

\label{fig:Wcompare}       
\end{figure}

As mentioned previously, the bound neutron is also off-mass-shell in addition to moving around in the deuteron. 
Even in the quasifree process $p^{\mu}_{i}$ is not equal to $p^{\mu}_{s}$, because the initial state proton $p_{i}$ is off mass shell while the outgoing ``spectator'' proton $p_{s}$ is on mass shell; see also Eq.~\eqref{eq:pdpim}. 
In this process the relation $\vec{p}_{i}=\vec{p}_{s}=-\vec{n}$ is not influenced by the off-mass-shell effects. The off-shell neutron four-momentum can be best approximated by  $n^{\mu}=(E_n,-\vec{p}_{s})$ with $E_{n}=\sqrt{(\vec{n})^{2}+(M_{off})^{2}}$. To avoid uncertainties due to the off-shell-ness of the target neutron when presenting the final cross section, it is better to choose the invariant mass as $W_{f}$, which is directly measured and well defined by $W_{f}^{2}=(p^{\mu}+(\pi^{-})^{\mu})^{2}$, rather than as $W_{i}$ defined by $W_{i}^{2}=(q^{\mu}+n^{\mu})^{2}$, which is affected by the off-shellness of the target nucleon. Regarding the ``spectator,'' in order to best conserve energy and momentum in the scattering process, we have set the off-shell mass of the neutron to


\begin{equation}
M_{off}=m_{n}-2\frac{p^{2}_{s}}{2m_{n}}-E_{B},
\label{eq:En}
\end{equation}
reestablishing $W_{i}=W_{f}$. Here $m_{n}$ is mass of the free neutron and $E_{B}$ is the binding energy of deuteron. Other possible choices for $M_{off}$ lead to shifted and/or smeared $W_i$ distributions when compared to $W_f$, as shown in Fig.~\ref{fig:Wcompare}.


\subsection{Final state interactions (FSI)}
\begin{figure}[!hbt]
\includegraphics[width=0.5\textwidth,height=0.35\textwidth]{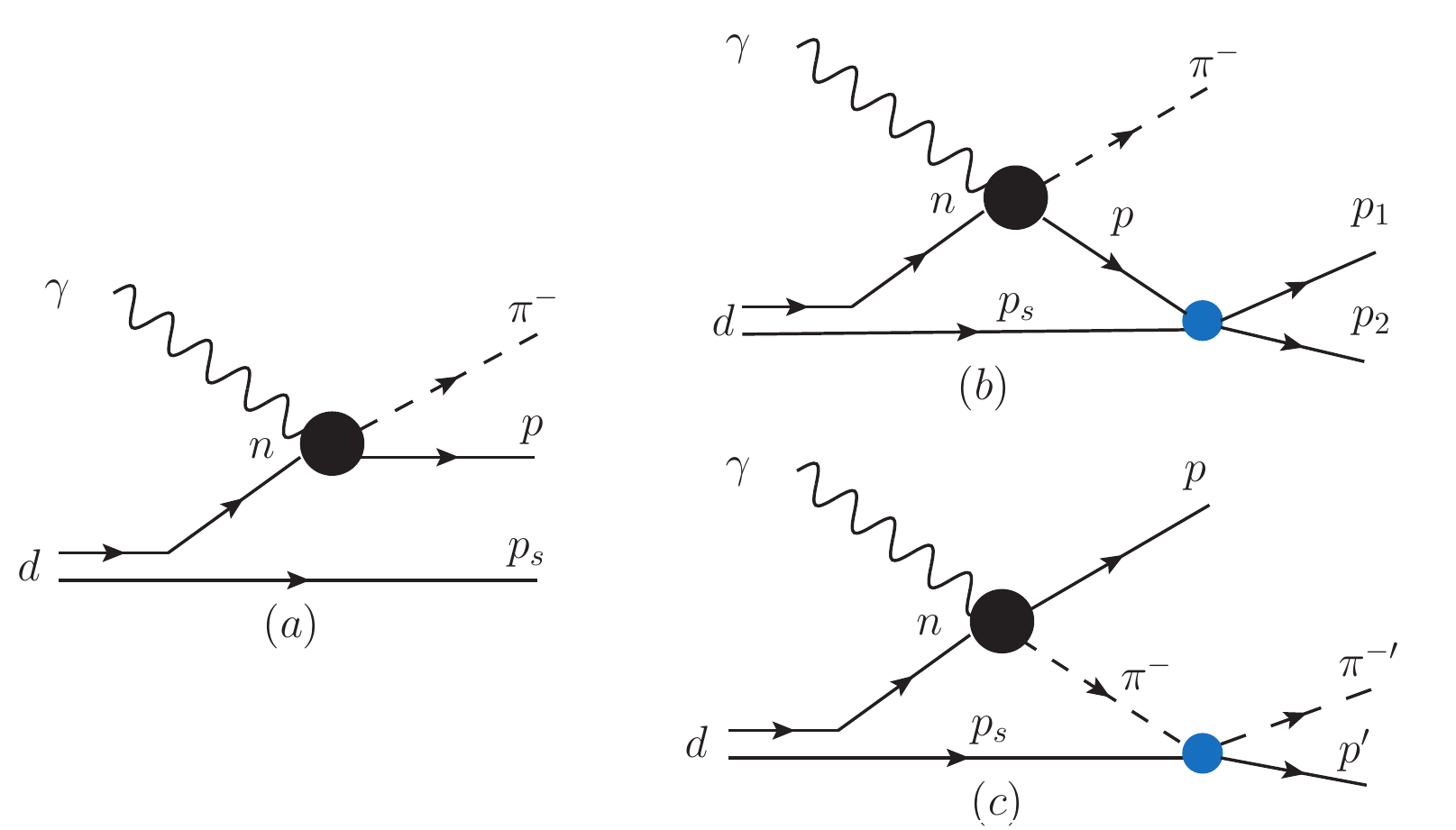}
\caption{\justify{\label{fig:KFSI}Kinematic sketch of the three leading terms in the ${\gamma}_v+D\rightarrow \pi^{-}+p+p$ process: (a) impulse approximation, (b) $pp$ rescattering, and (c) $\pi^- p$ rescattering. Diagrams (b) and (c) are the two main sources of final state interactions.}}      
\end{figure}

The full exclusive reaction process of interest is described by Eq.~\eqref{eq:pdpim}, but for $\lvert\vec{p}_{s}\lvert<200\;\text{MeV}$, the quasifree process, which is depicted by the impulse approximation diagram in Fig.~\ref{fig:KFSI}(a), is dominant (see Sec.~\ref{subsec:level4-2}). However, in the full exclusive process it is also possible to have final state interactions, such as $pp$ rescattering and $p\pi$ rescattering, shown in Figs.~\ref{fig:KFSI}(b) and~\ref{fig:KFSI}(c), respectively. These processes correspond to the situation in which the outgoing proton or $\pi^{-}$ interacts with the spectator proton ($p_{s}$). Thus, the four-momenta of the final state particles are changed due to these final state interactions. It is also possible to have other kinds of FSI in the pion production process off the deuteron, such as $\pi^{0}+n_{s}\rightarrow \pi^{-}+p$ and $\pi^{-}+p_{s}\rightarrow \pi^{0}+n$, which can increase or decrease the final state $\pi^{-}p$ production. In this paper, these kinds of final state interactions are not further quantified, since the main interest here focuses on the quasi-free cross section extraction. 

\subsection{Boosting of the kinematic variables}
In order to use the kinematic variables, directly corresponding to those describing pion electroproduction off free protons to present the final cross sections of ${\gamma}_v+D(n)\rightarrow \pi^{-}+p+p_{s}$, we first boost all particle four-momenta from the laboratory frame (deuterium rest frame) into the neutron rest frame with the boost vector $\vec{\beta}_{1}=-\vec{n}/E_{n}$, where $\vec{n}$ and $E_{n}$ are calculated as defined above. The four-momentum transfer $q^{\mu}= (\nu,\vec{q})$, and hence particularly the direction of $\vec{q}$ relative to which the angles $\theta_\pi$ and $\phi_\pi$ are defined, is then calculated in this frame. Therefore, the reported final $(W_{f}, Q^{2}, \cos\theta^\text{c.m.\!}_\pi, \phi^\text{c.m.\!}_\pi)$-dependent cross sections (where ``c.m.'' denotes when variables are calculated in the CM frame) are no longer influenced by the Fermi momentum of the initial state neutron in the deuteron, and the neutron rest frame coordinate system is defined by setting $\hat{z}$ parallel to the virtual photon direction and $\hat{y}$ perpendicular to the electron scattering plane with $\hat{x}$ staying in the electron scattering plane. Now we boost all particle four-momenta from the laboratory frame into the $\pi^{-}$p CM frame with the well defined boost vector $\vec{\beta}_{2}=-(\vec{p}+\vec{\pi}^{-})/(E_{p}+E_{\pi^{-}})$, and rotate $\hat{z}_\text{c.m.\!}$ back to $\hat{z}$ the virtual photon direction in the neutron rest frame to stay true to the proper polar and azimuthal angle definitions.

\begin{figure}[!hbt]
\centering
\includegraphics[width=0.5\textwidth,height=0.3\textwidth]{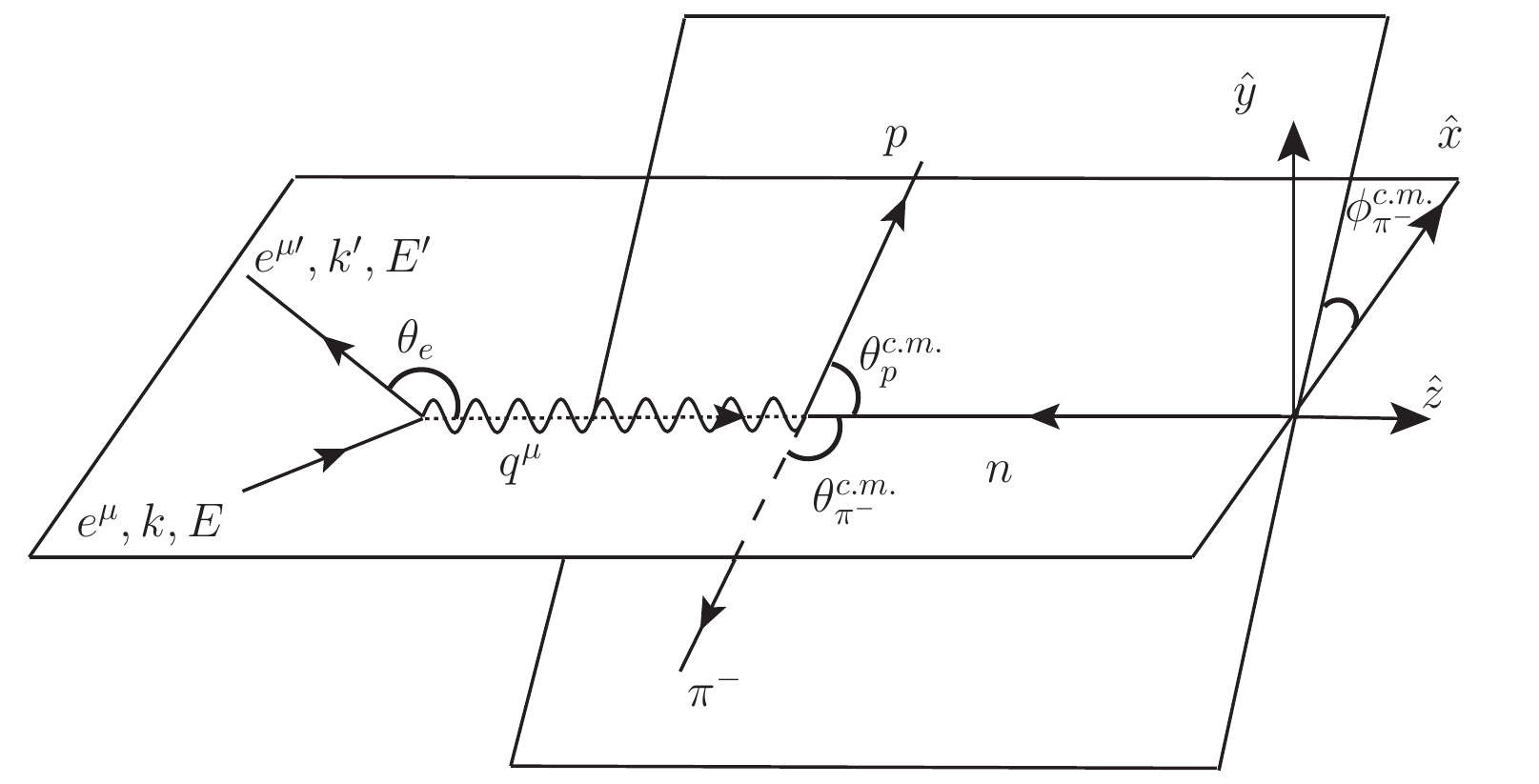}
\caption{\label{fig:kinematic}Schematics of $\pi^{-}$ electroproduction off a moving neutron.}      
\end{figure}

In summary, the coordinates are set by: 
\begin{enumerate}[label=(\arabic*)]
\item $\hat{z}$ in the direction of $\vec{q}$ in the $n$ rest frame, 
\item $\hat{x}$ in the $\vec{k}, \vec{k}^{'}$ plane of the $n$ rest frame and perpendicular to $\hat{z}$,
\item and $\hat{y}=\hat{z} \times \hat{x}$,  
\end{enumerate}
which are illustrated in Fig.~\ref{fig:kinematic}.

\section{\label{sec:level3}Experimental facility \protect\\ }

\begin{figure}[hbt!] 
\centering
\includegraphics[width=0.4\textwidth]{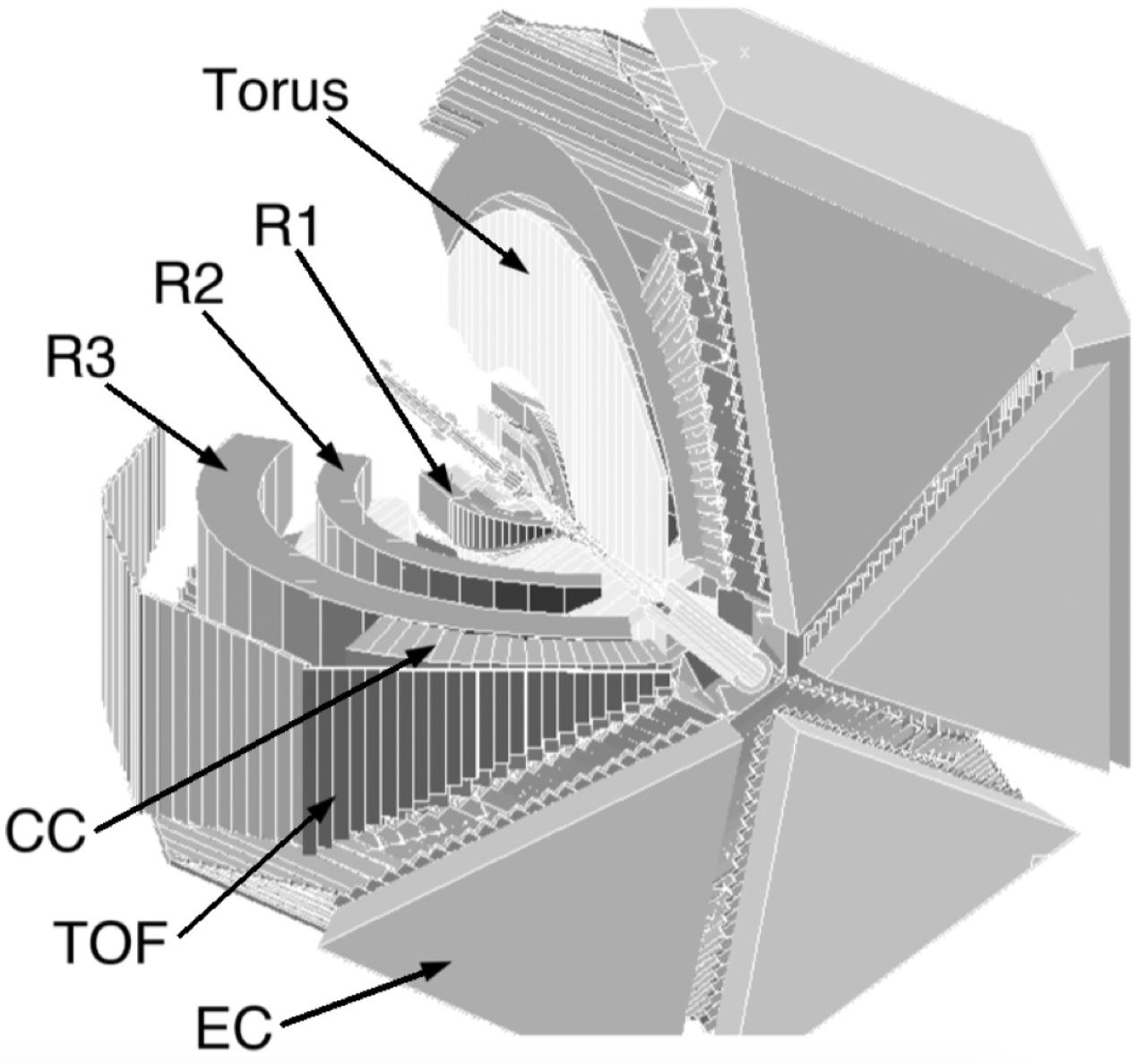}
\caption[Three-dimensional view of the CLAS detector]{\justify{Three-dimensional view of the CLAS detector cut along the beamline~\cite{Mecking:2003zu}, with EC--electromagnetic calorimeter, TOF--scintillation counter, CC--Cherenkov counter, three regions R1--R3 of drift chambers, and the torus magnet. 
}}
\label{fig:CLAS}       
\end{figure}

This experiment was carried out with the CEBAF Large Acceptance Spectrometer (CLAS)~\cite{Mecking:2003zu} (see Fig.~\ref{fig:CLAS}) in Hall B at Jefferson Laboratory.
The CLAS torus magnet coils naturally separated the detector into six identical and independent sectors. Each of the CLAS sectors was equipped with an identical set of detectors: three layers of drift chambers (DC) for charged particle tracking and momentum reconstruction, Cherenkov counters (CC) for electron identification and event triggering, scintillation counters (SC) for time-of-flight measurements and charged particle identification, and sampling-type electromagnetic calorimeters (EC) for refined electron identification and triggering.

\begin{figure}[hbt!] 
\centering
\includegraphics[width=0.4\textwidth]{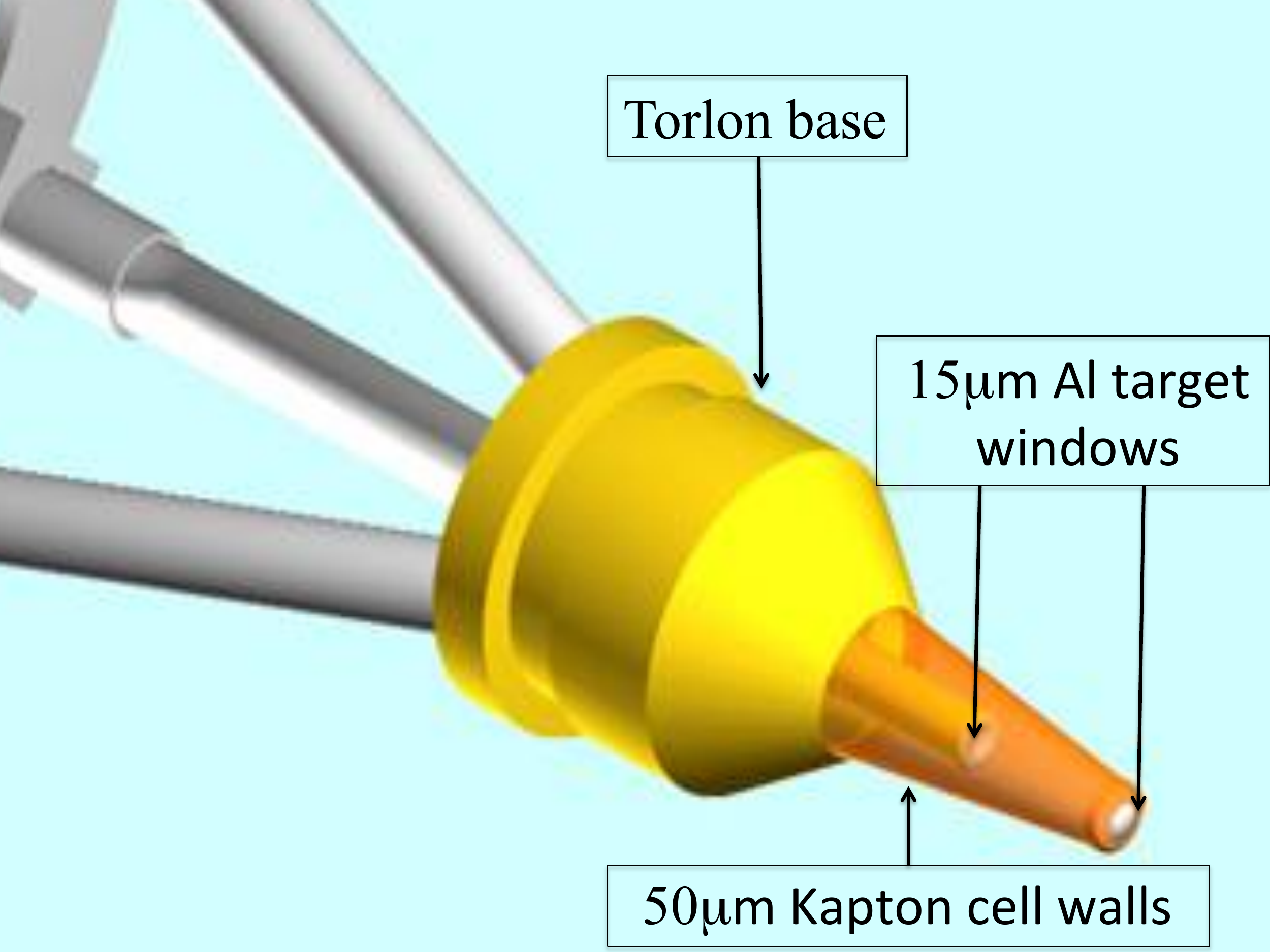}
\caption{A schematic diagram of the ``e1e'' target~\cite{e1etarget} indicating the target window positions.}
\label{fig:e1e}       
\end{figure}

This measurement was part of the ``e1e'' run period that started in the beginning of 2003. An electron beam with an energy of 2.039~GeV interacted with a 2-cm-long unpolarized liquid-deuterium target. The target had a conical shape with a diameter varying from 0.4 to 0.6~cm (see Fig.~\ref{fig:e1e}). 
Data were taken with a +2250~A torus current and +6000~A minitorus current (a small normal-conducting magnet to keep low momentum electrons produced by M{\o}ller scattering in the target from reaching the innermost drift chambers). Furthermore, empty-target runs were performed to measure contributions from all three target windows, which were used to subtract the contribution of the background events produced by the scattering of electrons on the 15~$\mu m$ target windows (see Fig.~\ref{fig:e1e}).

\begin{figure}[hbt!] 
\centering
\includegraphics[width=0.5\textwidth]{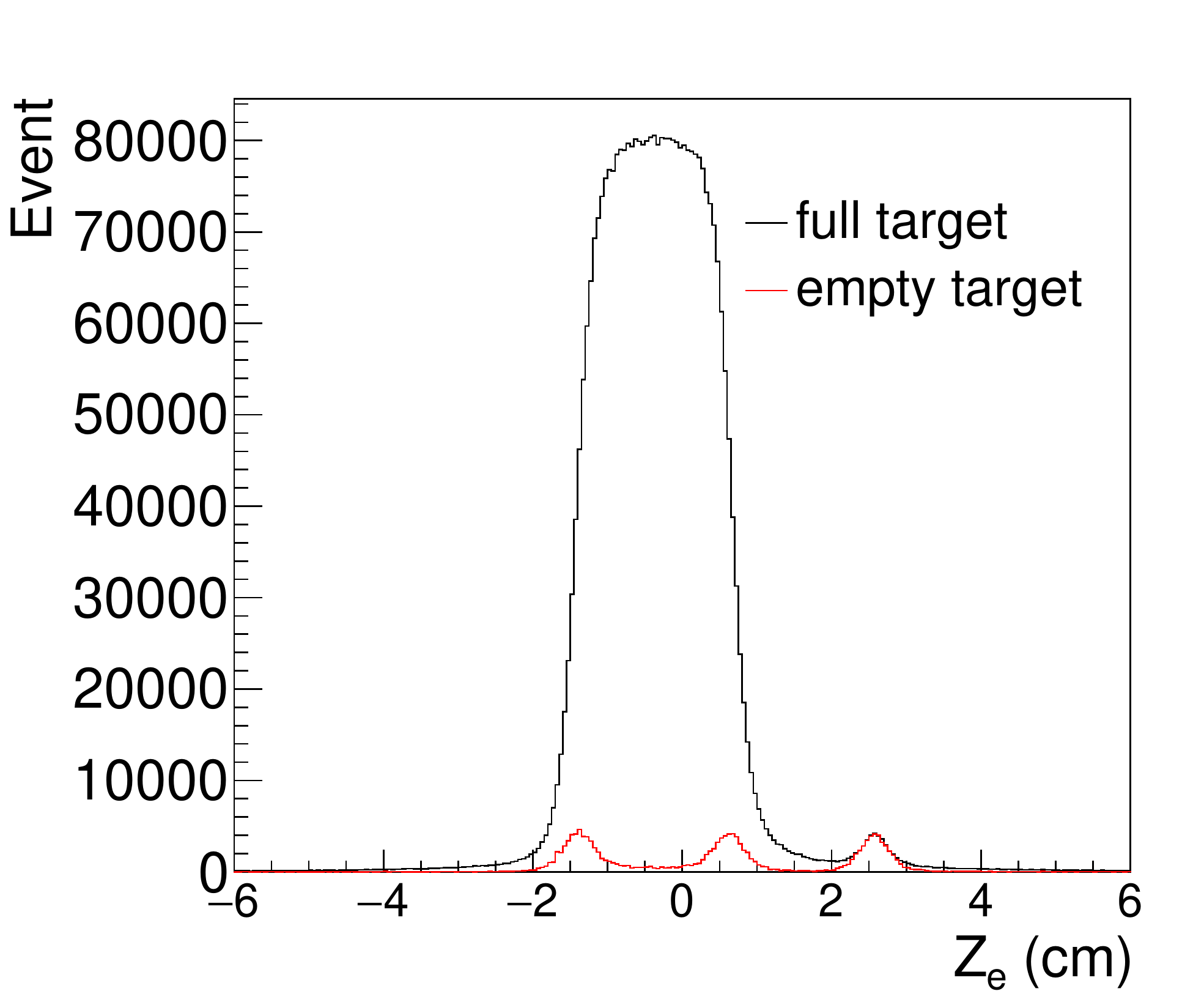}
\caption{Measured electron vertex ($Z_e$) distributions for full target events (black) and scaled empty target events (red).}
\label{fig:zvertexs}       
\end{figure}

The electron $z$-vertex distributions for full-target and empty-target events are compared, as shown in Fig.~\ref{fig:zvertexs}. Both distributions are normalized to the corresponding charge accumulated in the Faraday cup (FC), which
is located in the beam dump, $\approx 29$ meters downstream from the CLAS target. It completely stops the electrons and thus allows measurement of the accumulated charge of the incident beam.
There is a small peak at 2.58~cm due to the downstream 15-$\mu$m-thick aluminum foil of the target, which should be at the same position for both full-target and empty-target events neglecting thermal expansions  [
empty target $Z_e$ (red) presented in Fig.~\ref{fig:zvertexs}]. 

\section{\label{sec:data}Data Analysis }
\subsection{Electron identification}
An accepted electron candidates required geometrical matching of each negative DC track (inbending toward the beamline in the ``e1e" experiment) with the corresponding hits in the CC, SC, and EC detectors (see Fig.~\ref{fig:CLAS}). The overall EC energy resolution as well as uncertainties in the EC output from the summing electronics gave arise to the amplitude fluctuations of the EC response near the hardware threshold.
According to Ref.~\cite{note007}, the $P_e>461\;\text{MeV}$ cut was applied to the electron candidates to select reliable EC signals.
\begin{figure}[hbt!]
\includegraphics[width=0.5\textwidth]{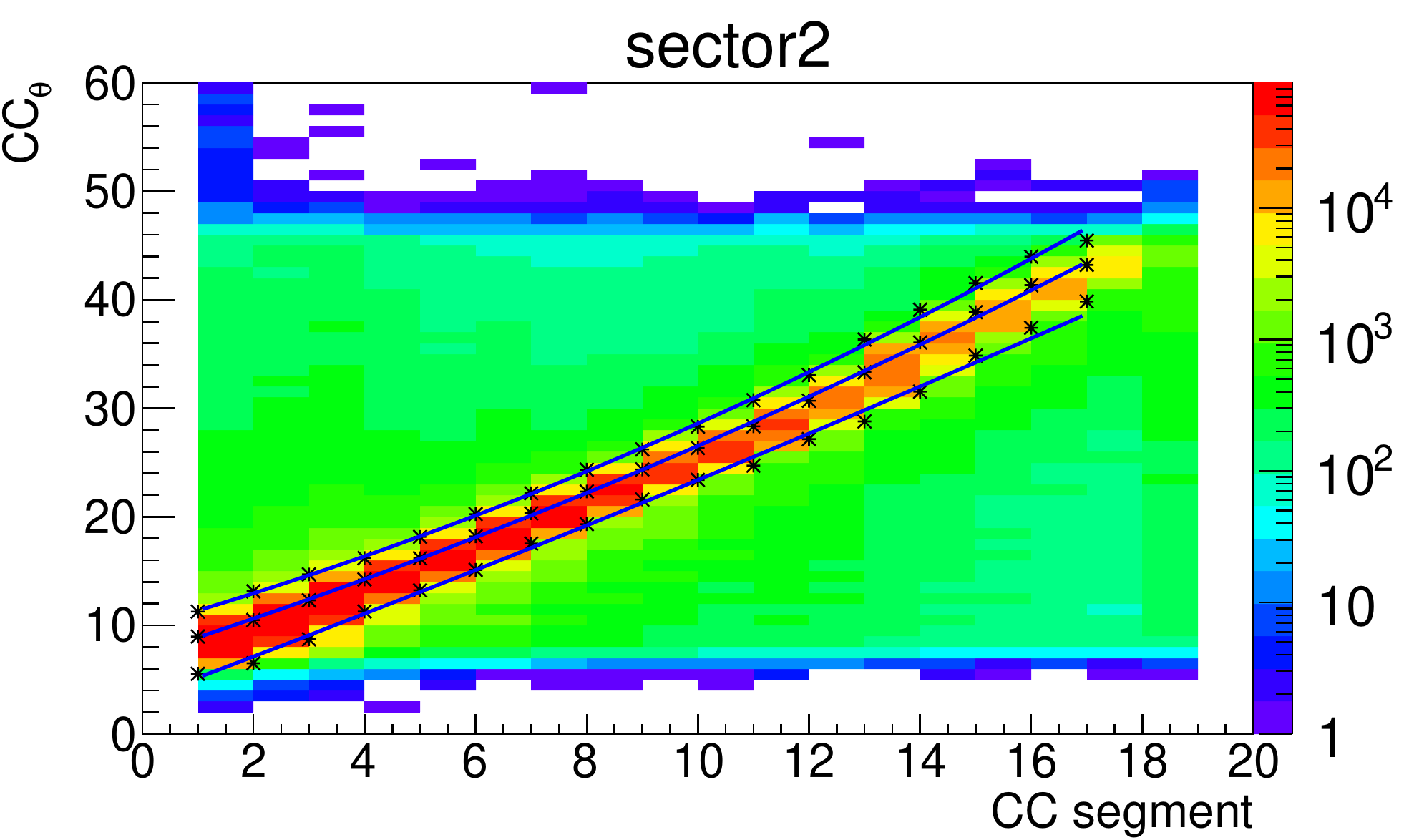}
\caption{\justify{$\theta_{CC}$ versus CC segment number histogram in sector 2, where $\mu$, $\mu+3\sigma$, and $\mu-4\sigma$ are marked as black stars and fit by second degree polynomial functions shown by the curves.}}
\label{fig:CCsegment}
\end{figure}

Furthermore, the torus magnetic field bent the electrons toward the beamline and the CC segments were placed radially relative to the CLAS polar angle, so there should be a one-to-one correspondence between $\theta_{CC}$ and the CC segment number for real electron tracks. The background and accidental tracks should not show such a correlation, as shown in Fig.~\ref{fig:CCsegment}. $\theta_{CC}$~\cite{referencenoteYe} can be calculated from
\begin{equation}
\theta_{CC}=\mathrm{arccos}\left(\frac{\mid{p_{z}}\mid}{\mid\vec{p}\mid}\right).
\label{eq:CCtheta}
\end{equation}
The $\theta_{CC}$ cuts shown by the outer lines in Fig.~\ref{fig:CCsegment} were applied to both experimental data and simulation.

\begin{figure}[htbp]
\centering
\includegraphics[width=0.5\textwidth]{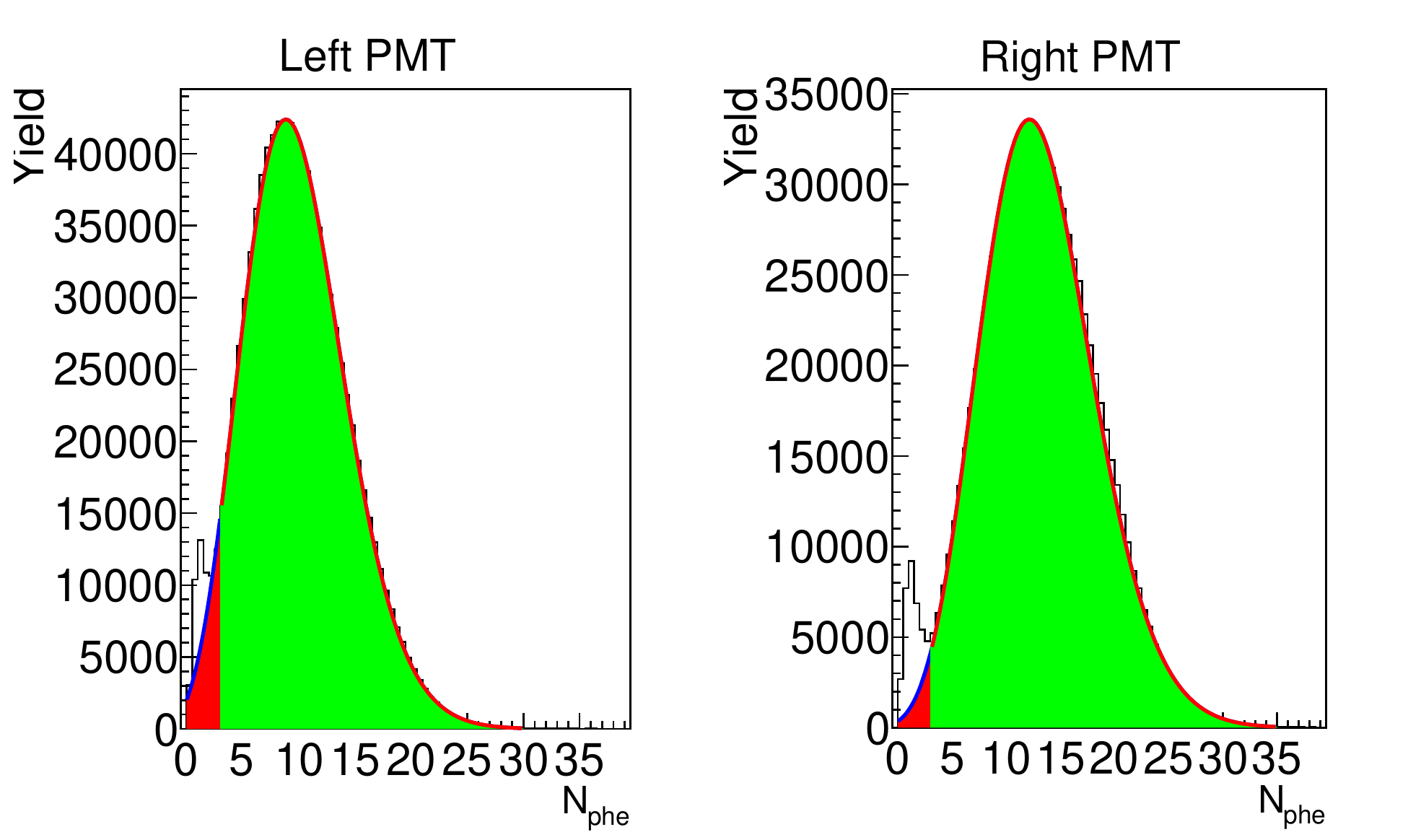}
\caption{\justify{$N_{phe}$ histograms of the left and right PMTs in the tenth CC segment of sector 2 plotted separately and fit by the Poisson function in Eq.~\eqref{eq:poisson} shown by the red curve.}}
\label{fig:Nphe}     
\end{figure}

In order to further reduce contributions from negative pions and other background tracks, cuts on the photoelectron yield $N_{phe}$ measured in the CC ($N_{phe}>3$)  
were applied on the electron candidates. In Fig.~\ref{fig:Nphe}, the green area under the Poisson fit function [from Eq.~\eqref{eq:poisson} shown by the red curve] corresponds to good electron candidates, and the small peak at $N_{phe}\approx 2$ contains not only background and negative pions, but also some good electron candidates beneath it. With the extrapolation of the fitted modified Poisson function, those lost candidates are quantified by the calculated red area, which can be accounted for by applying the correction factor ($N_{phe}^{correct}$) as a weight for each accepted event in this segment. The weight factor $N_{phe}^{correct}$ was calculated by
\begin{eqnarray}
N_{phe}^{correct}&&=\frac{green\;area}{red\;area+green\;area}\nonumber \\
&&=\frac{\int_{3}^{45}f(x)dx}{\int_{0}^{45}f(x)dx},
\label{eq:NpheC}
\end{eqnarray}
where $f(x)$ is the Poisson fit function (see red curves in Fig.~\ref{fig:Nphe}) defined as
\begin{equation}
f(x)=p_{0}\frac{p_{1}^{(\frac{x}{p_{2}})}e^{-p_{1}}}{\Gamma(\frac{x}{p_{2}}+1)},
\label{eq:poisson}
\end{equation}
where $p_{0}$, $p_{1}$, and $p_{2}$ are free fit parameters. 
The correction factor was determined by the $N_{phe}$ distributions of the left or right photomultiplier tube (PMT) in each CC segment. 

\begin{figure}[hbt!]
\centering
\includegraphics[width=1\linewidth]{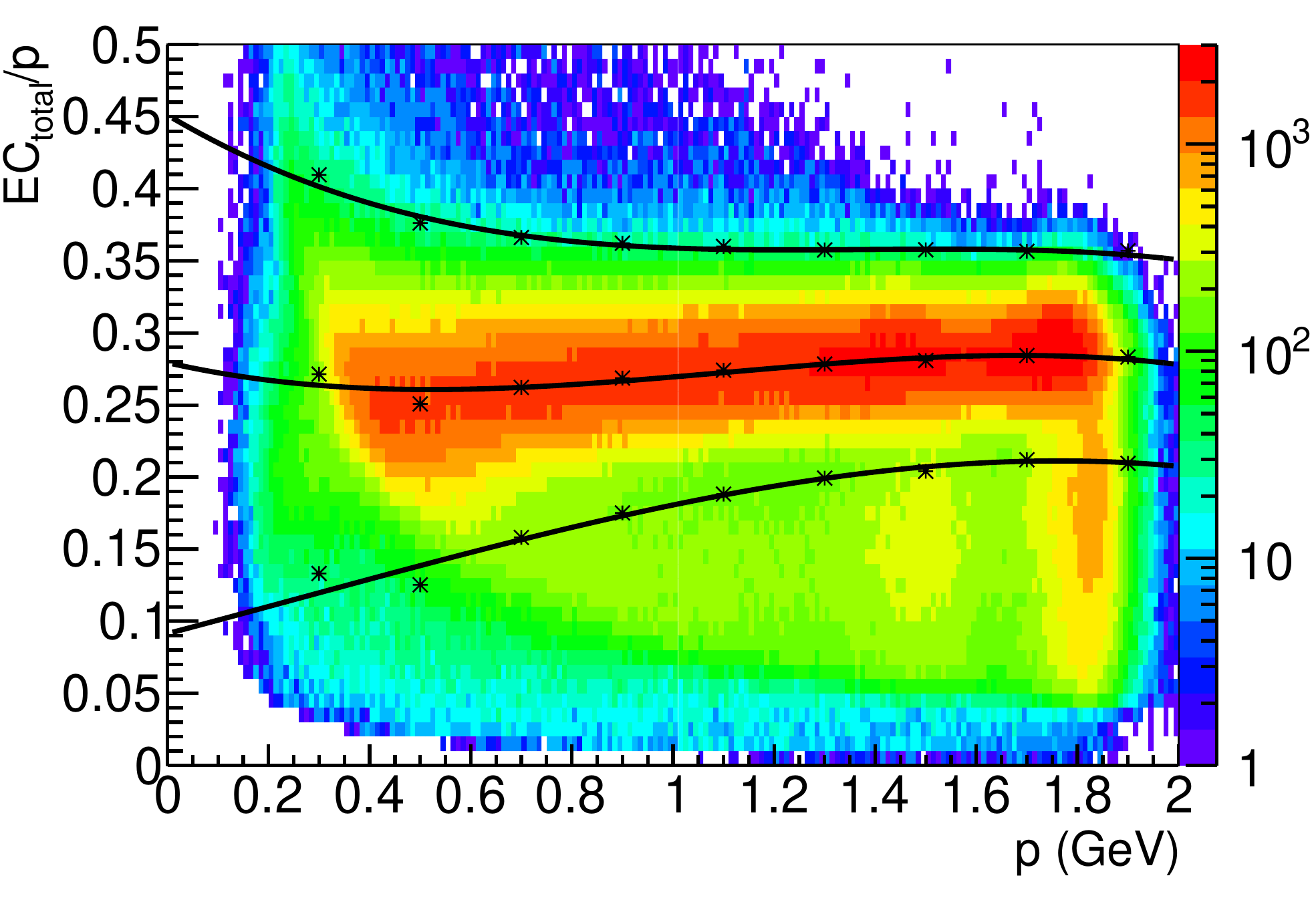}
\caption{$E_{total}/p$ versus $p$ distribution, where the black lines show the upper and lower $E_{total}/p$ cut limits. The events outside the cut limits correspond to minimum ionizing particles and background.}
\label{fig:sampF}
\end{figure}

The EC was used for separating the electrons from the fast-moving pions. Pions and electrons have different mechanisms of primary energy deposition in the EC. The energy deposition mechanism of an electron in the EC depends linearly on its momentum. Meanwhile, charged pions lose their energy largely due to ionization, which is not directly proportional to their momentum, resulting in much less energy deposited in the EC. Thus the measured deposited energy $E_{total}$ for showering electrons should be proportional to their momentum, resulting in a constant value of $E_{total}/p_e$ versus $p_e$. This sampling fraction (SF) for electrons in the EC is roughly 25\%, as shown in Fig.~\ref{fig:sampF}. In this analysis $\pm$3$\sigma$ cuts were placed on this distribution to select the scattered electrons, with separate cut limits determined for each sector of both data and simulation.

\subsection{Hadron identification}
The difference $\Delta T_{i}$ between the time calculated from the velocity and track length of the hadron candidate $i$ and the actual measured SC time $t_{i}^{SC}$ should naively peak at zero for the assumed hadron candidate. This time difference is defined by   
\begin{equation}
\Delta T_{i}  = \frac{l_{i}^{SC}}{\beta_{i}c}-t_{i}^{SC}+t_{0},
\label{eq:DeltaT}
\end{equation}
where $l_{i}^{SC}$ is the path length of the hadron candidate track from the vertex to the SC hit, and $\beta_{i}=\frac{v_{i}}{c}$ is the speed of the hadron candidate calculated from the momentum and the assumed rest mass $m_{i}$ of the hadron candidate given by
\begin{equation}
\beta_{i}  = \sqrt{\frac{p_{i}^{2}}{m_{i}^{2}c^{2}+p_{i}^{2}}},\
\label{eq:beta}
\end{equation}
and $t_{0}$ is the start time of each reconstructed event,
\begin{equation}
t_{0}  = t_{e}^{SC}-\frac{l_{e}^{SC}}{c}.
\label{eq:t0}
\end{equation}
Here $t_{e}^{SC}$ is the electron flight time measured from SC, $l_{e}^{SC}$ is the electron path length from the vertex to the SC hit, and $c$ is the speed of light. $t_{0}$ is used as the reference time for all remaining tracks in that event. 

\begin{figure}[hbt!]
\centering
\begin{subfigure}[b]{0.5\textwidth}
\centering
\includegraphics[width=1\linewidth]{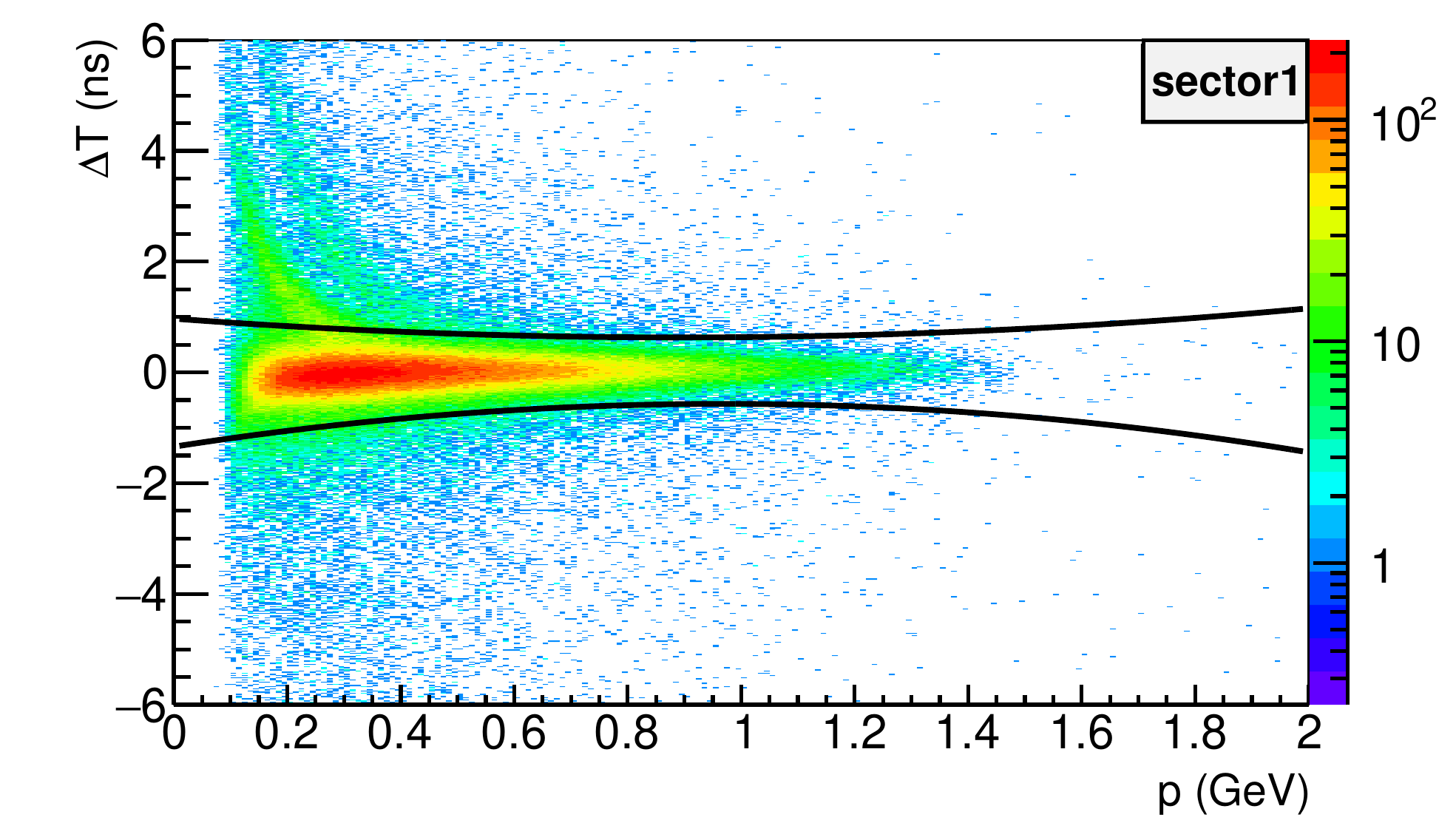}
\caption{}
\label{fig:pionDelTsector3}
\end{subfigure}
\begin{subfigure}[b]{0.5\textwidth}
\centering
\includegraphics[width=1\linewidth]{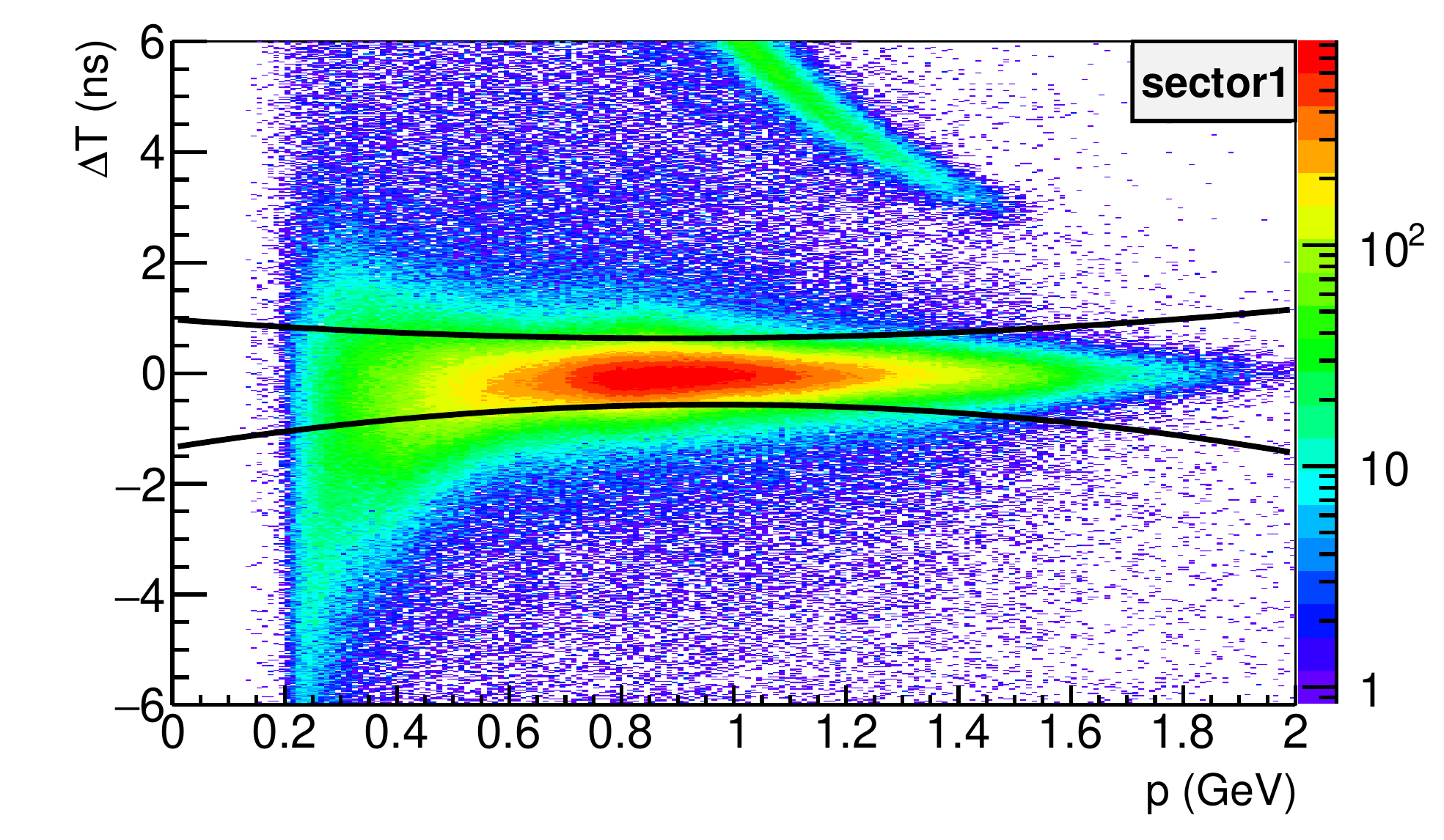}
\caption{}
\label{fig:protonDelTsector3}
\end{subfigure}
\caption{\justify{(a) Typical negative pion $\Delta T$ versus $p$ histogram for sector 1 with upper and lower $\Delta T$ cut limits. The bands above $\pi^{-}$ upper $\Delta T$ cut limit correspond to negative muons and electrons, separately. (b) Proton $\Delta T$ versus $p$ histogram for sector 1 with upper and lower $\Delta T$ cut limits. The band above proton upper cut limit corresponds to $\pi^{+}$.}}
\end{figure}

Figures~\ref{fig:pionDelTsector3} and~\ref{fig:protonDelTsector3} show typical distributions of $\Delta T_{i}$ versus momentum for $\pi^{-}$ and proton candidates, respectively. The solid black curves represent the corresponding $\Delta T$ cuts for hadron identiﬁcation, which were individually applied on the hadron candidates for each sector. 

During the ``e1e" run, some SC scintillation counters with low gain PMT
were removed from both experimental data and simulation. Additionally, it was found that some hadron candidates were shifted from the nominal position on the $\Delta T_{i}$ plots, which could be attributed to 
SC timing calibration inaccuracies. A special procedure was developed to correct the timing information for the affected SC counters~\cite{referencenoteYe}.

\subsection{\label{sec:level1}Kinematic corrections }
Due to our somewhat incomplete knowledge of the actual CLAS detector geometry and magnetic field distribution, which is therefore not precisely reproduced in the simulation process, a small momentum correction needs to be applied to the experimental data. From CLAS-Note 2003-012~\cite{note012}, it is known that momentum corrections are essential for only the highest momentum particles. For the ``e1e" run, with a beam energy of 2.039~GeV, the expected momentum corrections for hadrons are significantly less than for electrons and can be neglected. 
\begin{figure}[hbt!]
\centering
\includegraphics[width=0.45\textwidth]{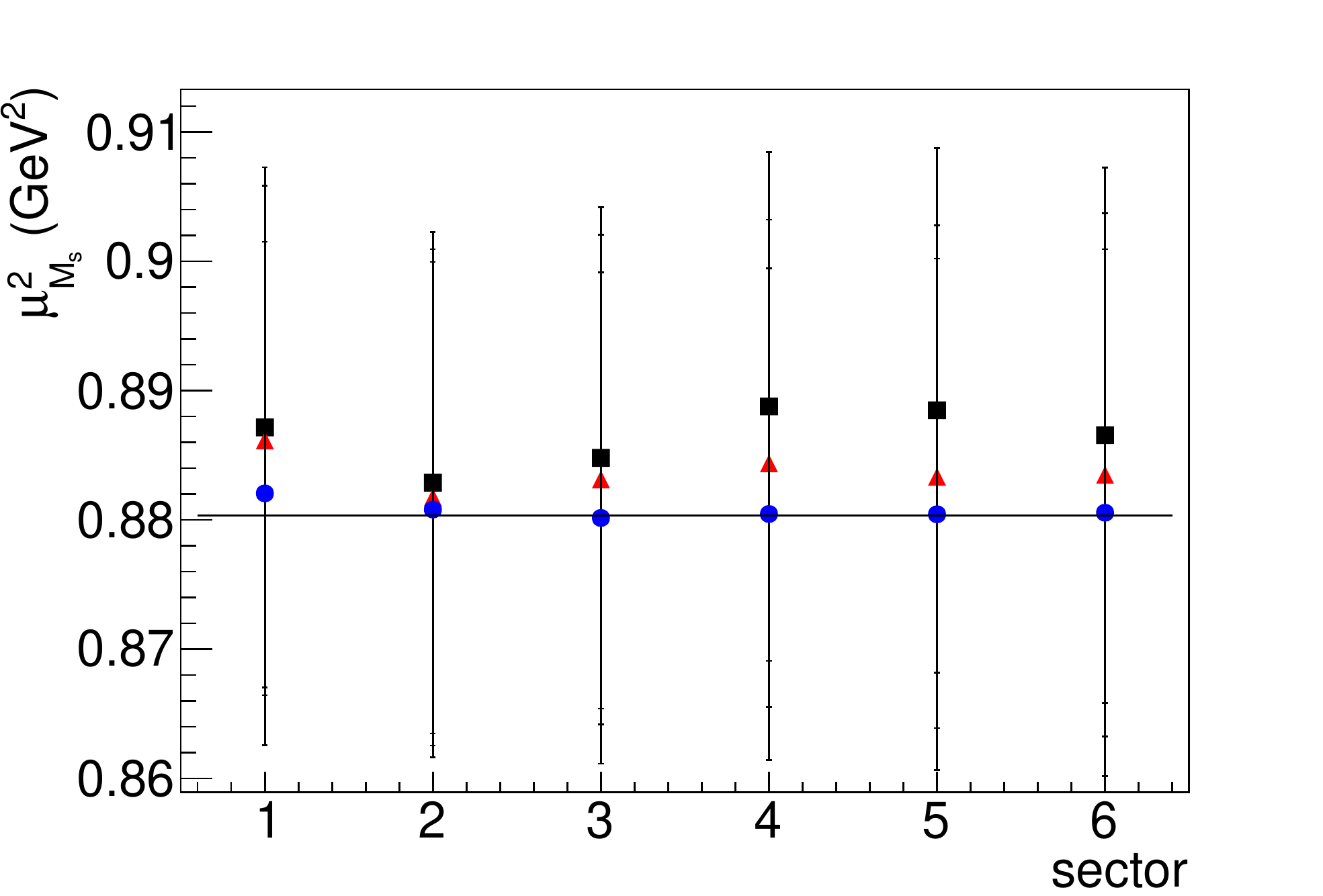}
\caption{\justifyy{The fitted mean values of the measured proton spectator missing mass squared $\mu_{M_s^2}$ versus detector sector without any kinematic corrections (black squares), with only electron momentum corrections (red triangles), and with both electron momentum and proton energy loss corrections (blue dots). The black line represents the squared proton rest mass $\approx$0.88~GeV$^2$.}}
\label{fig:mismefcor}       
\end{figure}
In the analysis, the reconstructed momentum is lower than the initial momentum at the vertex of the reaction, particularly for slow charged particles. This effect has much more influence on the heavy charged particles, which in this case are the low energy protons in the $\pi^{-}p$ channel. This effect is also reproduced in the simulation process. Therefore, energy loss corrections~\cite{referencenoteYe} have to be applied to the reconstructed proton momentum for both experimental data and simulation. 

The influence of these corrections on the spectator missing mass squared mean  $\mu_{M_s^2}$ [$M_s^2$ is defined by Eq.~\eqref{eq:mism2}] is shown in Fig.~\ref{fig:mismefcor}. Although the spectator proton is dependent on the selection cuts and therefore not always a true spectator, we keep this nomenclature for consistency throughout this paper. The corrections bring the position of the missing mass squared of the spectator proton closer to the proton mass squared for all six CLAS sectors.

\subsection{Fiducial cuts}
\begin{figure}[hbt!]
\centering
\begin{subfigure}[b]{0.5\textwidth}
\centering
\includegraphics[width=0.9\linewidth]{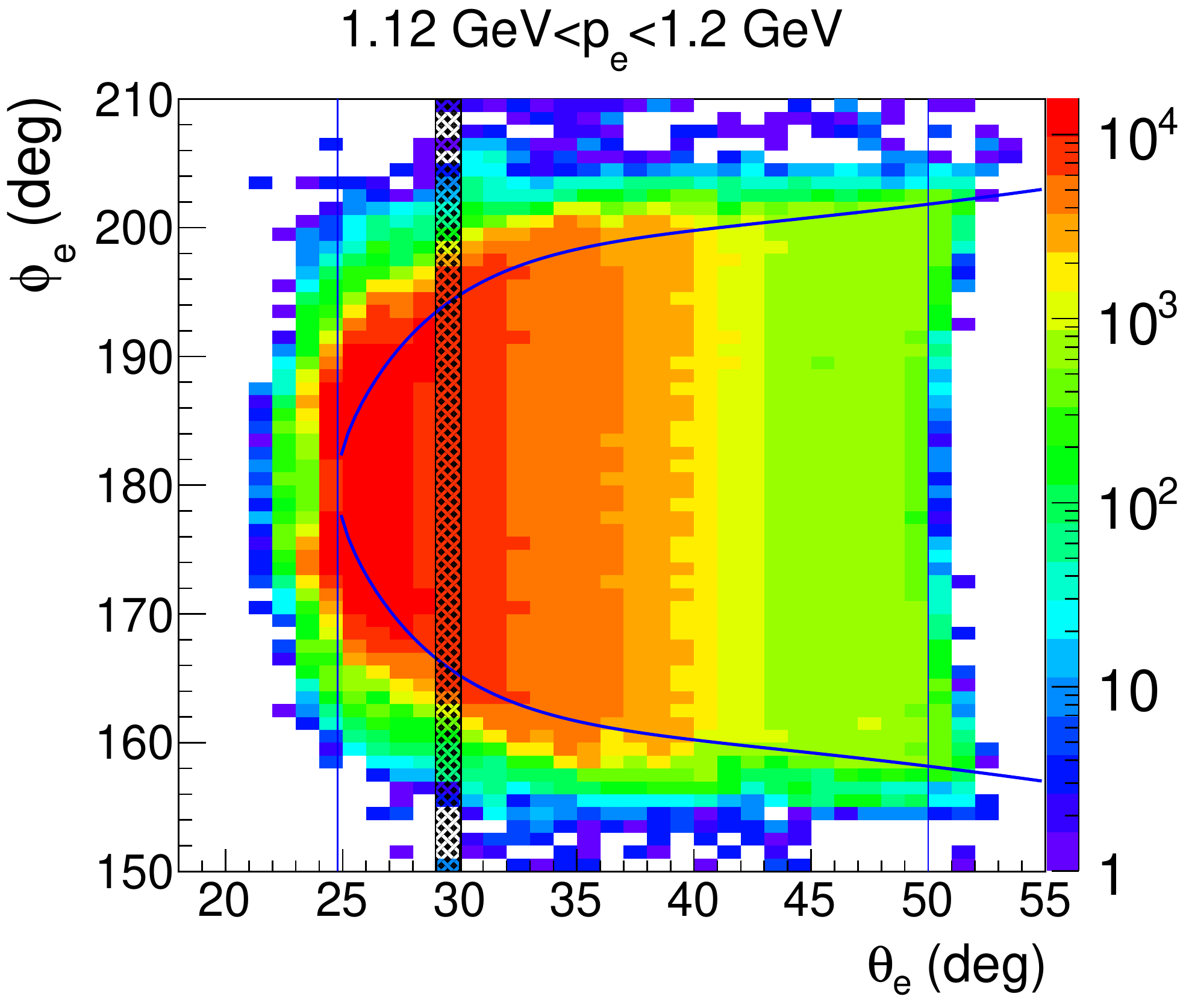}
\caption{}
\label{fig:noeffid}
\end{subfigure}
\begin{subfigure}[b]{0.5\textwidth}
\centering
\includegraphics[width=0.9\linewidth]{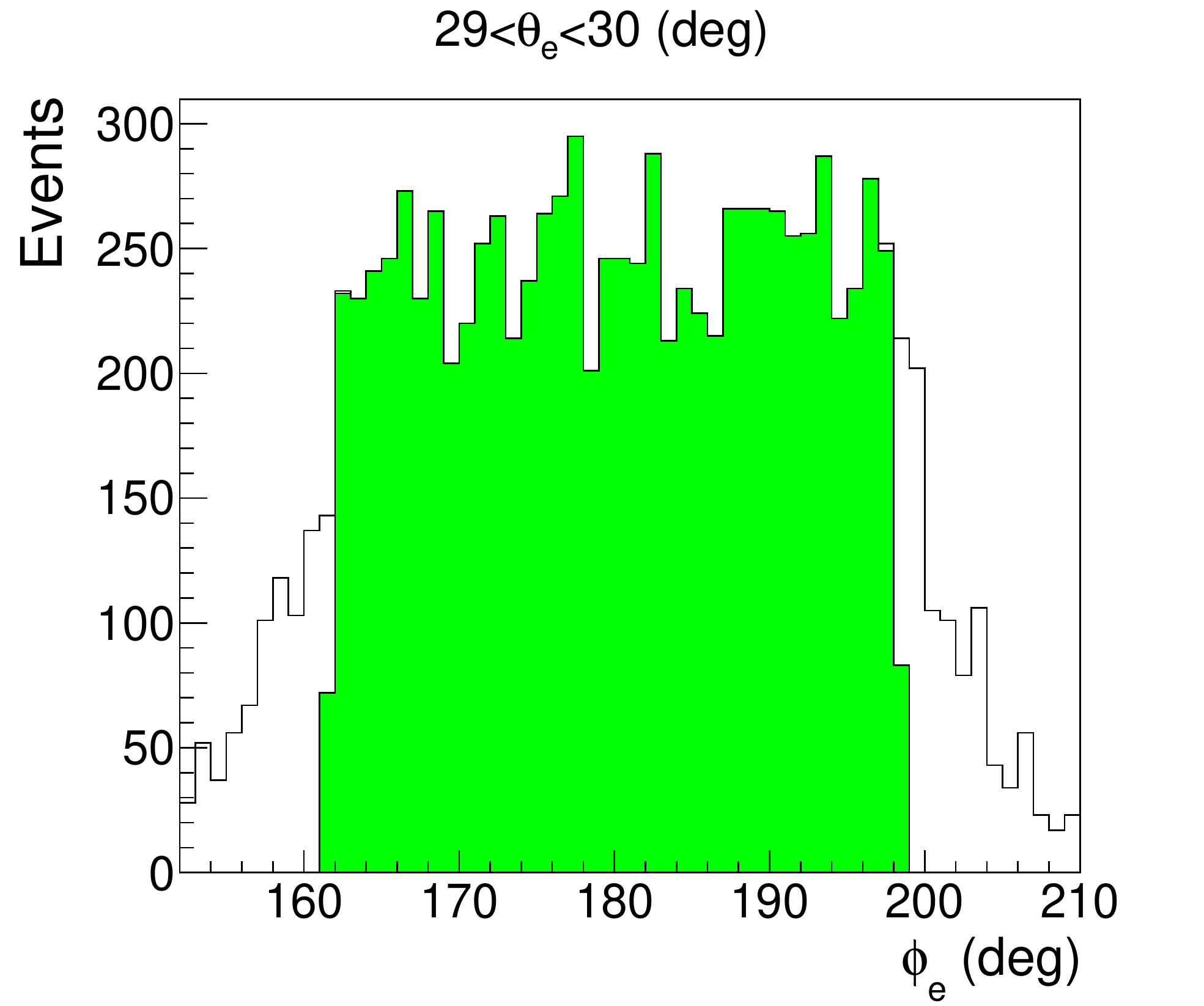}
\caption{}
\label{fig:noeffidphi}
\end{subfigure}
\caption{\justifyy{(a) $\phi_e$ versus $\theta_e$ distribution of electrons for sector 4 within the 1.12 $<\lvert\vec{p}_{e}\rvert<$ 1.2~GeV momentum interval. The blue lines show the fiducial cut boundaries for electrons. (b) $\phi_e$ distributions for the selected $\theta_e$ bin [$29^\circ <\theta_e<30^\circ$ shown as the vertical shaded band in (a)] for the same momentum bin. The green area in the center indicates the selected ﬁducial range.}}
\end{figure}

\begin{figure}[hbt!]
\centering
\begin{subfigure}[b]{0.5\textwidth}
\centering
\includegraphics[width=0.9\linewidth]{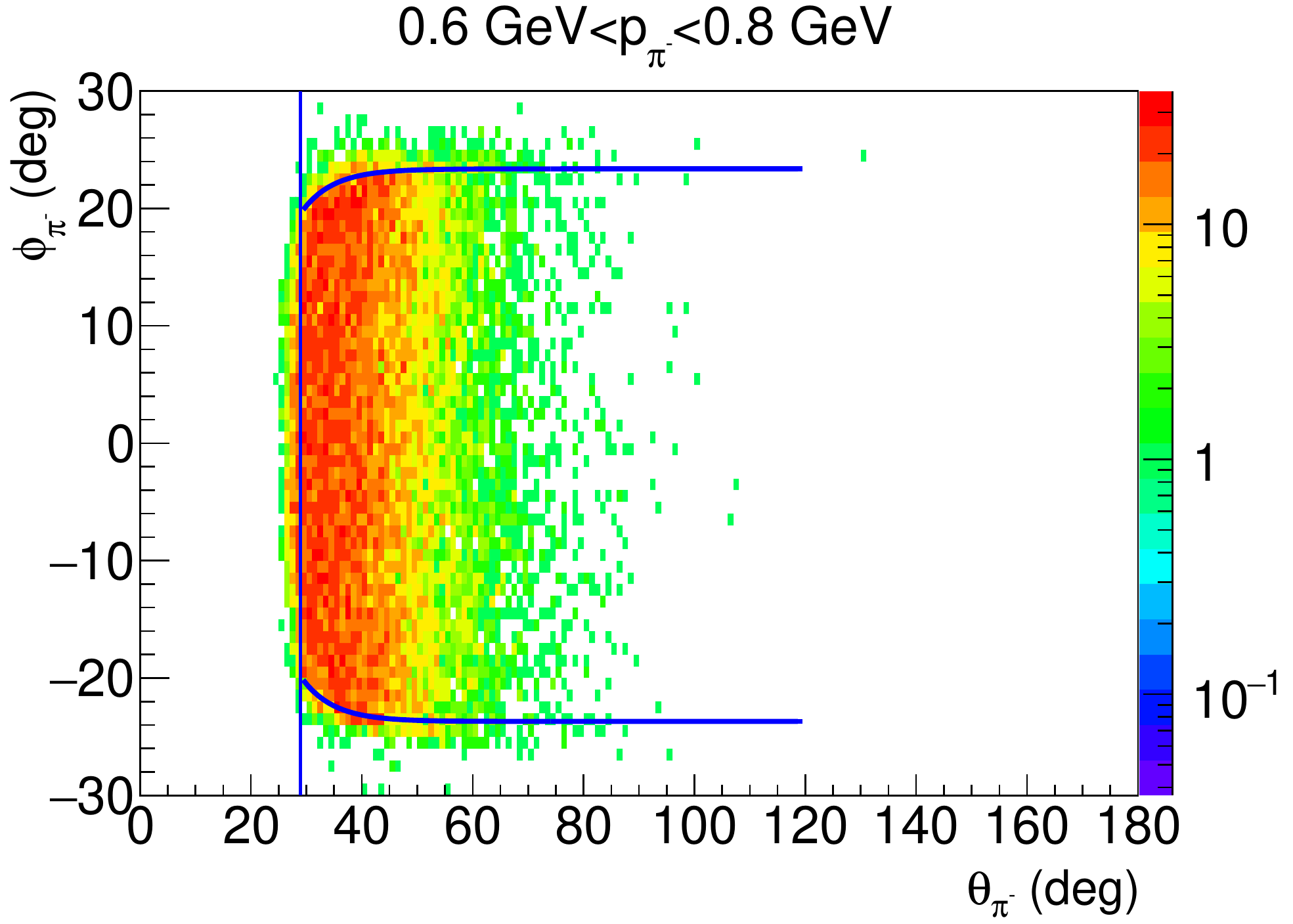}
\caption{}
\label{fig:piftrap1D}
\end{subfigure}
\begin{subfigure}[b]{0.5\textwidth}
\centering
\includegraphics[width=0.9\linewidth]{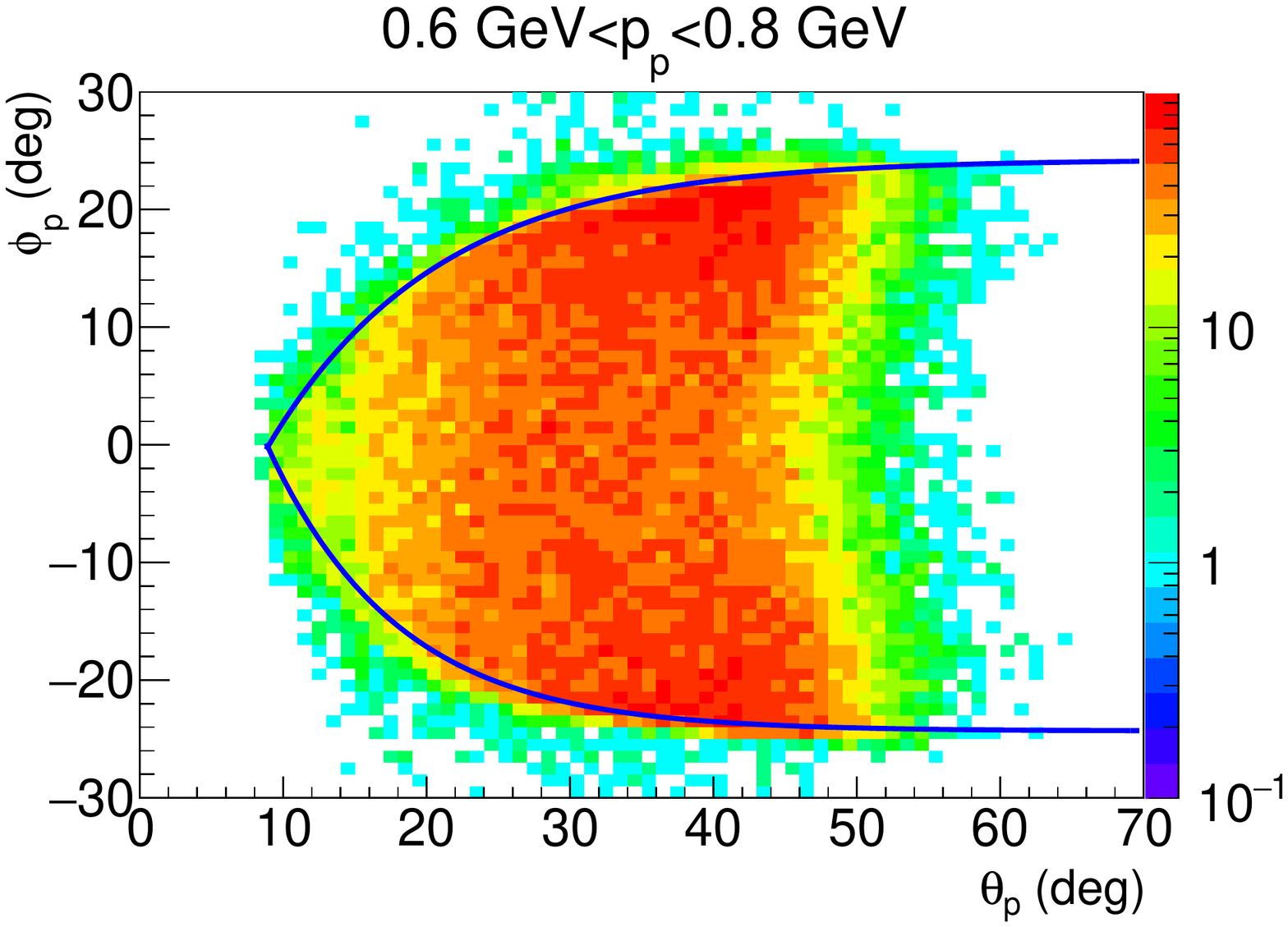}
\caption{}
\label{fig:piftrap2D}
\end{subfigure}
\caption{Typical $\phi$ versus $\theta$ distributions for $\pi^-$ (a) and protons (b) in sector 1 within the same momentum interval 0.6 $<$ $\lvert\vec{p}\rvert$ $<$ 0.8~GeV.}
\end{figure}

The active detection area of CLAS was limited by the torus field coils and the edge regions of the detectors. Therefore, fiducial volumes were defined to select the maximal phase space coverage with reliable detector efficiencies. These fiducial cut functions depend on azimuthal and polar angles, as well as momentum, and are different for different particles. For negatively charged particles ($e^{-}$ and $\pi^{-}$),  symmetrical momentum-dependent but sector-independent cuts were applied on both experiment and simulation reconstructed data. A typical example for the electron $\phi_{e}$ versus $\theta_{e}$ distributions in a specific momentum slice for sector 4 is shown in Fig.~\ref{fig:noeffid}.  The $\phi_{e}$ distribution for each $\theta_{e}$ and $p_{e}$ interval per sector is expected to be a flat distribution [see green regions in Fig.~\ref{fig:noeffidphi}] because the cross section is $\phi_{e}$ independent in the laboratory frame. The empirical shape of this kind of fiducial cut for the ``e1e" run is described in Ref.~\cite{glebclasnote}. For protons, which were outbending (bending away from the beamline), momentum-independent and slightly asymmetrical, sector-dependent fiducial cuts were established in the same way as for electrons and pions. Corresponding examples of the $\phi$ versus $\theta$ distributions in a specific momentum slice for sector 1 with the applied fiducial cuts are shown in Fig.~\ref{fig:piftrap1D} for $\pi^-$ and Fig.~\ref{fig:piftrap2D} for protons. 

\begin{figure}[hbt!]
\centering
\begin{subfigure}[b]{0.5\textwidth}
\centering
\includegraphics[width=0.9\linewidth]{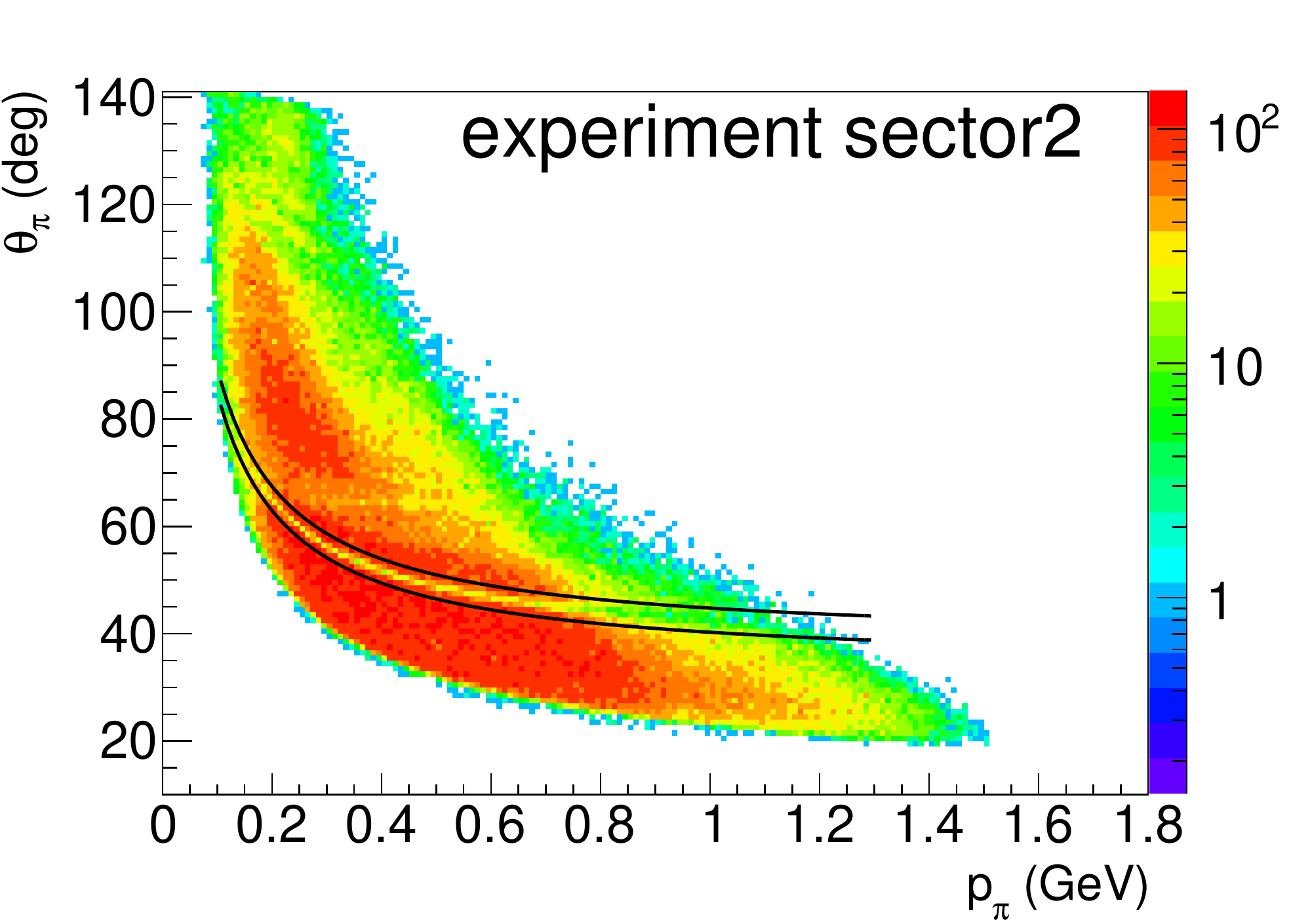}
\caption{}
\label{fig:pifthetavsp_data}
\end{subfigure}
\begin{subfigure}[b]{0.5\textwidth}
\centering
\includegraphics[width=0.9\linewidth]{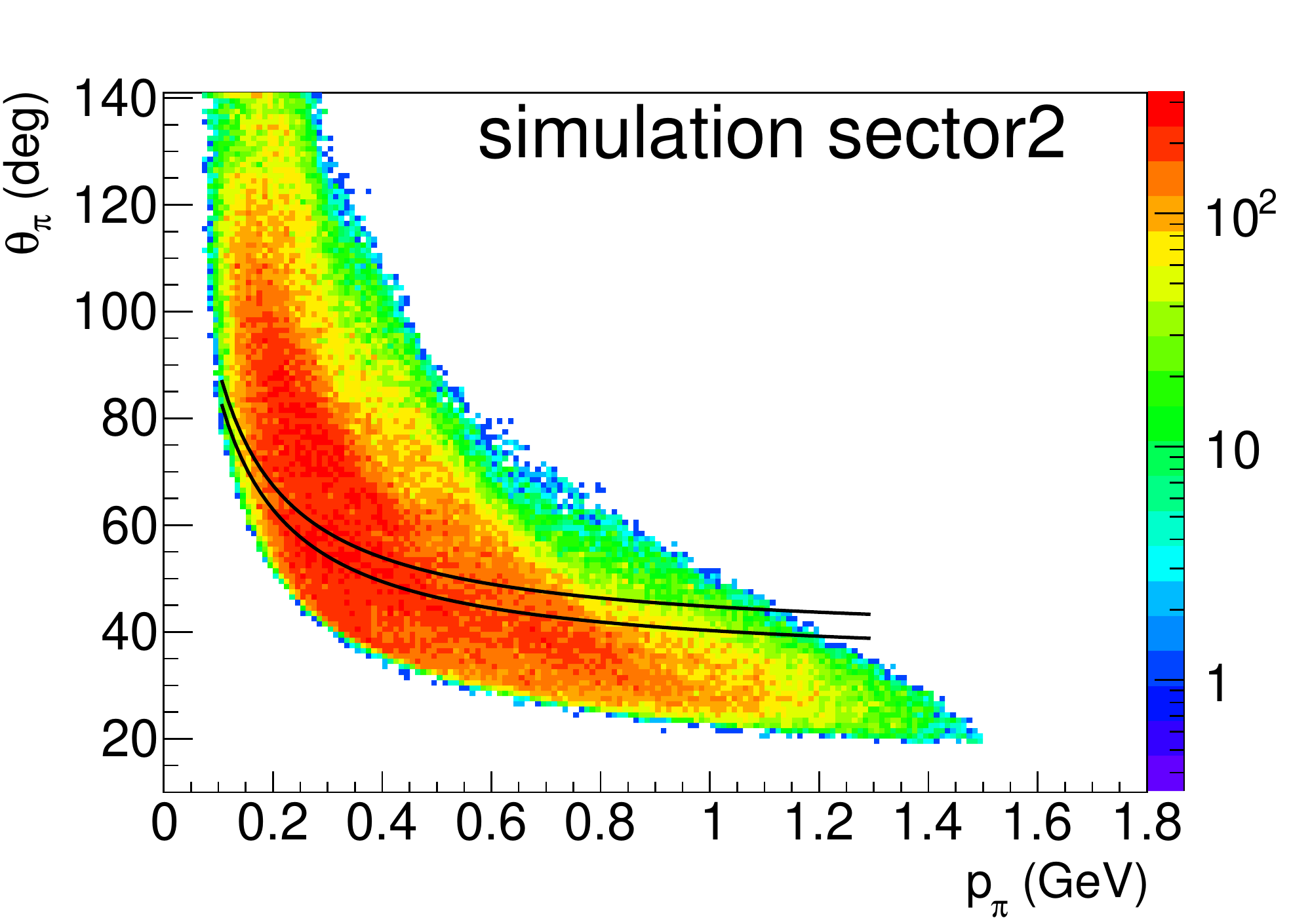}
\caption{}
\label{fig:pifthetavsp_sim}
\end{subfigure}
\caption{\justify{Typical $\theta$ versus $p$ histograms of $\pi^-$ in sector 2 are compared for experiment (a) and simulation reconstructed (b) data. The paired black lines show the corresponding removed low efficiency region defined from experiment data.}}
\end{figure}

Furthermore, there were additional low-efficiency regions due to dead wires of the DC and bad photomultiplier tubes in the SC. These regions, seen in the $\theta$ versus $p$ distributions for the particles in each sector, were cut out in both data and simulation. In Figs.~\ref{fig:pifthetavsp_data} and ~\ref{fig:pifthetavsp_sim}, the pairs of black lines represent the boundaries of a removed region in sector 2 for $\pi^{-}$, which was applied simultaneously to experiment and simulation reconstructed data.      

\section{Event Selection}
\subsection{Exclusive event selection}
\begin{figure}[hbt!]
\centering
\includegraphics[width=0.45\textwidth]{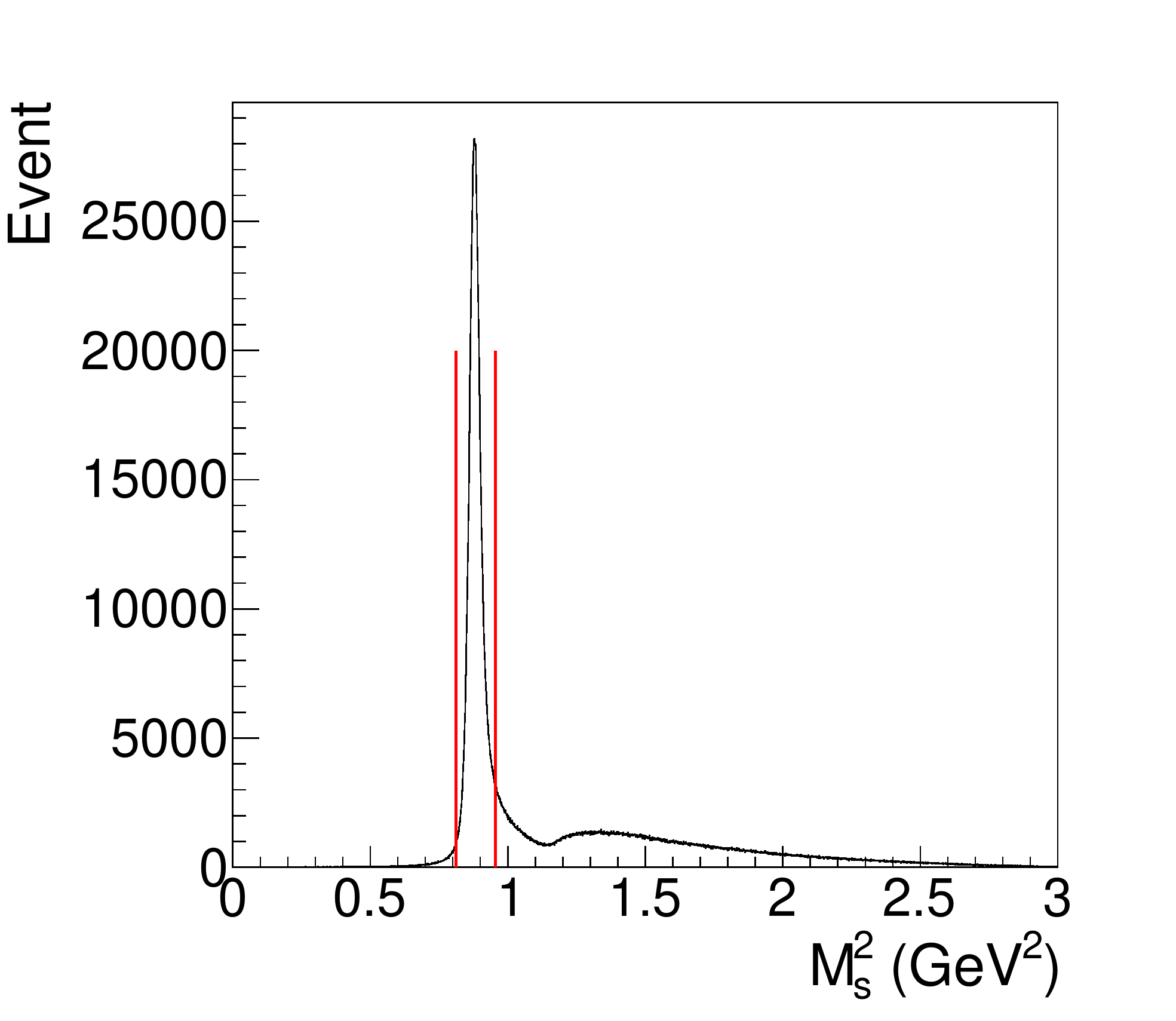}
\caption{\justifyy{$M_{s}^{2}$ distribution with the cut region represented by the red lines showing the exclusive event selection process. }}
\label{fig:misms2}       
\end{figure}

The ``spectator'' proton missing mass squared $M_{s}^{2}$ was determined by
\begin{equation}
M_{s}^{2}=(p_{e}^{\mu}-p_{e^{'}}^{\mu}+p_{D}^{\mu}-p_{\pi^{-}}^{\mu}-p_{p}^{\mu})^{2},
\label{eq:mism2}
\end{equation}
where $p_{e}^{\mu}$, $p_{e^{'}}^{\mu}$, $p_{D}^{\mu}$, $p_{\pi^{-}}^{\mu}$, and $p_{p}^{\mu}$ are the four-momenta of the corresponding particles. The 0.811 $<M_{s}^{2}<$ 0.955~GeV$^{2}$ cuts (see Fig.~\ref{fig:misms2}) were applied for both experiment and simulation reconstructed data (see Sec.~\ref{section:sim} for simulation details) to select the exclusive process ${\gamma}_vn\left(p\right) \rightarrow p {\pi}^{-}\left(p\right)$, where the contribution of any physical background, such as two-pion electroproduction, is negligible. See Sec.~\ref{subsection:bkg} for further details.

\subsection{\label{subsec:level4-2}Quasifree exclusive event selection}

\begin{figure}[htb!] 
\centering
\includegraphics[width=0.45\textwidth]{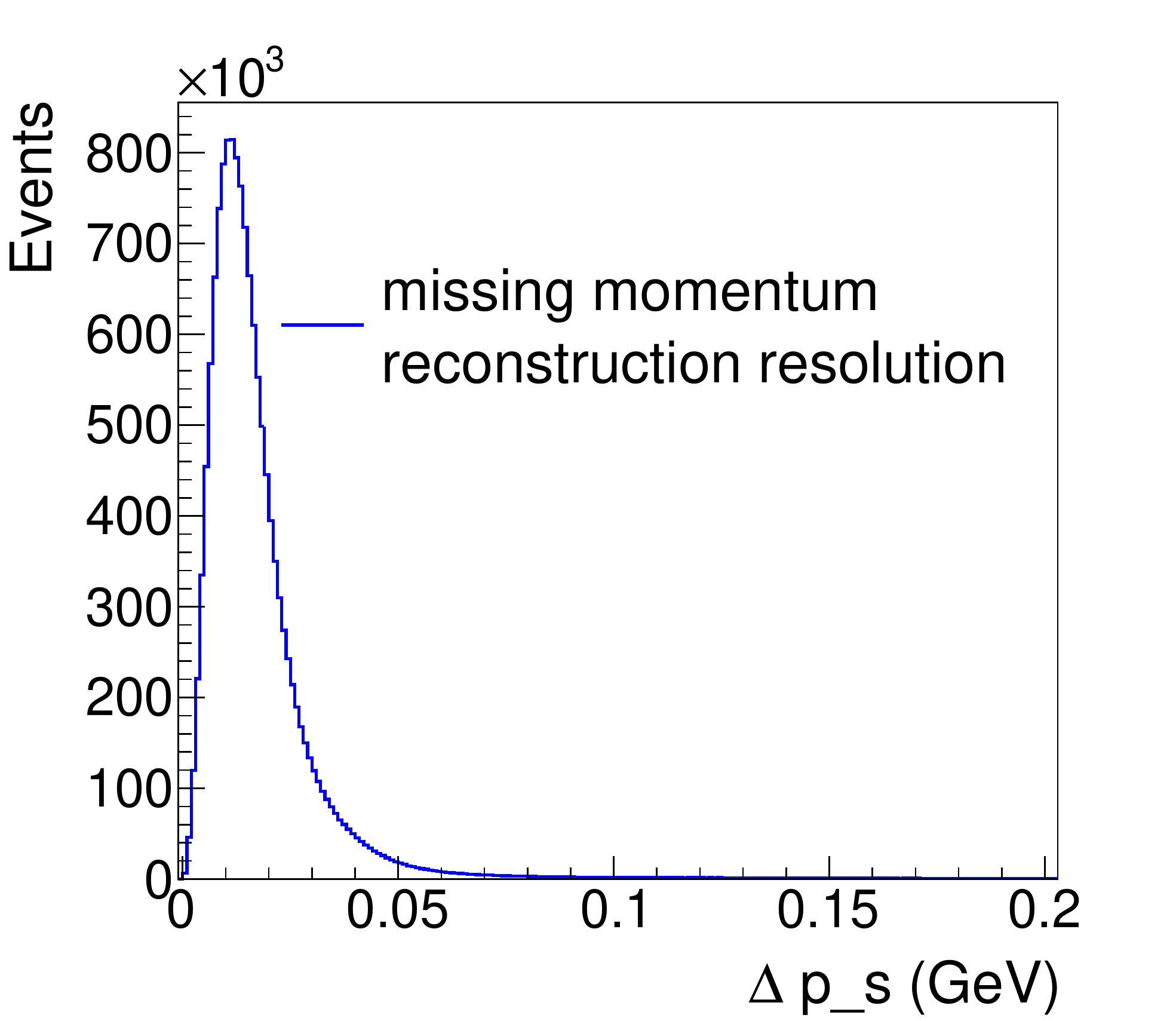}
\caption{Experimental momentum resolution $\Delta$p\_s, which is described by the difference between the simulation thrown and reconstructed momenta.}
\label{fig:Deltappotential}       
\end{figure}

\begin{figure}[hbt!]
\centering
\begin{subfigure}[b]{0.5\textwidth}
\centering
\includegraphics[width=1\linewidth]{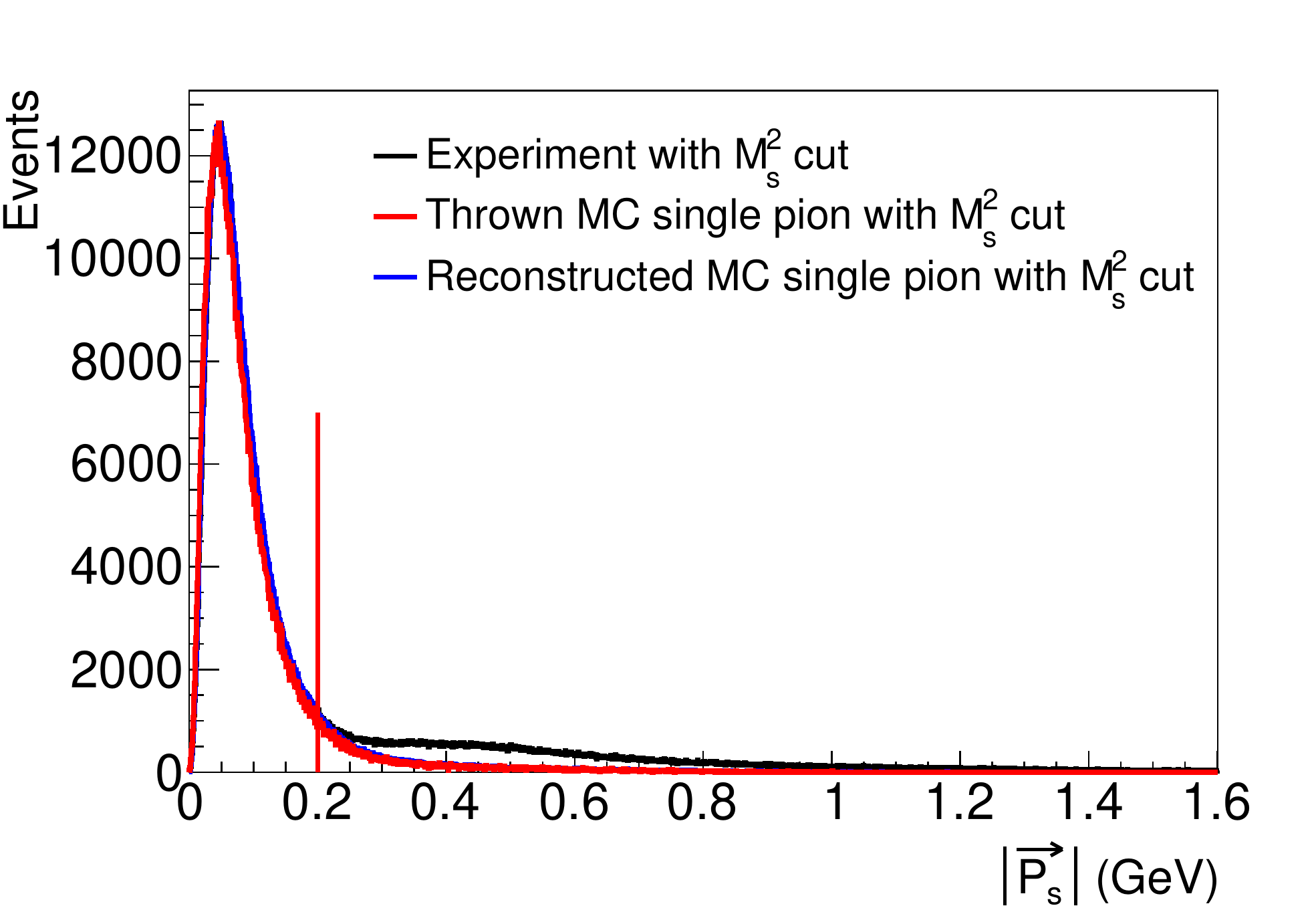}
\caption{}
\label{fig:misps}
\end{subfigure}
\begin{subfigure}[b]{0.5\textwidth}
\centering
\includegraphics[width=1\linewidth]{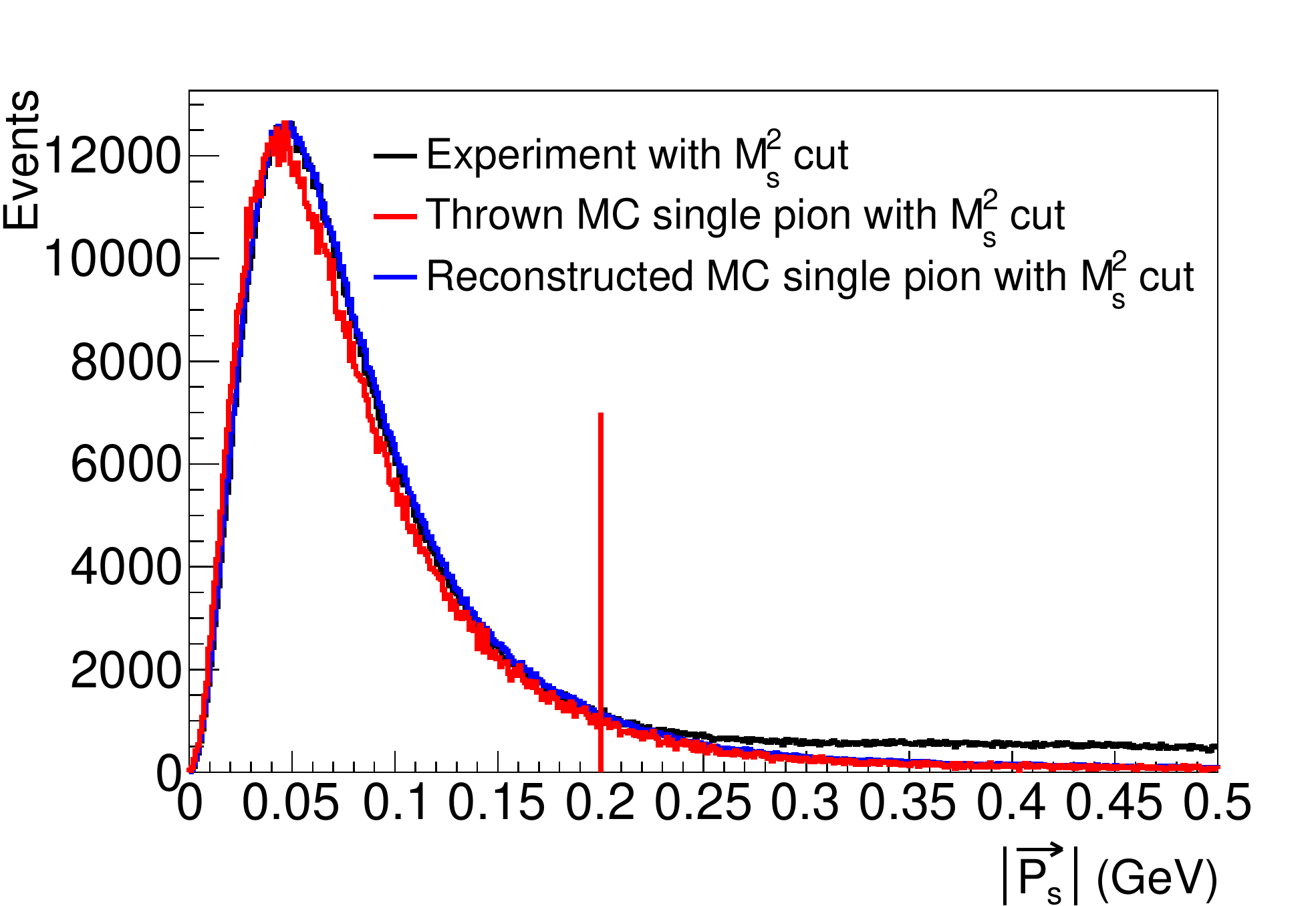}
\caption{}
\label{fig:mispszoom}
\end{subfigure}
\caption{\justify{The black histogram represents the missing momentum distribution ($\lvert\vec{p}_{s}\lvert$) of the unmeasured proton from experimental data. Based on the CD-Bonn potential~\cite{entem2003accurate}, the scaled Monte Carlo simulated (thrown) proton momentum distribution is shown by the red histogram and the detector-reconstructed Monte Carlo distribution by the blue histogram. (b) Zoomed in version of (a) to show this comparison at small more clearly. The red vertical line indicates the $200\;\text{MeV}$ missing momentum cut position.}}
\label{fig:main}
\end{figure}

\begin{figure}[htb!]
\centerline{\includegraphics[width=0.45\textwidth]{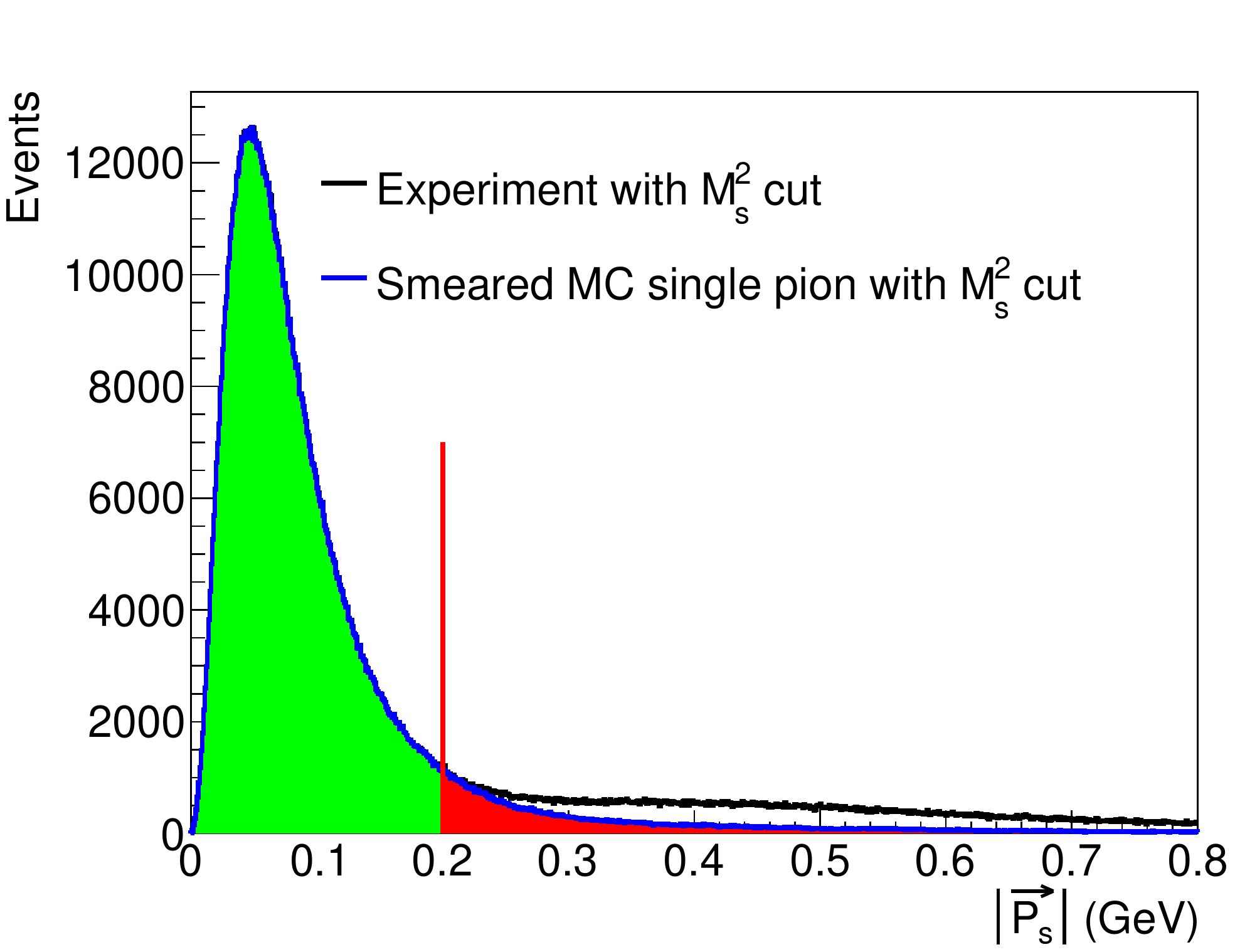}}
\caption{\justify{The missing momentum distributions of the ``spectator" proton $\vert\vec{p}_s\vert$ of experimental data (black histogram) and simulation (blue histogram) where the ``green'' and ``red'' filled areas represent the integral of the blue distribution from $0$ to $200\;\text{MeV}$ and above $200\;\text{MeV}$, respectively.}}
\label{f:r}
\end{figure}

Based on the exclusive events, an additional cut on the missing momentum of the ``spectator'' ($\lvert\vec{p}_{s}\lvert$) was applied to both experiment and simulation reconstructed data, as is shown in Fig.~\ref{fig:main}.
$\lvert\vec{p}_{s}\lvert$ is calculated by

\begin{equation}
\lvert\vec{p}_{s}\lvert=\mid\vec{p}_{e}-\vec{p}_{e^{'}}-\vec{p}_{\pi^{-}}-\vec{p}_{p}\mid.
\label{eq:misp2}
\end{equation}
Figure~\ref{fig:misps} shows the missing momentum of the spectator proton for experimental data (black histogram), simulated thrown data (red histogram), and simulated data smeared by the experimental resolution for the reconstructed measured missing momentum (blue histogram). As expected from an adequate detector simulation, the simulated missing momentum distribution that is smeared according to the experimental detector resolution (shown in Fig.~\ref{fig:Deltappotential}) should match in the absence of FSI the measured missing momentum distribution almost perfectly. No meaningful difference between the reconstructed simulated (blue histogram) and measured (black histogram) missing momentum distributions up to $\vert\vec{p}_s\vert=200$~MeV is visible in Fig.~\ref{fig:mispszoom}. There is a significant difference between the simulated thrown (red histogram) and measured (black histogram) missing momentum distributions at low momenta. Therefore, any final state interaction with a momentum transfer between the spectator proton and any other hadron that is on average larger than $10$ MeV (corresponding to an energy transfer larger than $50$ KeV) would cause a comparable additional broadening of the measured distribution beyond the broadening due to experimental detector resolution. This reveals that the quasifree process is absolutely dominant in the $\lvert\vec{p}_{s}\lvert<200\,\text{MeV}$ region, up to potential FSIs with less than $50$~KeV energy transfer. 
Hence, the quasifree process can be kinematically isolated by applying the $\lvert\vec{p}_{s}\lvert<200\,\text{MeV}$ cut. 
Meanwhile, some good quasifree events were cut as well. Here ``$r$" denotes the factor to correct for good quasifree events outside the $\lvert\vec{p}_{s}\lvert<200\,\text{MeV}$ cut, which is calculated from the reconstructed simulation data by
\begin{equation}
\begin{split}
&r(W, Q^{2}, \cos\theta^\text{c.m.\!}_\pi, \phi^\text{c.m.\!}_\pi)\\
&=\frac{N_{simu}^{\lvert\vec{p}_{s}\lvert<200\,\text{MeV}}(W, Q^{2}, \cos{\theta_{\pi}^{\text{c.m.\!}}}, \phi_{\pi}^{\text{c.m.\!}})}{N_{simu}^{qf}(W, Q^{2}, \cos{\theta_{\pi}^{\text{c.m.\!}}}, \phi_{\pi}^{\text{c.m.\!}})}\\
&=\frac{green}{green+red}\;(\text{in~Fig.}~\ref{f:r}),
\label{eq:r}
\end{split}
\end{equation}
where $N_{simu}^{qf}$ represents the simulated exclusive quasifree yields in each kinematic bin and $N_{simu}^{\lvert\vec{p}_s\lvert<200\,\text{MeV}}$ are the simulation yields in each kinematic bin after applying the $\lvert\vec{p}_s\lvert<200\,\text{MeV}$ cut. The corresponding green and red areas shown in Fig.~\ref{f:r} represent the integrals of the $\lvert\vec{p}_s\lvert$ distribution below and above the $200\;\text{MeV}$ cut, respectively. 

\section{\label{section:sim}Simulation}
\begin{figure}[hbt!] 
\centering
\includegraphics[width=0.5\textwidth]{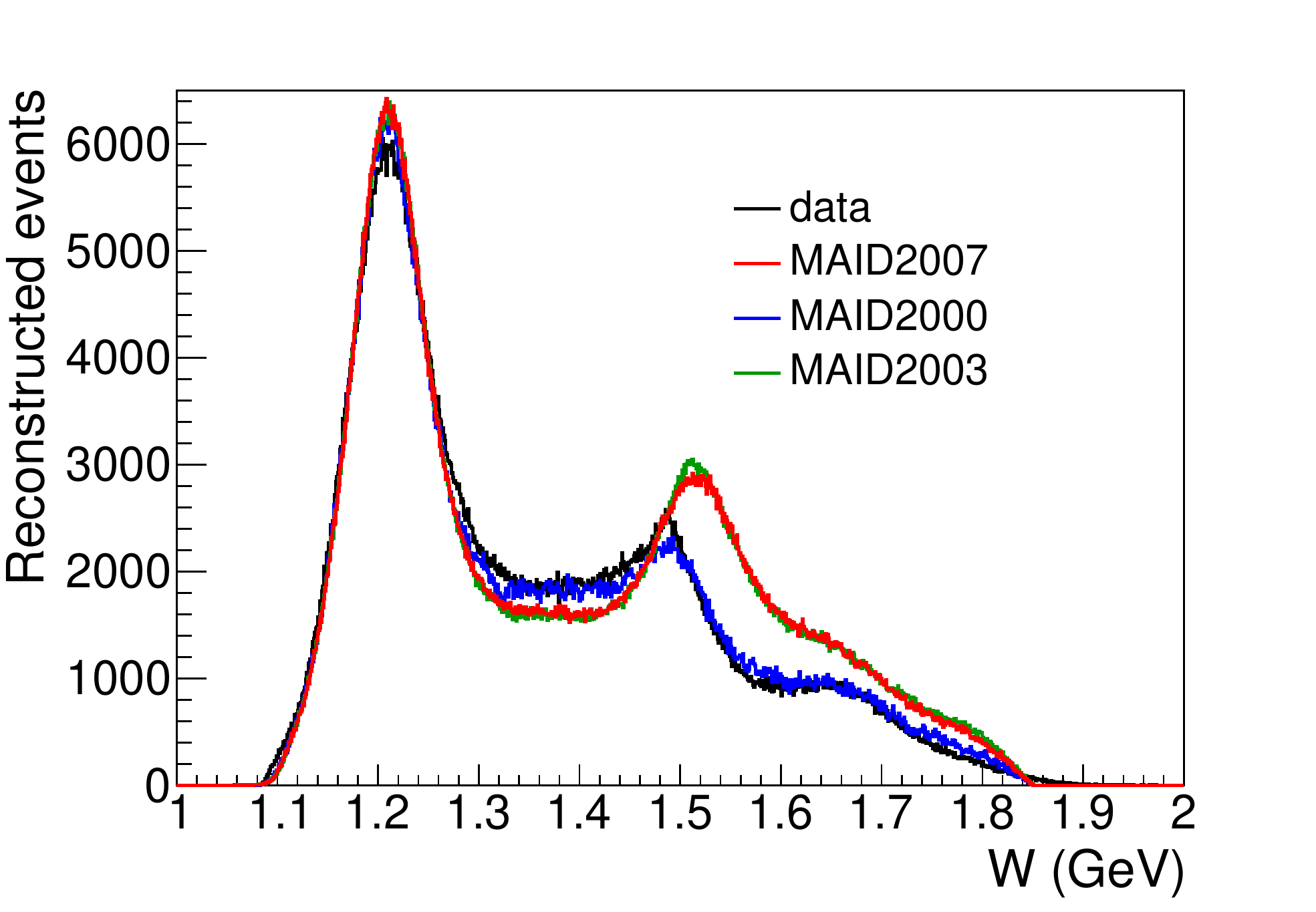}
\caption{\justify{Experimental $W$ distribution of quasifree exclusive event yields in comparison to various MAID models.}}
\label{fig:wmodel}       
\end{figure}

MAID is a unitary isobar model for partial wave analysis on the world data of pion photo and electroproduction in the resonance region. After comparison of our measured quasifree exclusive event yields with different MAID versions, the electromagnetic multipole table~\cite{refMAID2007} of the MAID2000 model~\cite{refMAID2000} was chosen as input for the event generator. Besides the MAID2000 version, there are MAID98, MAID2003, and MAID2007 versions~\cite{refMAID2007} also available in the ``$aao\_rad$'' package~\cite{refCVS}. In order to determine which version describes the experimental data best, the $W$ distributions of the quasifree exclusive reconstructed events from different MAID versions and the data were plotted, as shown in Fig.~\ref{fig:wmodel}. Even though MAID2007 is the latest version, the second resonance peak from this version is shifted relative to the experimental neutron data. However, the resonance peaks from the MAID2000 version match with that from the neutron data best, which is important for the radiative and bin centering corrections.

Simulated $en \to e'p\pi^-$ events with radiative effects, according to the prescription of Mo and Tsai~\cite{mo1969radiative}, were generated by a modified version of the available ``$aao\_rad$'' software package~\cite{refCVS}. The initial state neutron mass was set to the neutron rest mass. An additional proton was generated with its Fermi momentum calculated from the CD-Bonn potential~\cite{entem2003accurate} and its mass set to the proton rest mass. In this way, the generated proton behaved like a spectator ($p_{s}$).  After adding the ``spectator'' proton in the event generator, the simulated physics process could be treated in the same way as the exclusive quasifree process of the experimental data. 

\section{\label{sec:cros}CROSS SECTION EXTRACTION}
\label{sec-extr}
\subsection{Kinematic binning}
The kinematic variables $W=W_{f}$, $Q^{2}$, $\cos\theta^\text{c.m.\!}_\pi$, and $\phi^{\text{c.m.\!}}_{\pi}$ are used to present the final cross sections. The binning choices are listed in Tables~\ref{kinmaticbin} and~\ref{kinmaticthetaphibin}.
 Due to the low detector acceptance for $\pi^{-}$, even in the highest statistics $W$ and $Q^{2}$ bins there are empty kinematic phase space cells at very small and very large $\phi_{\pi}^{\text{c.m.\!}}$ angles. To mitigate this problem, various $\phi_{\pi}^{\text{c.m.\!}}$ bin sizes were studied as listed in Table~\ref{kinmaticthetaphibin}. This allows for the optimization of the individual bin statistics and kinematic coverage and serves as a consistency check for the cross sections.
\begin{table}[ht]
\centering
\caption{$W$ and $Q^{2}$ binning of the analysis.}
\begin{tabular}{cccc}
\hline
 \hline
Variable & Lower limit & Upper limit & Bin size\\
 \hline
$W$ & 1.1 & 1.825 & 0.025 GeV \\
$Q^{2}$& 0.4 & 1.0 & 0.2 $\text{GeV}^{2}$\\
\hline
 \hline
 \end{tabular}
\label{kinmaticbin}
\end{table}

\begin{table}[ht]
\centering
\caption{$\cos\theta^\text{c.m.\!}_\pi$ and $\phi^\text{c.m.\!}_\pi$ binning of the analysis.}
\begin{tabular}{cccc}
\hline
\hline
Variable & Lower limit & Upper limit & Bin size\\
\hline
$\cos\theta^\text{c.m.\!}_\pi$& -1 & 1 & 0.2 \\

$\phi^\text{c.m.\!}_\pi$& $0^{\circ}$ & $360^{\circ}$ & $40^{\circ}$, $45^{\circ}$, $60^{\circ}$\\


\hline
\hline
\end{tabular}
\label{kinmaticthetaphibin}
\end{table}

\subsection{Bin-centering corrections}
Because of the possibly nonlinear behavior of the cross section across a bin, the average cross section value does not necessarily correspond to the center of the bin. Hence, presenting the extracted cross section at the center of the bin might not be accurate. To account for this effect, a correction was applied to the cross sections for each four-dimensional $(W,Q^2,\cos\theta^\text{c.m.\!}_\pi,\phi^\text{c.m.\!}_\pi)$ bin. This bin-centering correction ($R_{BC}$) was calculated as

\begin{equation}
R_{BC}(W,Q^{2},\cos\theta^\text{c.m.\!}_\pi,\phi^{\text{c.m.\!}}_{\pi})=\frac{\sigma^{model}_{center}}{\sigma^{model}_{average}},
\label{eq:RBC}
\end{equation}
where $\sigma^{model}_{center}$ is the cross section calculated by using the parametrization function of the MAID2000 model at the center of each kinematic bin and $\sigma^{model}_{average}$ is
\begin{equation}
\sigma^{model}_{average}=\frac{\int_{x_{1}}^{x_{2}}\sigma(x)dx}{\Delta W \Delta Q^{2} \Delta\cos\theta^\text{c.m.\!}_\pi \Delta \phi^\text{c.m.\!}_\pi},
\label{eq:modelaverage}
\end{equation}
where $ x $ represents the kinematic bin $(W,Q^2,\cos\theta^\text{c.m.\!}_\pi,\allowbreak\phi^\text{c.m.\!}_\pi)$, $x_{1}$ and $x_{2}$ are the limits of the bin, and $\sigma(x)$ is the MAID2000 model cross section function within the bin. 
    
\subsection{Acceptance corrections}
\begin{figure}[hbt!]
\centering
\includegraphics[width=0.5\textwidth]{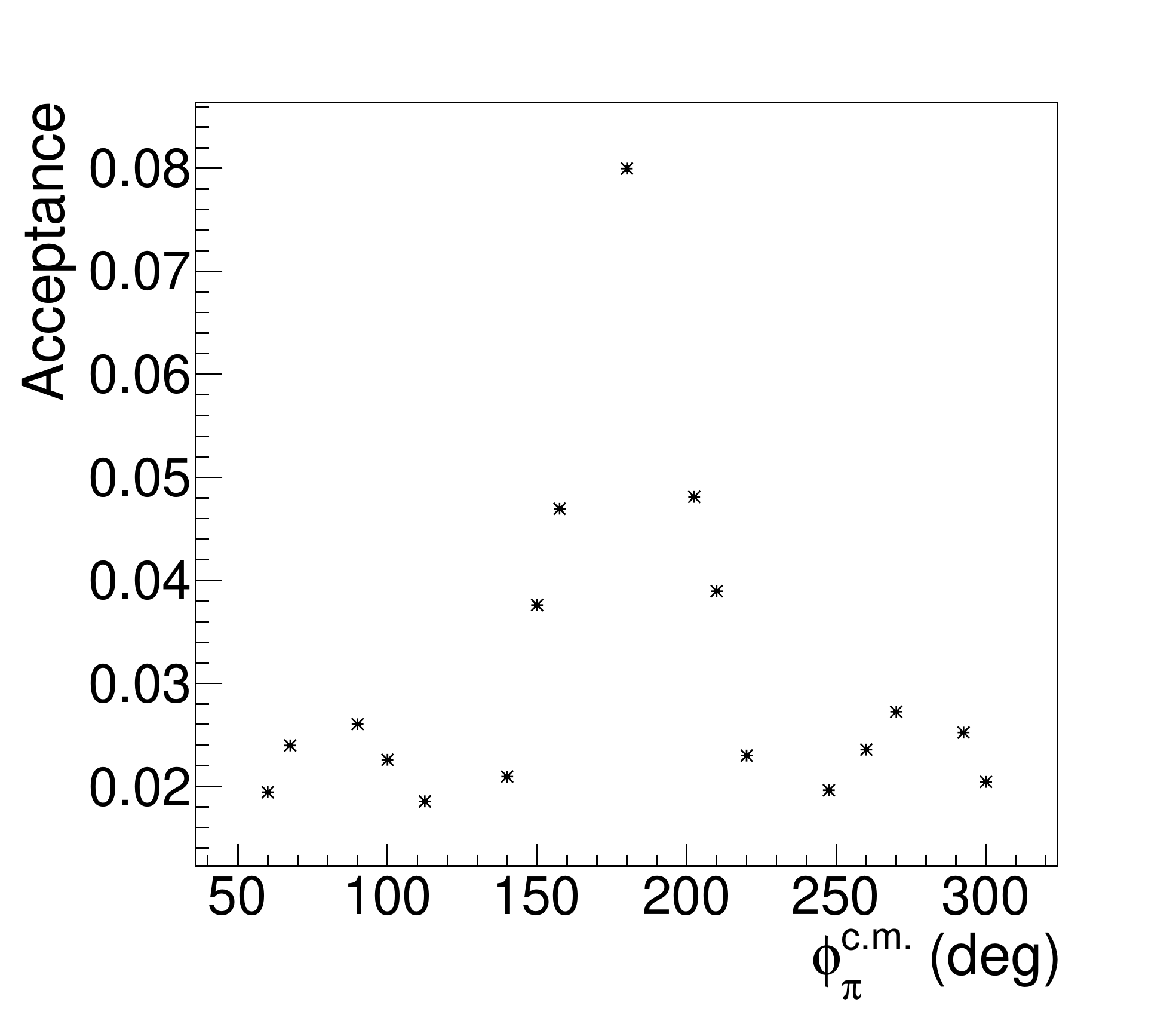}
\caption{Typical of acceptance correction factors as a function of $\phi^{c.m.}_{\pi^{-}}$ for $W=$ 1.2125~GeV and $Q^{2}=$ 0.5~GeV$^2$, and $\cos\theta^{c.m.}_{\pi^{-}}=$ 0.1.}
\label{fig:Acc}       
\end{figure}

Acceptance-correction factors ($A^{Rad}$) were calculated using the Monte Carlo simulated events (total 8$\times$10$^{9}$ events to avoid statistical fluctuations) for each four-dimensional bin by
\begin{equation}
\begin{split}
&A^{Rad}(W,Q^{2},\cos{\theta^{\text{c.m.\!}}_\pi},\phi^\text{c.m.\!}_\pi)\\
&=\frac{N^{Rad}_{rec}(W,Q^{2},\cos{\theta^{\text{c.m.\!}}_\pi},\phi^{\text{c.m.\!}}_\pi)}{N^{Rad}_{thrown}(W,Q^{2},\cos{\theta^{\text{c.m.\!}}_\pi},\phi^{\text{c.m.\!}}_\pi)},
\label{eq:acceptance}
\end{split}
\end{equation}
where $N^{Rad}_{thrown}(W,Q^2,\cos\theta^\text{c.m.\!}_\pi,\phi^\text{c.m.\!}_\pi)$ represents the number of events that were generated by the physics event generator ``$aao\_rad$'' with MAID2000 and radiative effects turned on in each kinematic bin. $N^{Rad}_{rec}$ denotes the number of events in the same kinematic bin that have gone through the entire simulation and reconstruction process passing all of the analysis cuts described above. A typical $\phi$-dependent acceptance distribution for one $(W,Q^2,\cos\theta^\text{c.m.\!}_\pi)$ bin is presented in Fig.~\ref{fig:Acc}.
\begin{figure}[hbt!]
\centerline{\includegraphics[width=0.5\textwidth]{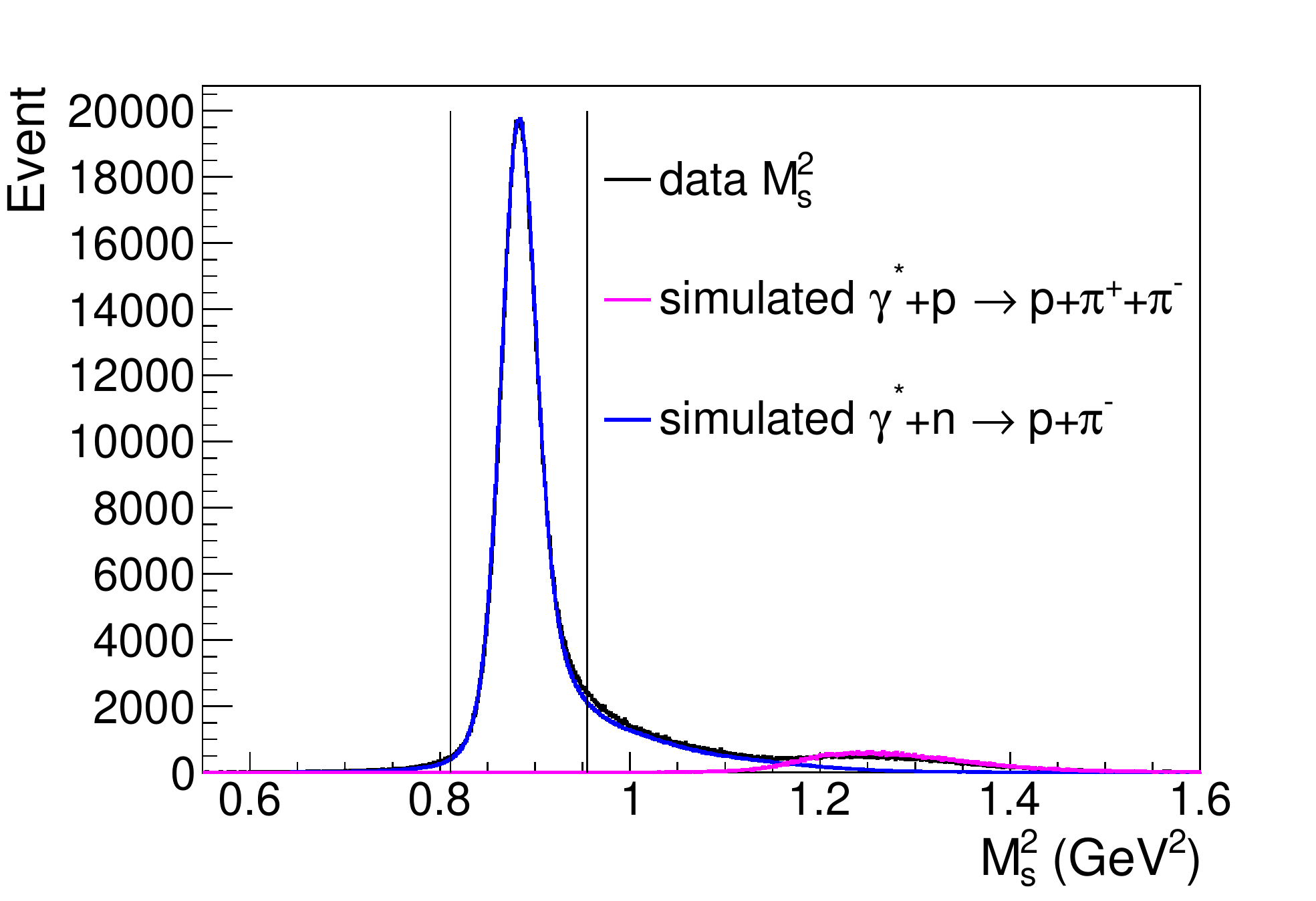}}
\caption{\justify{$M_{s}^{2}$ distributions for the measured (black histogram) and simulated ${\gamma}_vn\left(p\right) \rightarrow p {\pi}^{-}(p)$ (blue histogram) data, as well as the simulated ${\gamma}_vp \rightarrow p {\pi}^{-}\pi^{+}$ events (magenta histogram). The $M_s^2$ cut region is shown by the vertical lines.}}
\label{f:MISMbacg}
\end{figure}
\subsection{Radiative corrections}
The incoming and outgoing scattered electrons can change their energy (emit unobserved photons) due to the radiative effects. Although those effects do not influence the kinematic variable $W_{f}$ $[W_{f}= p^{\mu}+(\pi^{-})^{\mu}]$, they can influence the variables $W_{i}$ and  $Q^{2}$. 
The approach developed by Mo and Tsai~\cite{mo1969radiative} was used to correct the final results. The same number of $en \to e'p\pi^-$ events with and without radiative effects were generated by the available ``$aao\_rad$'' and ``$aao\_norad$'' software packages~\cite{refCVS}, respectively, by using the same electromagnetic multipole table from the MAID2000 model. The radiative correction factor RC was calculated by
\begin{equation}
\begin{split}
\text{RC}&(W,Q^{2},\cos{\theta^{\text{c.m.\!}}_\pi},\phi^{\text{c.m.\!}}_\pi)\\
&=\frac{N^{Rad}_{thrown}(W,Q^{2},\cos{\theta^{\text{c.m.\!}}_\pi},\phi^{\text{c.m.\!}}_\pi)}{N^{noRad}_{thrown}(W,Q^{2},\cos{\theta^{\text{c.m.\!}}_\pi},\phi^{\text{c.m.\!}}_\pi)},
\label{eq:RCcorrect}
\end{split}
\end{equation}
where $N^{noRad}_{thrown}\left(W,Q^2,\cos\theta^\text{c.m.\!}_\pi,\phi^\text{c.m.\!}_\pi\right)$ is the number of events without radiative effects generated by the ``$aao\_norad$" software package~\cite{refCVS} in each kinematic bin, and $N^{Rad}_{thrown}(W,Q^2,\cos\theta^\text{c.m.\!}_\pi,\phi^\text{c.m.\!}_\pi)$ is the same quantity as used in Eq.~\eqref{eq:acceptance}. The order of the RC corrections is less than $4\%$ over all coverage W region. Finally, RC was combined with the acceptance corrections factor $A^{Rad}$ (see Eq.~\eqref{eq:acceptance}) to calculate the radiative-corrected acceptance $A_{RC}$, represented by

\begin{equation}
\begin{split}
&A_{RC}(W,Q^{2},\cos{\theta^{\text{c.m.\!}}_\pi},\phi^{\text{c.m.\!}}_\pi)\\
&=A^{Rad}(W,Q^{2},\cos{\theta^{\text{c.m.\!}}_\pi},\phi^{\text{c.m.\!}}_\pi)\text{RC}(W,Q^{2},\cos{\theta^{\text{c.m.\!}}_\pi},\phi^{\text{c.m.\!}}_\pi)\\
&=\frac{N^{Rad}_{rec}(W,Q^{2},\cos{\theta^{\text{c.m.\!}}_\pi},\phi^{\text{c.m.\!}}_\pi)}{N^{noRad}_{thrown}(W,Q^{2},\cos{\theta^{\text{c.m.\!}}_\pi},\phi^{\text{c.m.\!}}_\pi)}.
\end{split}
\label{eq:RDacceptance}
\end{equation}

\subsection{\label{subsection:bkg}Background subtraction}
The events of the ${\gamma}_vp \rightarrow p \pi^+\pi^-$ process, considered to be the main source of possible physics background, were simulated by the double-pion electron scattering event generator (``genev''~\cite{refgenev}) and reconstructed with the same analysis procedure. Figure~\ref{f:MISMbacg} shows the resulting and properly scaled $M_{s}^{2}$ distributions in comparison to the ``e1e'' run experimental data and the ${\gamma}_vn(p) \rightarrow p \pi^-\left(p\right)$ simulation events. Inside the 0.811 $<M_{s}^{2}<$ 0.955~GeV$^2$ cut region, there is no ${\gamma}_vp \rightarrow p \pi^+\pi^-$ background contribution. Furthermore, the $M_{s}^{2}$ distributions for experimental events were compared bin by bin [kinematic bin ($W,Q^2,\cos\theta^\text{c.m.\!}_\pi,\phi^\text{c.m.\!}_\pi$)] with the simulated ${\gamma}_vn\left(p\right) \rightarrow p \pi^-\left(p\right)$ events to check the variation in the background contribution~\cite{referencenoteYe}. In summary, there is no need to apply any background subtraction for the exclusive ${\gamma}_vn\left(p\right) \rightarrow p \pi^-\left(p\right)$ process in the ``e1e'' analysis.

\subsection{Systematic uncertainties}
The characteristic parameters corresponding to each step in the data analysis procedure have been varied to quantify their influence on the final cross sections and structure functions on a bin-by-bin basis. A summary of all sources studied and the magnitudes of the assigned systematic uncertainties are listed in Table~\ref{table_syserr}. The total average systematic uncertainty of the cross sections is $8.6\%$, calculated as the quadrature sum of the individual contributions. The individual systematic uncertainties are reported for each data point in the CLAS Database~\cite{CLAS:DB}.  

The biggest source of systematic uncertainties is the yield normalization. A comparison of the measured inclusive cross sections and Osipenko's world-data parametrization~\cite{osipenko2006measurement} was carried out, and the ratios deviate from 1 by no more than $5\%$~\cite{referencenoteYe}. Due to the model dependence of the Osipenko event generator, we also cross-checked against the systematic uncertainty of quasi-elastic scattering cross section of nucleons in nuclei~\cite{referencenote}. We found that the world data and the normalized ``e1e" data agree to the $5\%$ level with these parametrizations, which is consistent with our Osipenko-derived uncertainty. 
\vspace{4mm}
\begin{table}[hbt]
\begin{center}
\caption{Summary of sources of the average systematic uncertainty. Further information on the systematic uncertainties due to different boost vectors and different deuteron potentials can be found in~\cite{referencenoteYe} }
\begin{tabular}{cc}
\hline
\hline
Source &
\multicolumn{1}{c}{Uncertainty ($\%$)}\\
\hline
Electron $\theta_{CC}$ cut  &     0.78\\
Electron sampling fraction cut &     1.26\\
Electron fiducial cut  &   2.10\\
Proton $\Delta T$ cut  &    1.39\\
Proton fiducial cut  &    2.39\\
Pion $\Delta T$ cut  &    1.78\\
Pion fiducial cut  &    1.73\\
$M_{s}^{2}$ cut &    2.29\\
$p_{s}$ cut & 2.21\\
Boosts into neutron rest frame~\cite{referencenoteYe}  &    2.12\\
Choice of deuteron potential~\cite{referencenoteYe}  &    3.2\\
Bin centering correction & 0.55 \\
Radiative correction & 2.0 \\
Normalization  &    5.0\\
\hline
Total  &    8.6\\
\hline
\hline
\end{tabular}
\label{table_syserr}
\end{center}
\end{table}
\subsection{Full exclusive cross section}
The exclusive cross section of the ${\gamma}_vn\left(p\right) \rightarrow p {\pi}^{-}\left(p\right)$ process can be calculated from the acceptance-corrected yield of the exclusive events as 
\begin{widetext}
\begin{equation}
\begin{split}
\frac{d^{2}\sigma^{ex}}{d\Omega_{\pi}^{\text{c.m.\!}}}&=\frac{1}{\Gamma_{\upsilon}\left(W,Q^{2}\right)}\frac{d^{4}\sigma}{dWdQ^{2}d\Omega_{\pi^{\text{c.m.\!}}}}=\frac{(\Delta N_{full}\left(W,Q^{2},\cos{\theta^{\text{c.m.\!}}_{\pi}},\phi^{\text{c.m.\!}}_{\pi}\right)-S_{ratio}\Delta N_{empty}\left(W,Q^{2},\cos{\theta^{\text{c.m.\!}}_{\pi}},\phi^{\text{c.m.\!}}_{\pi}\right))R_{BC}}{\Gamma_{\upsilon}\left(W,Q^{2}\right)A_{RC}(W,Q^{2},\cos{\theta^{\text{c.m.\!}}_{\pi}},\phi^{\text{c.m.\!}}_{\pi})\Delta W \Delta Q^{2} \Delta \cos{\theta^{\text{c.m.\!}}_{\pi}} \Delta \phi^{\text{c.m.\!}}_{\pi}\mathcal{L}_{int}},
\end{split}
\label{eq:diffcrossex}
\end{equation}
\end{widetext}
where $\Delta N_{full}$ and $\Delta N_{empty}$ represent the number of exclusive events inside each four-dimensional bin ($W,Q^{2},\cos{\theta^{\text{c.m.\!}}_{\pi}},\phi^{\text{c.m.\!}}_{\pi}$) for the target with (full) and without (empty) LD$_{2}$ (liquid deuterium), respectively. The virtual photon flux $\Gamma_{\upsilon}\left(W,Q^{2}\right)$ is defined in Appendix \ref{appendix:a}. $A_{RC}(W,Q^{2},\cos{\theta^{\text{c.m.\!}}_{\pi}},\phi^{\text{c.m.\!}}_{\pi})$ was calculated from Eq.~\eqref{eq:RDacceptance}, and $S_{ratio}$ is the integrated Faraday cup ratio between the target with and without LD$_{2}$, which was calculated to be
\begin{equation}
S_{ratio}=\frac{Q_{total}}{Q_{empty}}=\frac{4.420\;\text{mC}}{0.467\;\text{mC}}=9.465,
\label{eq:scalefactor}
\end{equation}
where $Q_{total}$ and $Q_{empty}$ are the live-time accumulated charge in the Faraday cup for runs with LD$_{2}$ and empty target, respectively. Live time refers to the fact that the signal is integrated when the DAQ system is live, and the charge is corrected for the DAQ dead time. In addition, the bin-centering correction factor $R_{BC}$ was calculated from Eq.~\eqref{eq:RBC}. 
$\Delta W$, $\Delta Q^{2}$, $\Delta \cos{\theta^{\text{c.m.\!}}_{\pi}}$, and $\Delta \phi^{\text{c.m.\!}}_{\pi}$ are the bin widths of the corresponding kinematic variables. $\mathcal{L}_{int}$ is the integrated luminosity calculated by 
\begin{equation}
\begin{split}
\mathcal{L}_{int}&=N_{e}N_{d}=\left(\frac{Q_{total}}{e}\right)\times\left(\frac{N_{A}d_{T}l_{T}}{M_{d}}\right)\\
&=2.6788\times10^{39}\;\text{cm}^{-2},
\label{eq:luminosity}
\end{split}
\end{equation}
where $e$ is the elementary charge, $d_{T}$ is the density of the liquid deuterium, $l_{T}$ is the target length, $N_{A}$ is Avogadro's number, and $M_{d}$ is the molar mass of deuterium.

\subsection{Exclusive quasifree cross section}
The exclusive quasifree cross section was calculated by 
\begin{equation}
\frac{d^{2}\sigma^{qf}}{d\Omega_{\pi}^{\text{c.m.\!}}}=\frac{d^{2}\sigma^{cut}}{d\Omega_{\pi}^{\text{c.m.\!}}}\frac{1}{r\left(W,Q^{2},\cos{\theta^{\text{c.m.\!}}_\pi},\phi^{\text{c.m.\!}}_\pi\right)},
\label{eq:diffcrossqf}
\end{equation}
where $\frac{d^2\sigma^{cut}}{d\Omega_\pi^{\text{c.m.\!}}}$ is the cross section calculated after applying the $\lvert\vec{p}_{s}\lvert<200\;\text{MeV}$ cut and $r\left(W,Q^{2},\cos{\theta^{\text{c.m.\!}}_\pi},\phi^{\text{c.m.\!}}_\pi\right)$ obtained from Eq.~\eqref{eq:r} denotes the factor to correct for the good quasifree events outside the $\lvert\vec{p}_{s}\lvert<200\;\text{MeV}$ cut. Based on the yield of the events surviving this cut, the cross section was extracted via
\begin{widetext}
\begin{equation}
\frac{d^{2}\sigma^{cut}}{d\Omega_{\pi}^{\text{c.m.\!}}}=\frac{(\Delta N^{cut}_{full}\left(W,Q^{2},\cos{\theta^{\text{c.m.\!}}_\pi},\phi^{\text{c.m.\!}}_\pi\right)-S_{ratio}\Delta N^{cut}_{empty}\left(W,Q^{2},\cos{\theta^{\text{c.m.\!}}_\pi},\phi^{\text{c.m.\!}}_\pi\right))R_{BC}}{\Gamma_{\upsilon}\left(W,Q^{2}\right)A^{cut}_{RC}(W,Q^{2},\cos{\theta^{\text{c.m.\!}}_\pi},\phi^{\text{c.m.\!}}_\pi)\Delta W \Delta Q^{2} \Delta \cos{\theta^{\text{c.m.\!}}_\pi} \Delta \phi^{\text{c.m.\!}}_\pi\mathcal{L}_{int}}\; ,\\
\label{eq:diffcroscut}
\end{equation}
\end{widetext}
where ``cut" designates the corresponding quantities that were calculated with the $\lvert\vec{p}_{s}\lvert<200\;\text{MeV}$ cut. For the quasifree events, the radiative-corrected acceptance $A^{cut}_{RC}(W, Q^{2}, \cos\theta^{\text{c.m.\!}}_\pi, \phi^{\text{c.m.\!}}_\pi)$ was calculated analogously to Eq.~(\ref{eq:RDacceptance}) by
\begin{equation}
\begin{split}
&A^{cut}_{RC}(W, Q^{2}, \cos\theta^{\text{c.m.\!}}_\pi, \phi^{\text{c.m.\!}}_\pi)\\
&=\frac{N^{(\lvert\vec{p}_{s}\lvert<200\;\text{MeV}) Rad}_{rec}(W, Q^{2}, \cos\theta^{\text{c.m.\!}}_\pi, \phi^{\text{c.m.\!}}_\pi)}{N^{(\lvert\vec{p}_{s}\lvert<200\;\text{MeV}) noRad}_{thrown}(W, Q^{2}, \cos\theta^{\text{c.m.\!}}_\pi, \phi^{\text{c.m.\!}}_\pi)}.
\label{eq:efficiency}
\end{split}
\end{equation}


\section{RESULTS AND DISCUSSION }
\label{results}
In this section we quantify the kinematically identified FSI contributions and present the extracted twofold differential cross sections for the reaction $\gamma_vn(p)\rightarrow p\pi^{-}(p)$ together with their quasifree contributions. From these results, we have determined the exclusive $\pi^-p$ electroproduction structure functions $\sigma_T + \epsilon \sigma_L$, $\sigma_{TT}$, and $\sigma_{LT}$ and their Legendre moments and explore their sensitivity to contributions from particular excited states of the nucleon. 

\subsection{\label{qf} Kinematically defined quasifree contributions}
The comparison of the missing momentum $\lvert\vec{p}_{s}\lvert $ distributions of the experimental and simulated data shows that the quasifree process is absolutely dominant and can hence be kinematically isolated in the $\lvert\vec{p}_{s}\lvert<200\;\text{MeV}$ region (see Sec.~\ref{subsec:level4-2}). However, for $\lvert\vec{p}_{s}\lvert>$200~MeV, FSI contributions appear and become larger with increasing $\vert\vec{p}_{s}\vert$. Beyond the extraction of the full exclusive and quasifree differential cross sections, this comparison allows us to calculate the final-state-interaction contribution factor $R_{FSI}$ for each four-dimensional bin $(W, Q^{2}, \cos\theta^\text{c.m.\!}_\pi, \phi^\text{c.m.\!}_\pi)$, which is kinematically defined based only on the experimental data. Hence, this factor provides information on the fraction of kinematically identified final state interactions in the fully exclusive process defined by
\begin{equation}
R_{FSI}(W, Q^{2}, \cos\theta^\text{c.m.\!}_{\pi}, \phi_\pi^\text{c.m.\!})
=\frac{\frac{d^{2}\sigma^{qf}}{d\Omega_{\pi}^{\text{c.m.\!}}}}{\frac{d^{2}\sigma^{ex}}{d\Omega_{\pi}^{\text{c.m.\!}}}}.
\label{eq:RFSI}
\end{equation}

\begin{figure*}[ht!]  
\centerline{\includegraphics[width=0.9\textwidth]{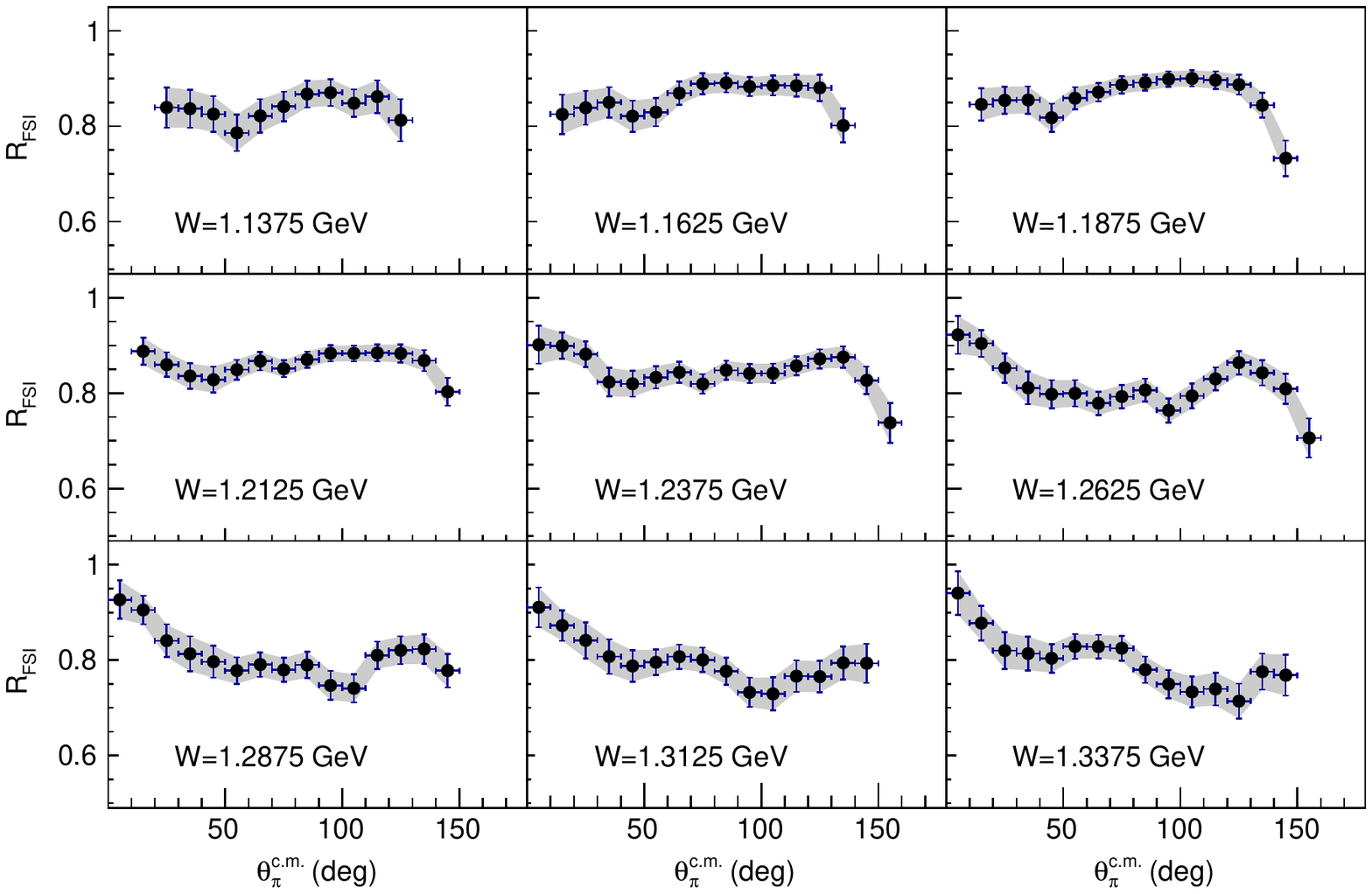}}
\caption{\justify{ The final-state-interaction contribution factor $R_{FSI}$ determined from experiment data which account for the FSI in deuteron target as a function of $\theta_{\pi}^{\text{c.m.\!}}$ for individual $W_{f}$ bins in 0.025~GeV increments in the range of 1.1375 $<$ $W$ $<$ 1.3375~GeV for 0.4 $<Q^{2}<$ 0.6~GeV$^2$. The gray shaded regions represent the corresponding systematic uncertainties.}}
\label{f:R1}
\end{figure*}

\begin{figure*}[hbt!]
\centerline{\includegraphics[width=0.9\textwidth]{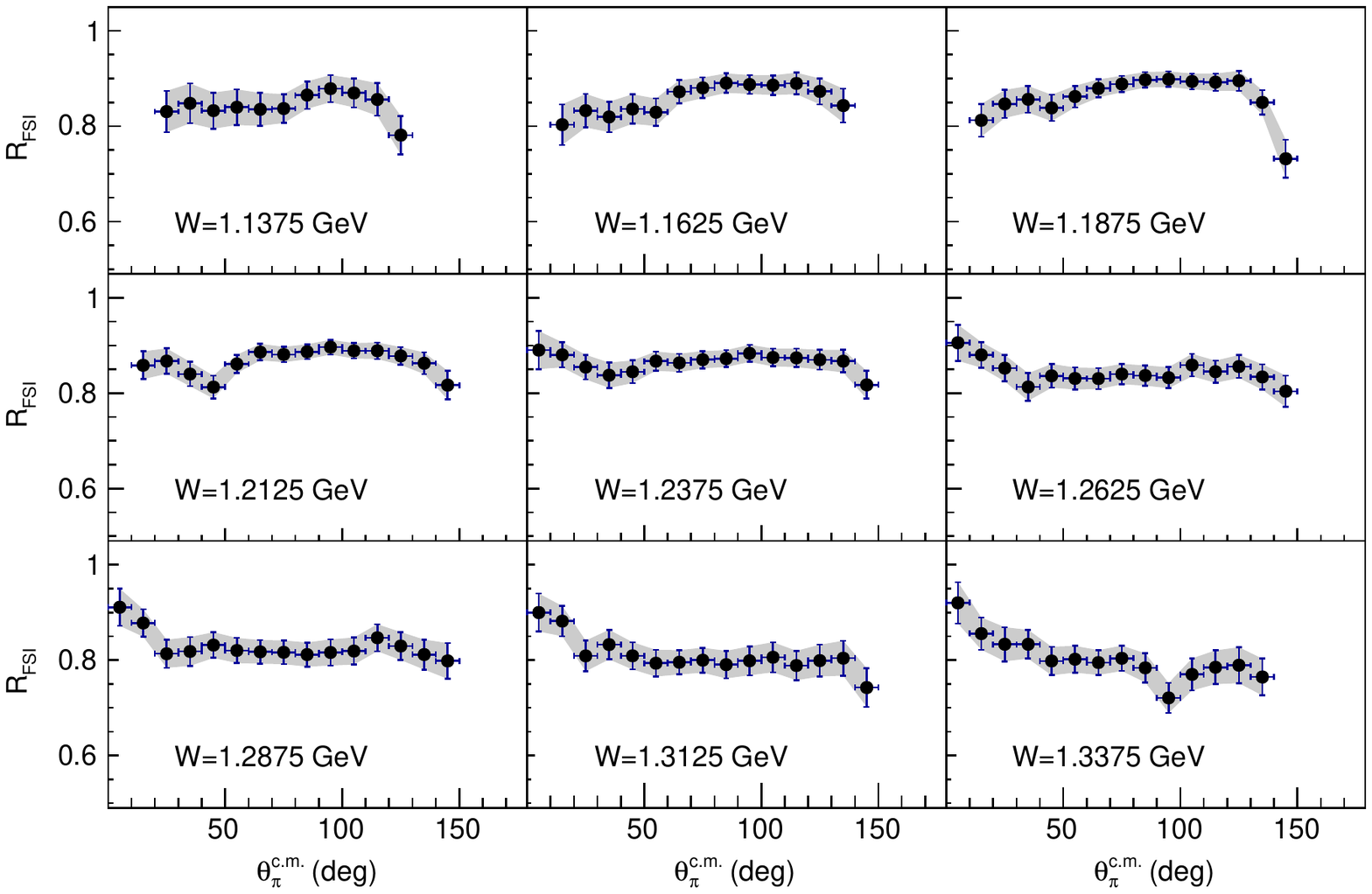}}
\caption{\justify{The final-state-interaction contribution factor $R_{FSI}$ determined from experiment data which account for the FSI in deuteron target as a function of $\theta_{\pi}^{\text{c.m.\!}}$ for individual $W_{f}$ bins in 0.025~GeV increments in the range of 1.1375 $<$ $W$ $<$ 1.3375~GeV for 0.6 $<Q^{2}<$ 0.8~GeV$^2$. The gray shaded regions represent the corresponding systematic uncertainties.}}
\label{f:R2}
\end{figure*}

\begin{figure*}[hbt!]
\centerline{\includegraphics[width=0.9\textwidth]{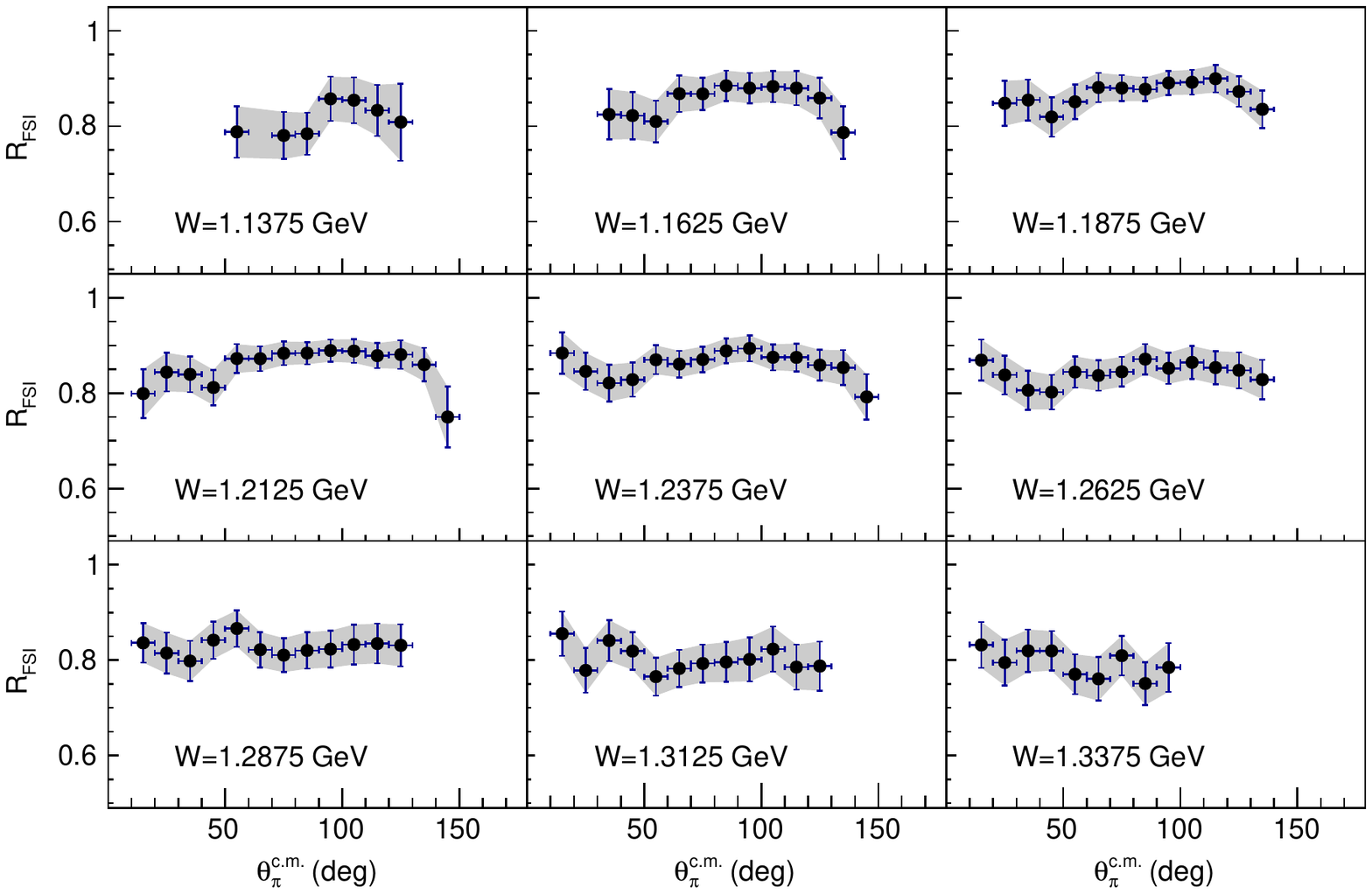}}
\caption{\justify{The final-state-interaction contribution factor $R_{FSI}$ determined from experiment data which account for the FSI in deuteron target as a function of $\theta_{\pi}^{\text{c.m.\!}}$ for individual $W_{f}$ bins in 0.025~GeV increments in the range of 1.1375 $<W<$ 1.3375~GeV for 0.8 $<Q^{2}<$ 1.0~GeV$^{2}$. The gray shaded regions represent the corresponding systematic uncertainties.}}
\label{f:R3}
\end{figure*}

In order to present the most meaningful $R_{FSI}$ values possible, the exclusive events were binned in $W_{f}=p^{\mu}+(\pi^{-})^{\mu}$ to be consistent with the binning of quasifree events, even though $W_{f}$ for exclusive events with final state interactions is different from the true $W=W_{i}$.
 Typical $\phi_{\pi}^{\text{c.m.\!}}$ integrated $R_{FSI}$ versus $\theta_{\pi}^{\text{c.m.\!}}$ distributions are plotted for different $W$ and $Q^{2}$ bins and shown in Figs.~\ref{f:R1}-\ref{f:R3}. It turns out that the kinematically defined final-state-interaction contribution for the reaction $\gamma_vn(p)\rightarrow p\pi^{-}(p)$ and the ``e1e" kinematics is on average about $10\%-20\%$. These results are of interest for reaction models that describe FSIs for the $\pi^-p$ final state with a deuteron target  \cite{Tarasov:2011ec, Nakamura:2018cst, Kamano:2016bgm, Anisovich:2017afs}.

\subsection{Differential $\gamma_vn(p)\rightarrow p\pi^{-}(p)$ cross sections}
\begin{figure*}[hbt!]
\centering
\includegraphics[width=1.0\textwidth]{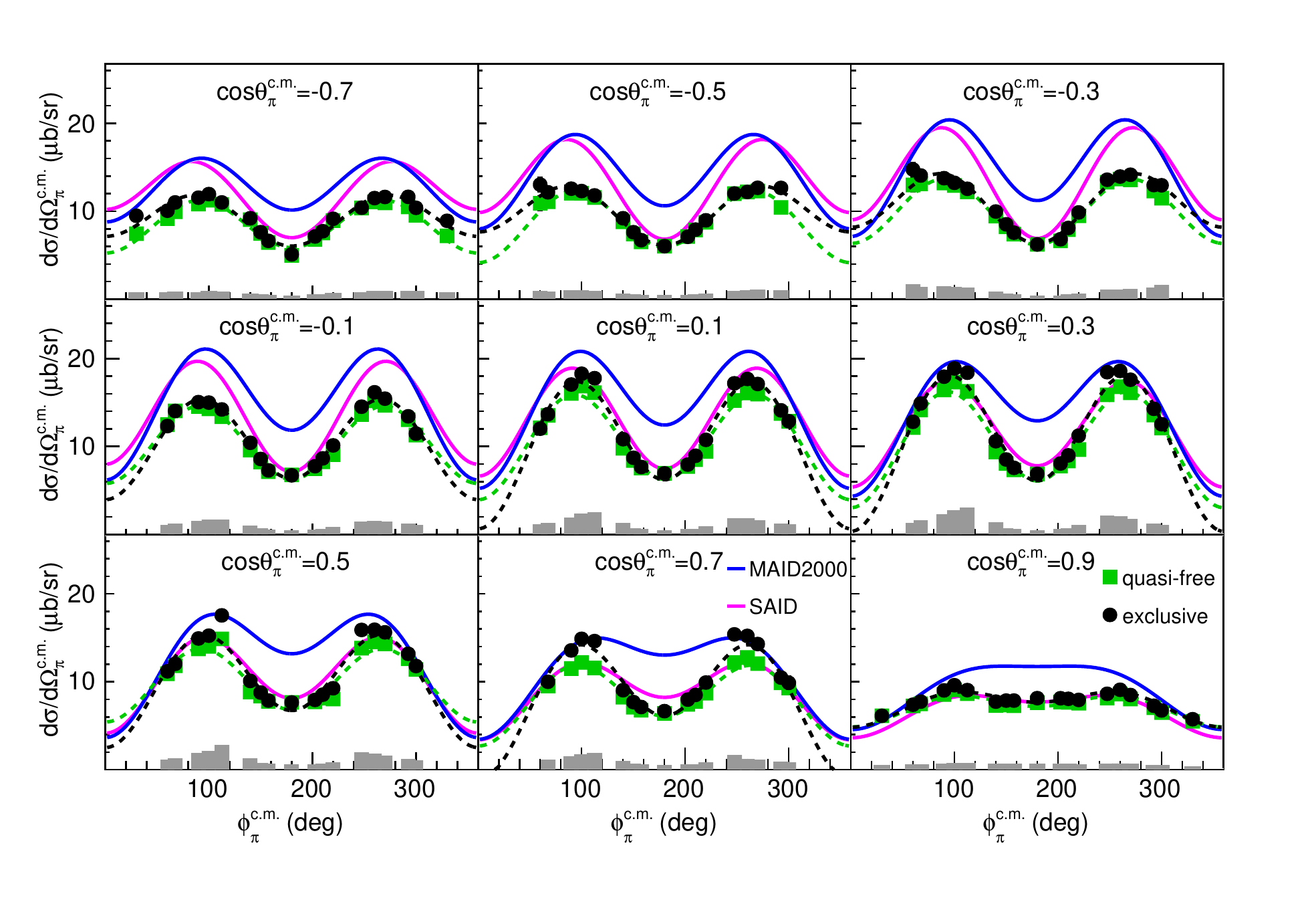}
\caption{\justify{Fully exclusive (black points) and quasi-free (green squares) cross sections in $\mu$b/sr for $W=$ 1.2125~GeV and $Q^{2}=$ 0.5~GeV$^2$. The $\phi_{\pi}^{\text{c.m.\!}}$-dependent cross sections are shown in each $\cos{\theta^{\text{c.m.\!}}_{\pi}}$ bin and the color-matched dashed lines represent the fits to the cross sections by the function $a+b\cos2\phi^{\text{c.m.\!}}_{\pi}+c\cos\phi^{\text{c.m.\!}}_{\pi}$. The magenta and blue solid lines show the SAID ~\cite{refSAID,Arndt:2009nv} and MAID2000~\cite{refMAID2000} model predictions, respectively. The gray bars at the bottom of each subplot quantify the systematic uncertainties of each cross section point and the statistical uncertainties are typically smaller than the data point markers.}}
\label{fig:secR1}       
\end{figure*}

\begin{figure*}[!hbt]
\centering
\includegraphics[width=1.0\textwidth]{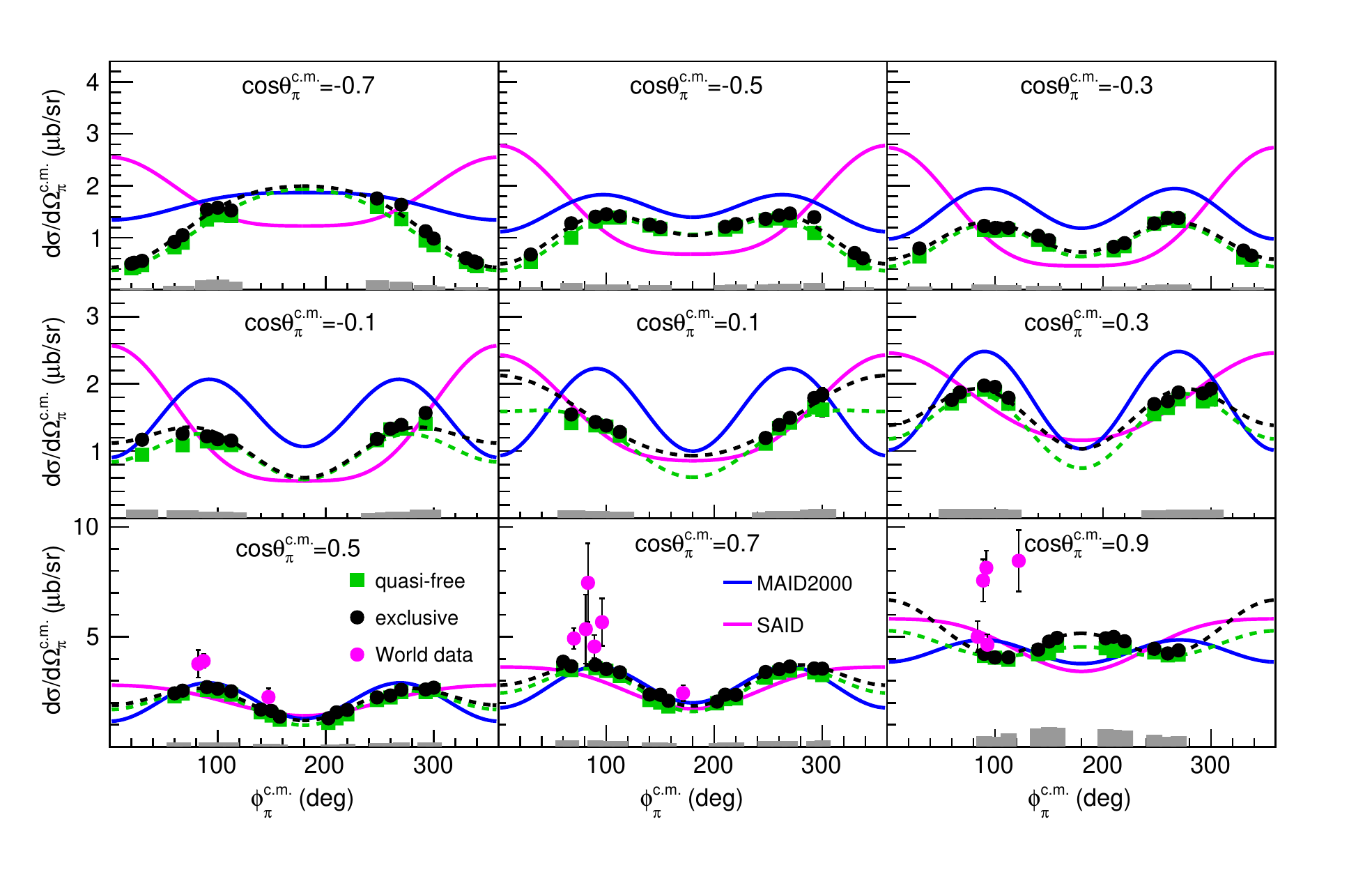}
\caption{\justify{Full exclusive (black points) and quasifree (green squares) cross sections in $\mu$b/sr are presented for $W$ $=$ 1.5125~GeV and $Q^{2}$ $=$ 0.5~GeV$^2$. The $\phi_{\pi}^{\text{c.m.\!}}$-dependent cross sections are shown in each $\cos{\theta^{\text{c.m.\!}}_{\pi}}$ bin and the color-matched dashed lines represent the fits to the cross sections by the function $a+b\cos2\phi^{\text{c.m.\!}}_{\pi}+c\cos\phi^{\text{c.m.\!}}_{\pi}$. The magenta and blue solid lines show the SAID~\cite{refSAID,Arndt:2009nv} and MAID2000~\cite{refMAID2000} model predictions, respectively. The magenta points show the previous available world data~\cite{Morris:1979mj,Wright:1980ff, Gaskell:2001fn}. The gray bars at the bottom of each subplot quantify the systematic uncertainties of each cross section point and the statistical uncertainties are typically smaller than the data point markers.}}
\label{fig:secR2}       
\end{figure*}

\begin{figure*}[!hbt]
\centering
\includegraphics[width=1.0\textwidth]{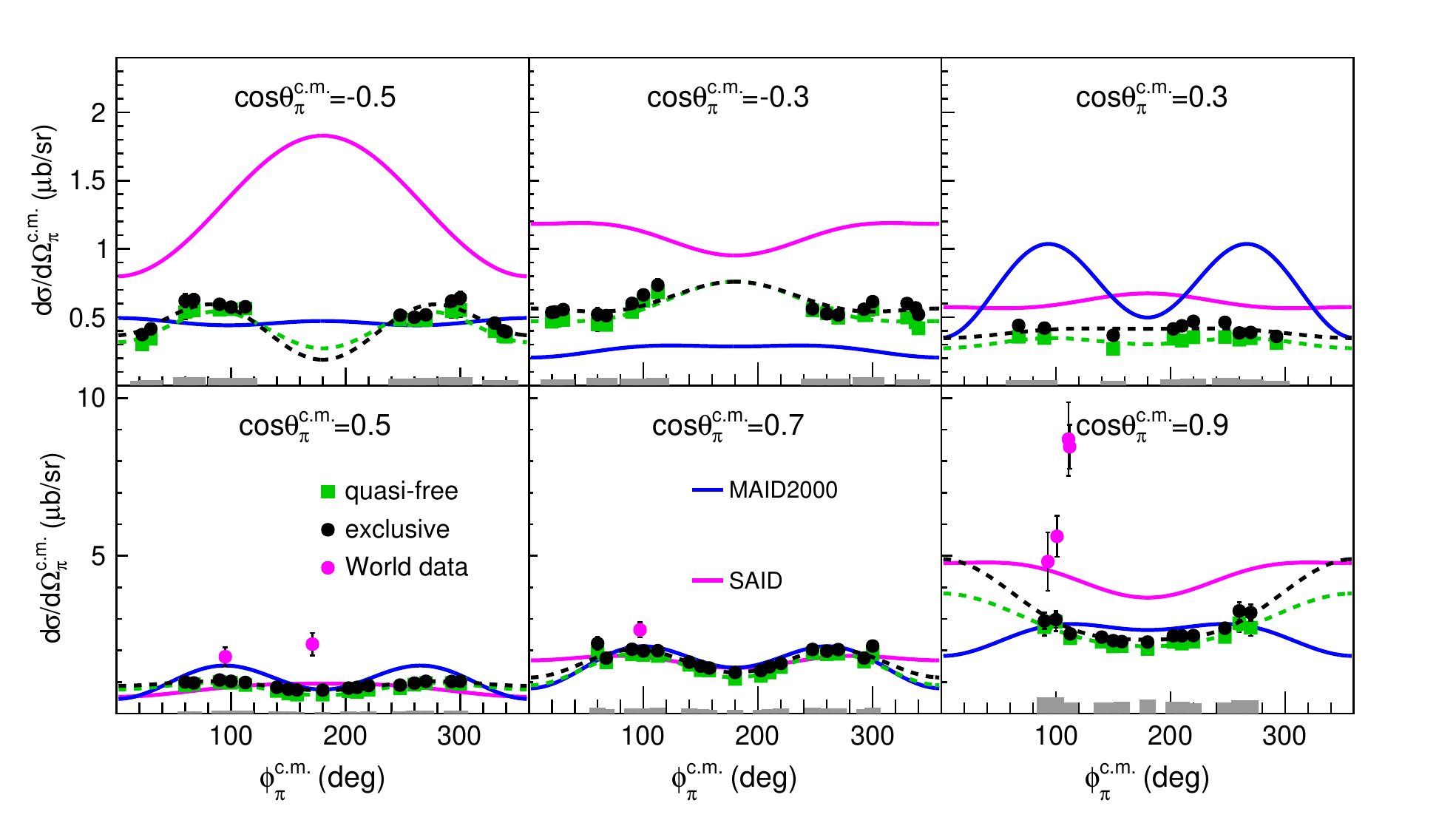}
\caption{\justify{Full exclusive (black points) and quasifree (green squares) cross sections in $\mu$b/sr are presented for $W$ $=$ 1.6625~GeV and $Q^{2}$ $=$ 0.5~GeV$^2$. The $\phi_{\pi}^{\text{c.m.\!}}$-dependent cross sections are shown in each $\cos{\theta^{\text{c.m.\!}}_{\pi}}$ bin (the bins with reasonable statistics are shown) and the color-matched dashed lines represent the fits to the cross sections by the function $a+b\cos2\phi^{\text{c.m.\!}}_{\pi}+c\cos\phi^{\text{c.m.\!}}_{\pi}$. The magenta and blue solid lines show the SAID~\cite{refSAID,Arndt:2009nv} and MAID2000~\cite{refMAID2000} model predictions, respectively. The magenta points show the previous available world data~\cite{Morris:1979mj,Wright:1980ff, Gaskell:2001fn}. The gray bars at the bottom of each subplot quantify the systematic uncertainties of each cross section point and the statistical uncertainties are typically smaller than the data point markers.}}
\label{fig:secR3}       
\end{figure*}

Fully exclusive differential $\gamma_vn(p)\rightarrow p\pi^- (p)$ cross sections off bound neutrons and their quasifree contributions estimated as described in Sec.~\ref{qf} are now available for $W<$ 1.825~GeV and 0.4 $<Q^2<$ 1.0~GeV$^2$. The numerical results can be found in the CLAS Physics Database (CLAS DB) \cite{CLAS:DB}. Figures~\ref{fig:secR1}-\ref{fig:secR3} show representative examples for the exclusive differential $\pi^- p$ electroproduction differential cross sections and quasifree contributions in the $Q^2$ bin from 0.4 to 0.6~GeV$^2$ and selected $W$ bins corresponding to the first, second, and third resonance regions. Our results are compared with the expectations from the SAID \cite{refSAID} and MAID2000 \cite{refMAID2000} model predictions.    

In Figs.~\ref{fig:secR1}-\ref{fig:secR3}, the exclusive and quasi-free cross sections are represented by the black filled circles and the filled green squares with error bars, respectively, and the corresponding systematic uncertainties are represented by the gray shaded bars at the bottom of each plot. The previously available data that were obtained with small-acceptance detectors \cite{Morris:1979mj,Wright:1980ff, Gaskell:2001fn} can also be seen in the 1.51 and 1.66~GeV $W$ bins. The very limited coverage in $\phi_{\pi}^\text{c.m.\!}$ of these older data, together with their substantial uncertainties, prevented the extraction of structure functions from these data. Our measurements extend the $\phi^\text{c.m.\!}_\pi$ coverage considerably. In most bins of $W$, $Q^2$, and $\cos \theta^\text{c.m.\!}_\pi$, nearly complete coverage over the azimuthal $\phi^\text{c.m.\!}_\pi$ angle has been achieved. However, there are not enough data in some $(W, Q^2, \cos \theta^\text{c.m.\!}_\pi)$ bins to provide overall statistically meaningful cross sections, particularly at very low and high $\phi^\text{c.m.\!}_\pi$ angles. These $\phi^{\text{c.m.\!}}_\pi$ areas can be best identified by excluding the phase space where the relative acceptance uncertainties determined by the simulation are smaller than 2\%. 

According to the results in Figs.~\ref{fig:secR1}-\ref{fig:secR3}, the fully exclusive cross sections in all ($W, Q^{2}, \cos\theta^{\text{c.m.\!}}_\pi$) bins is always larger than the quasifree cross section. In fact, the quasifree cross sections represent the contribution to the full cross section after contributions with FSI have been removed. As expected, the measured yields and the corresponding cross sections shown in Figs.~\ref{fig:secR1}-\ref{fig:secR3} decrease with increasing $Q^{2}$ and are symmetrically distributed with respect to $\phi_{\pi}^{\text{c.m.\!}}=180^{\circ}$.
The measured cross sections are compared with the predictions of two models, SAID~\cite{refSAID} and MAID2000~\cite{refMAID2000}, which successfully describe the cross sections of single-pion production off the free proton in the first and second resonance regions, but their comparison with the $\pi^-p$ electroproduction cross sections from the measurements reported here demonstrates substantial differences between the expectations from both models and our data, as well as between the predictions from SAID and MAID2000 themselves. Therefore, our $\pi^-p$ electroproduction measurements off bound neutrons provide new information on the dynamics of the $\gamma_vn(p)\rightarrow p\pi^{-}(p)$ reaction that was not captured by the SAID~\cite{refSAID} or the MAID2000~\cite{refMAID2000} reaction models, which predict cross sections off the free neutron. 

\subsection{\label{subsec:sf}Structure functions}

The exclusive structure functions $\sigma_T + \epsilon \sigma_L$, $\sigma_{TT}$, and $\sigma_{LT}$ for $\pi^-p$ electroproduction were determined assuming the one-photon-exchange approximation 
\cite{refMAID2000,refMAID2007} and by fitting the $\phi^{\text{c.m.\!}}_{\pi}$ angular distributions in each bin of $W$, $Q^2$, and $\cos\theta^\text{c.m.\!}_\pi$ according to

\begin{equation}
\begin{split}
&\frac{d^2\sigma}{d\Omega_{\pi}^{\text{c.m.\!}}}=a+b\cos 2\phi^{\text{c.m.\!}}_{\pi}+c\cos\phi^{\text{c.m.\!}}_{\pi},\\
&a=\sigma_{T}+\epsilon\sigma_{L},\\
&b={\sin}^{2}\theta_{\pi}\epsilon\sigma_{TT},\; \text{and}\\
&c=\sin\theta_{\pi}\sqrt{2\epsilon\left(1+\epsilon\right)}\sigma_{LT},
\label{eq:strucF}
\end{split}
\end{equation}
where $\epsilon$, defined in Appendix~\ref{appendix:a}, is the degree of transverse polarization of the virtual photon, $T$ and $L$ represent the transverse and longitudinal, $TT$ the transverse-transverse, and $LT$ the transverse-longitudinal interference structure functions. The $\phi^{\text{c.m.\!}}_{\pi}$ angular dependence expressed in Eq.~(\ref{eq:strucF}) is a direct consequence of the single-photon-exchange approximation for exclusive electroproduction dynamics. The good quality of our data description achieved using Eq.~(\ref{eq:strucF}), shown for representative bins in Figs.~\ref{fig:secR1}-\ref{fig:secR3} by the black and green dashed lines, supports a reliable extraction of the full exclusive as well as the quasifree $\pi^-p$ electroproduction cross sections off bound neutrons.

\begin{figure*}[!hbt]
\centering
\includegraphics[width=0.95\textwidth]{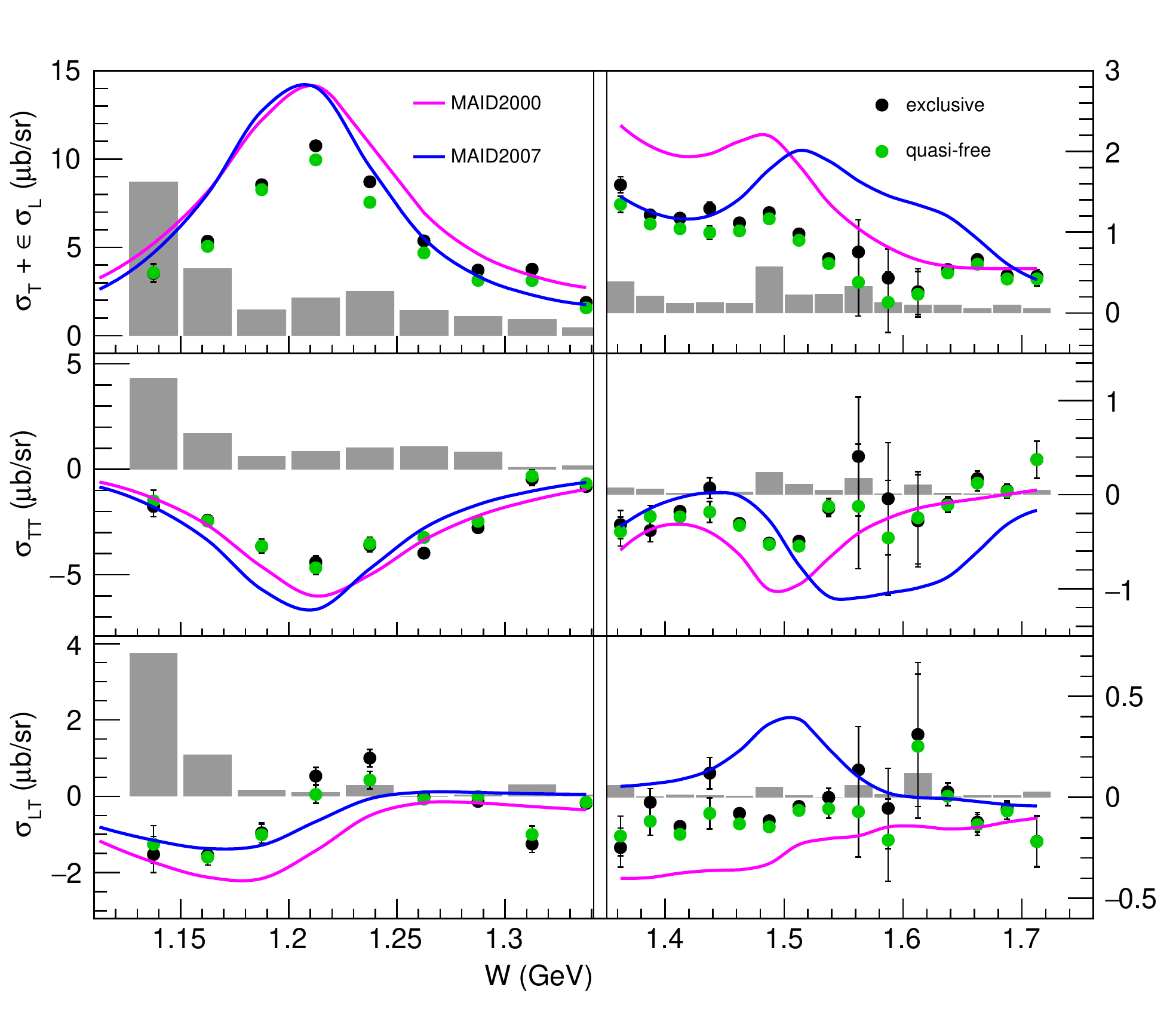}
\caption{Example of the $W$-dependent $\sigma_{T}+\epsilon\sigma_{L}$, $\sigma_{TT}$, and $\sigma_{LT}$ structure functions at $\cos{\theta^{\text{c.m.\!}}_{\pi}}=-0.3$ and $Q^{2}=$ 0.5~GeV$^2$ that were extracted from the fully exclusive (black points) and quasifree (green squares) cross sections. For $W>$ 1.35~GeV, the rightmost $y$-axis scale is used. The data are compared with the MAID2000~\cite{refMAID2000} (magenta line) and MAID2007~\cite{refMAID2007} (blue line) models. The gray bars represent the corresponding systematic uncertainties. The origin of the large gray bars is described in Sec.~\ref{subsec:sf}.}
\label{fig:SF_costheta3}    
\end{figure*}

Numerical data on the structure functions $\sigma_T + \epsilon \sigma_L$, $\sigma_{TT}$, and $\sigma_{LT}$, as determined from our results, can be found in the CLAS Physics Database \cite{CLAS:DB} along with the differential cross sections
(6475 differential cross section data points have been added in total). 
Representative examples for the $W$ dependencies of these structure functions are shown for the $Q^2$ bin from 0.4 to 0.6~GeV$^2$ in Fig. \ref{fig:SF_costheta3}, together with the expectations from the SAID \cite{refSAID} and MAID \cite{refMAID2000,refMAID2007} models. The larger systematic uncertainties (gray error bars) in Fig. \ref{fig:SF_costheta3} appear due to the uncertainties of the fit parameters 
in Eq. (\ref{eq:strucF}). Since there are not enough differential cross section data points in some $(W, Q^2, \cos \theta_\pi^\text{c.m.\!})$ bins at very low and high $\phi_{\pi}^{\text{c.m.\!}}$ angles, the structure function fits are not well constrained, which can lead to large uncertainties of the fit parameters.
The most prominent feature is the substantial contribution from the $\Delta(1232)3/2^+$ resonance, which is clearly exhibited in the $W$ dependencies of all three exclusive structure functions in the first resonance region. Beyond that, all structure functions exhibit shoulders in the second resonance region near $W$ $\approx$ 1.5~GeV, which is suggestive of interferences between nucleon excitations and nonresonant contributions or coupled-channels effects related
to the hadronic interaction of the $\pi^-p$ final state with the $\eta n$ channel that opens at $W>$ 1.5~GeV. The predictions from the MAID2000 \cite{refMAID2000} and MAID2007 \cite{refMAID2007} models are close in the first resonance region, where they are also in a reasonable agreement with our data. However, in the 1.45 $<W<$ 1.65~GeV regions the prediction of these models are very different and far from the data (see Fig. \ref{fig:SF_costheta3}), suggesting that our results provide new information on the $\pi^-p$ electroproduction dynamics, which so far has not been captured by either the SAID \cite{refSAID} or MAID \cite{refMAID2000,refMAID2007} models.      

\begin{figure*}[!hbt]
\centering
\includegraphics[width=1.0\textwidth, height=0.75\textwidth]{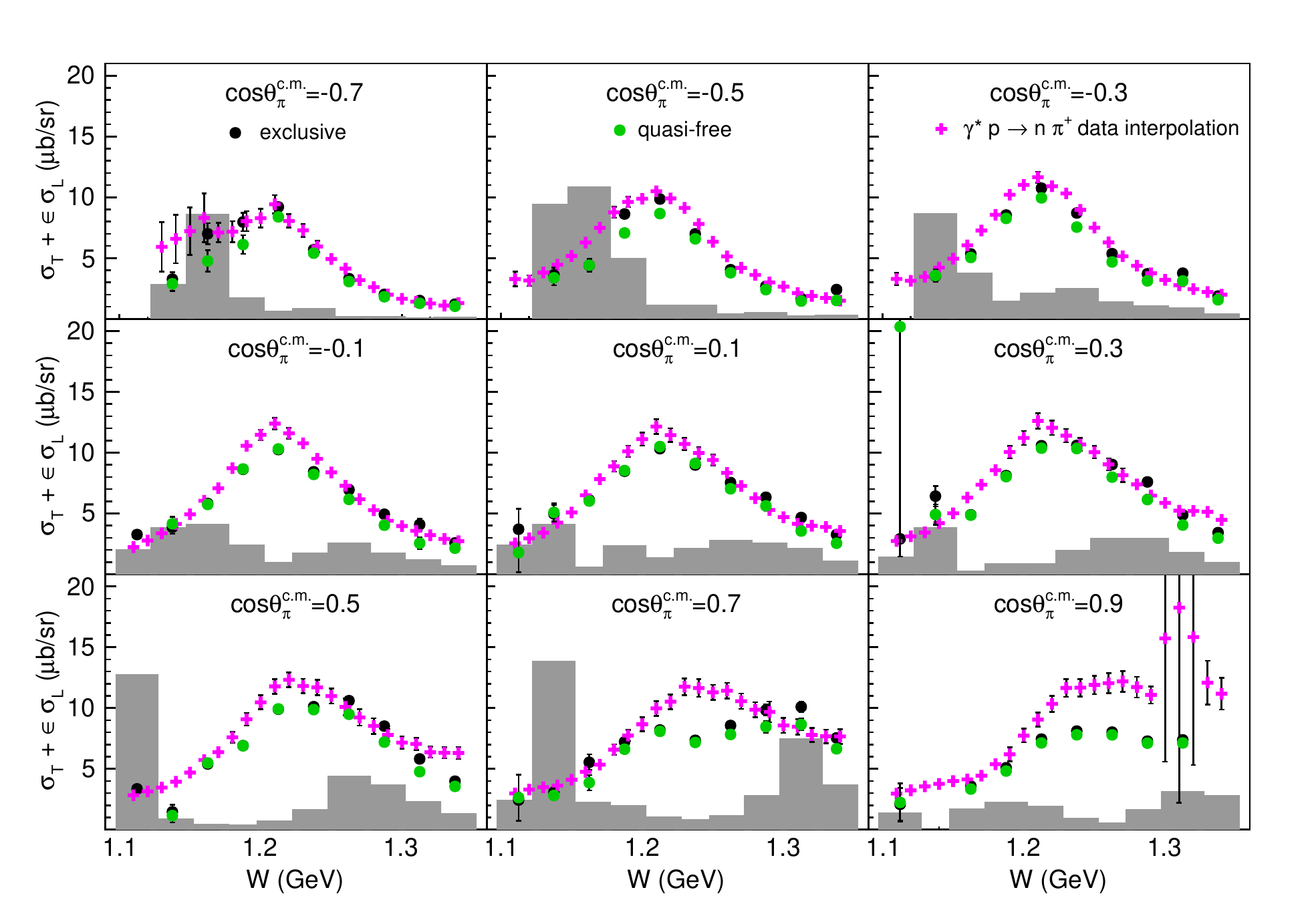}
\caption{\justify{Examples of the $W$-dependent $\sigma_{T}+\epsilon\sigma_{L}$ structure function at $Q^{2}=$ 0.5~GeV$^2$ for various $\cos{\theta^{\text{c.m.\!}}_{\pi}}$ that were extracted from the fully exclusive (black points) and quasifree (green squares) cross sections up to $W<$ 1.35~GeV and compared with the model-independent interpolation of all available CLAS $\pi^+n$ electroproduction cross section results (magenta points). The gray bars at the bottom represent the corresponding systematic uncertainties of the fully exclusive data (black points). The origin of the large gray bars is described in Sec.~\ref{subsec:sf}.}}
\label{fig:SF_alltheta}      
\end{figure*}

The $W$ dependencies of the so-called unpolarized $\sigma_T + \epsilon \sigma_L$ structure function for exclusive $\pi^-p$ electroproduction and their respective quasifree contributions are shown in Fig.~\ref{fig:SF_alltheta} compared to the interpolated results on the unpolarized structure function of $\pi^+n$ electroproduction off free protons \cite{Bulgakov:2021brd} in the $Q^2$ bin from 0.4 to 0.6~GeV$^2$ and for $W<$ 1.35~GeV, which corresponds to the first resonance region. Here, the $\pi N$ electroproduction amplitudes at forward angles are driven by both resonant and nonresonant parts, while with increasing pion CM angles the $\Delta(1232)3/2^+$ resonance contribution becomes dominant \cite{refMAID2000,refMAID2007,Aznauryan:2009mx}. Since this isospin $I$=3/2 resonance can only be excited through the isovector component of the electromagnetic current, the $\Delta(1232)3/2^+$ electroexcitation amplitudes off the free proton and the free neutron should be equal owing to the isospin invariance of the strong interaction. Therefore, in the kinematic areas where the  $\Delta(1232)3/2^+$ resonance dominates, the unpolarized structure function of quasifree $\pi^-p$ electroproduction off neutrons should be equal to that of $\pi^+n$ electroproduction off free protons. This expectation is reflected in the results of our measurements as demonstrated in Fig.~\ref{fig:SF_alltheta}. At $\cos\theta_{\pi}^{\text{c.m.}}<0.5$, the unpolarized structure functions for quasifree $\pi^-p$ electroproduction off neutrons are consistent within their uncertainties with the interpolated values of the unpolarized structure function for $\pi^+n$ electroproduction off free protons, which further supports the proper extraction of the quasi-free $\pi^-p$ cross sections from the measured data. Within the range of the small pion CM emission angles ($\cos\theta_{\pi}^{\text{c.m.}}>0.5$), sizable contributions from the nonresonance $t$-channel processes are responsible for the differences between $\pi^+n$ and  $\pi^-p$ cross sections.


\begin{figure*}[!hbt]
\centering
\includegraphics[width=1.0\textwidth, height=0.8\textwidth]{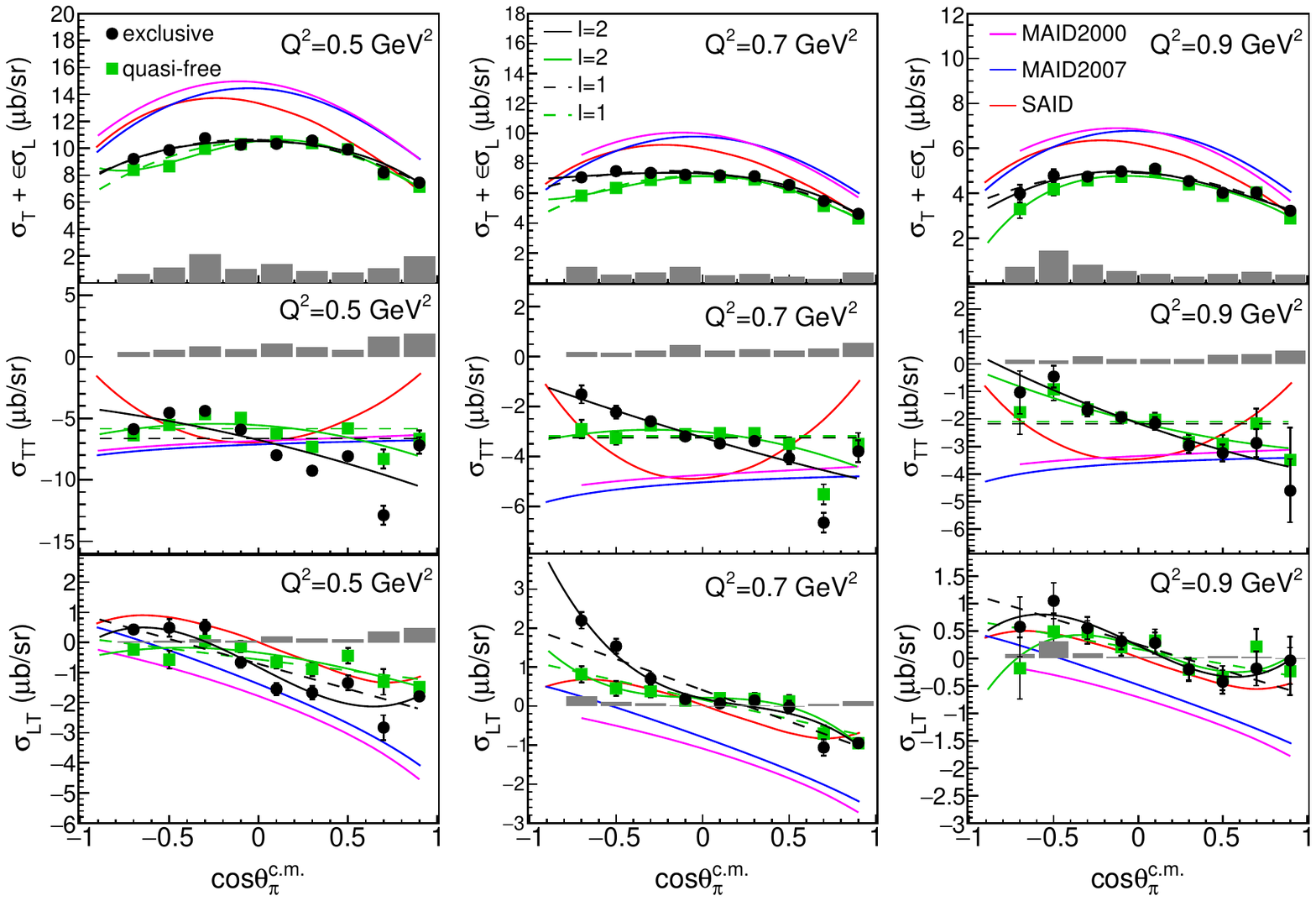}
\caption{\justify{Example of the $\cos\theta^{\text{c.m.\!}}_{\pi}$-dependent structure functions $\sigma_{T}+\epsilon\sigma_{L}$ (top row), $\sigma_{TT}$ (middle row), and $\sigma_{LT}$ (bottom row) for $W=$ 1.2125~GeV at $Q^{2}=$ 0.5~GeV$^2$ (left column), $Q^{2}=$ 0.7~GeV$^2$ (middle column), and $Q^{2}=$ 0.9~GeV$^2$ (right column) that were extracted for the exclusive (black points) and quasifree (green squares) cross sections and compared with the predictions of the SAID SM08 (red line), MAID2000 (magenta line), and MAID2007 (blue line) models. The solid gray bars represent the corresponding systematic uncertainties of the data. The Legendre polynomial expansions were fitted to the corresponding structure function data for $\pi^{-}p$ orbital momenta up to $l_{max}$=1 (black dashed lines) and $l_{max}$=2 (black solid lines).}}
\label{fig:legendre}      
\end{figure*}

\section{\label{sec:legendre}Legendre polynomial expansion}
In order to further explore the sensitivity of our data to the contributions from particular excited nucleon states, the angular dependencies  of the structure functions  in each ($W, Q^{2}$) bin were decomposed by the Legendre polynomials $P_{l}(\cos\theta^{\text{c.m.\!}}_{\pi}$). A general form of Legendre polynomial expansion can be written as 
\begin{equation}
\sigma_{T}+\epsilon\sigma_{L}\\
=\sum_{i=0}^{2l}A_{i}P_{i}(\cos\theta^{\text{c.m.\!}}_{\pi})
\label{eq:sigma0_gen}
\end{equation}
\begin{equation}
\sigma_{TT}=\sum_{i=0}^{2l-2}B_{i}P_{i}(\cos\theta^{\text{c.m.\!}}_{\pi})
\label{eq:sigmaLT2_gen}
\end{equation}
\begin{equation}
\sigma_{LT}=\sum_{i=0}^{2l-1}C_{i}P_{i}(\cos\theta^{\text{c.m.\!}}_{\pi}),
\label{eq:sigmaLT3_gen}
\end{equation}
where the Legendre moments $A_{l}(W)$, $B_{l}(W)$, and $C_{l}(W)$ have been obtained for the $\sigma_T + \epsilon \sigma_L$, $\sigma_{TT}$, and $\sigma_{LT}$ structure functions. $l$ is the orbital angular momentum of the $\pi^-$ relative to the proton.

A representative example for the angular dependencies of the structure functions at $W=$ 1.21~GeV within the covered $Q^2$ range is shown in Fig.~\ref{fig:legendre} in comparison with the MAID2000~\cite{refMAID2000}, MAID2007~\cite{refMAID2007}, and SAID~\cite{refSAID} reaction model predictions and a Legendre polynomial expansion up to $l=2$. Substantial differences between these model expectations and the experimental data, seen in all structure functions, emphasize again the value of the new results presented here.
The Legendre moments $A_{l}$, $B_{l}$, and $C_{l}$ can be associated with the magnetic ($M_{l\pm}$), electric ($E_{l\pm}$), and scalar ($S_{l\pm}$) $\pi N$ multipoles~\cite{Aznauryan:2011qj, raskin1989polarization}.
In the first resonance region, a Legendre moment decomposition truncated at $l_{max}=1$ already provides a reasonable description of the unpolarized structure function for $\pi^-p$, 
but fails to describe the angular dependencies of both the $\sigma_{TT}$ and $\sigma_{LT}$ structure functions (see Fig.~\ref{fig:legendre}). The angular dependencies of all structure functions are well reproduced with an $l_{max}=2$ truncation of the $\pi^-p$ orbital momenta. 

\begin{figure*}[!hbt]
\centering
\includegraphics[width=1.0\textwidth, height=0.75\textwidth]{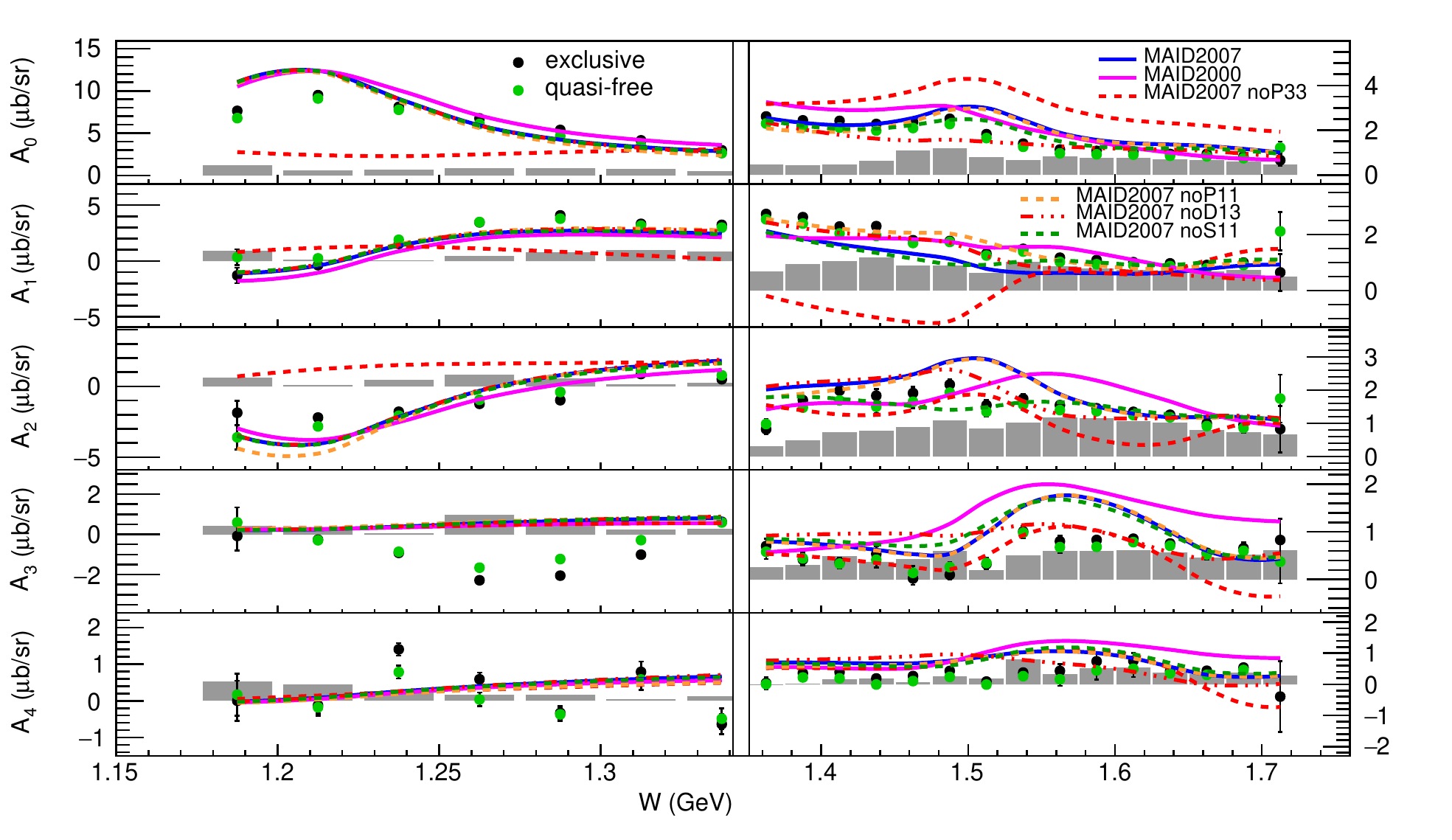}
\caption{The $W$-dependent Legendre moments $A_{i}$ of $\sigma_{T}+\epsilon \sigma_{L}$ in Eq.~\eqref{eq:sigma0_gen} up to $l_{max}=$~2 at $Q^{2}=$ 0.5~GeV$^2$ that are extracted from the exclusive (black points) and quasifree (green squares) cross sections. For $W>$ 1.35~GeV, the rightmost $y$-axis scale is used. The data are compared with the MAID2007~\cite{refMAID2007} and MAID2000~\cite{refMAID2000} models. The solid lines represent the full model calculations. The  dashed lines correspond to the MAID2007 model with specific resonant helicity amplitudes turned off [{\it e.g.}, noP33 indicates turning off the $\Delta(1232)3/2^+$ contributions]. The gray bars represent the corresponding systematic uncertainties of the data.}
\label{fig:LM_A}      
\end{figure*}

\begin{figure*}[!hbt]
\centering
\includegraphics[width=1.0\textwidth, height=0.7\textwidth]{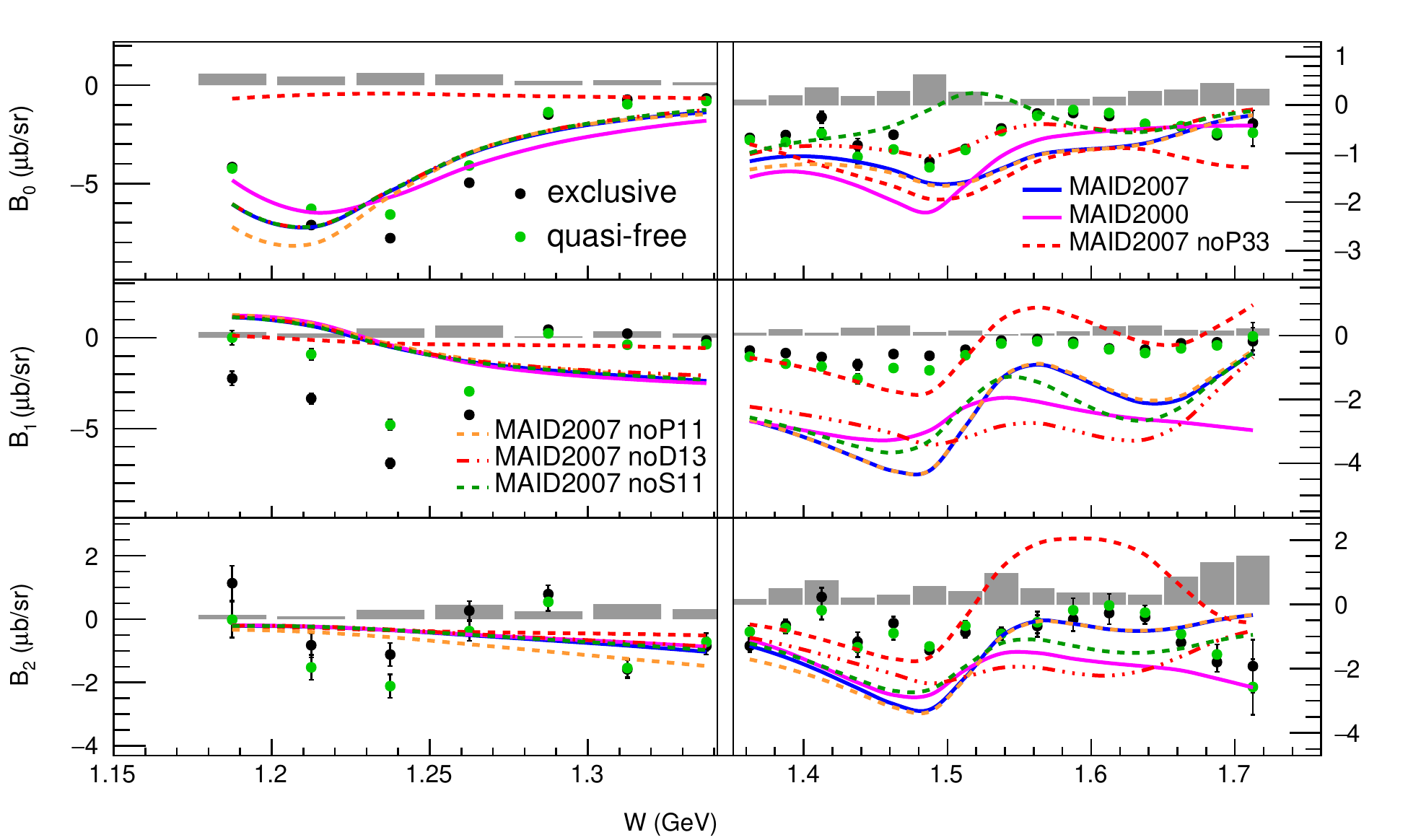}
\caption{\justify{The $W$-dependent Legendre moments $B_{i}$ of $\sigma_{TT}$ in Eq.~\eqref{eq:sigmaLT2_gen} up to $l_{max}=$ 2 at $Q^{2}=$ 0.5~GeV$^2$ that are extracted from the exclusive (black points) and quasifree (green squares) cross sections. For $W>$ 1.35~GeV, the rightmost $y$-axis scale is used. The data are compared with the MAID2007~\cite{refMAID2007} and MAID2000~\cite{refMAID2000} models. The solid lines represent the full model calculations. The dashed lines correspond to the MAID2007 model with specified resonant helicity amplitudes turned off [{\it e.g.}, noP33 indicates turning off the $\Delta(1232)3/2^+$ contributions]. The gray bars represent the corresponding systematic uncertainties of the data.}}
\label{fig:LM_B}      
\end{figure*}

\begin{figure*}[!hbt]
\centering
\includegraphics[width=1.0\textwidth, height=0.75\textwidth]{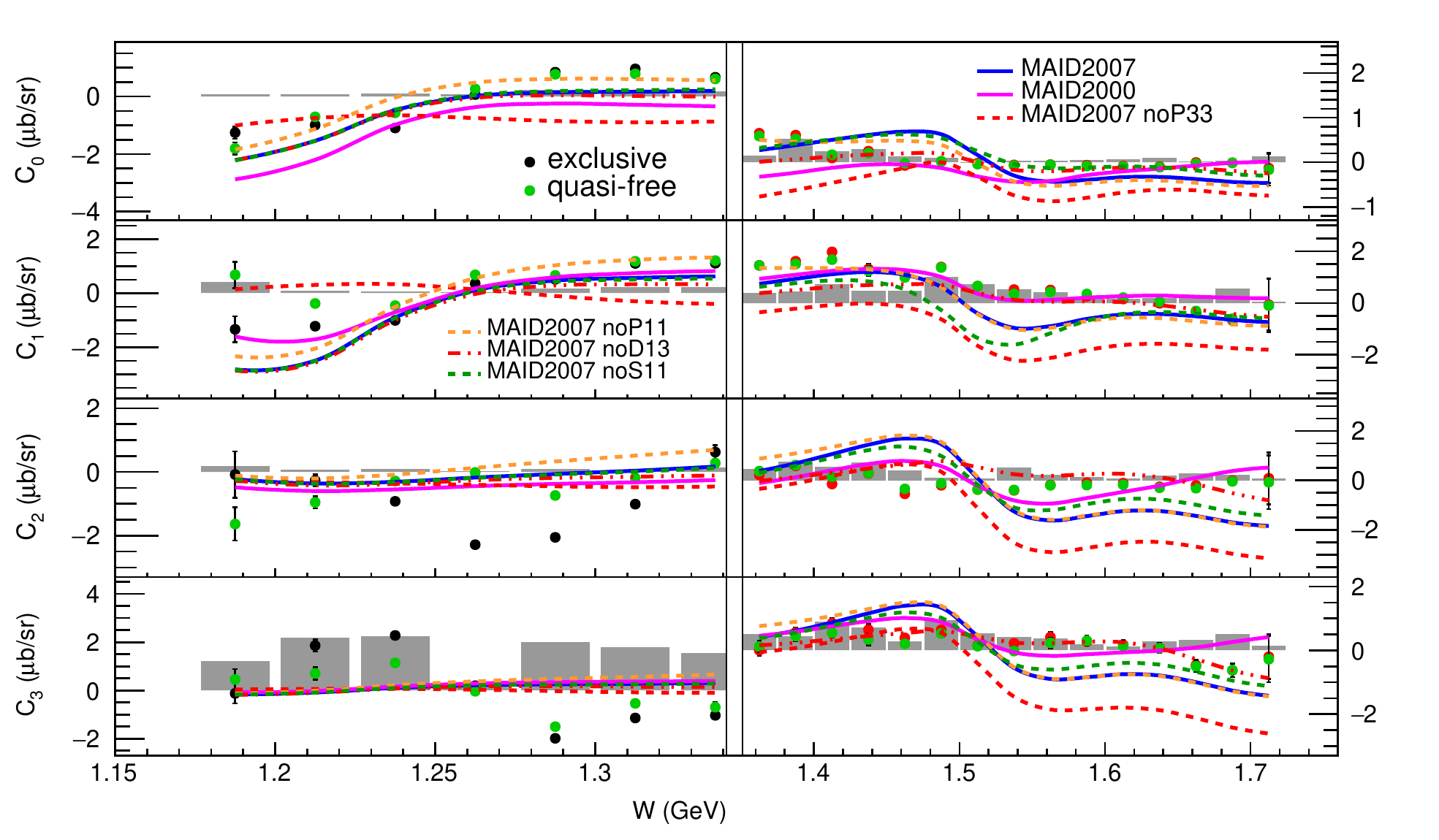}
\caption{\justify{The $W$-dependent Legendre moments $C_{i}$ of $\sigma_{LT}$  in Eq.~\eqref{eq:sigmaLT3_gen} up to $l_{max}=$ 2 at $Q^{2}=$ 0.5~GeV$^2$ that are extracted from the exclusive (black points) and quasifree (green squares) cross sections. For $W>$ 1.35~GeV, the rightmost $y$-axis scale is used. The data are compared with the MAID2007~\cite{refMAID2007} and MAID2000~\cite{refMAID2000} models. The solid lines represent the full model calculations. The dashed lines correspond to the MAID2007 model with specified resonant helicity amplitudes turned off [{\it e.g.}, noP33 indicates turning off the $\Delta(1232)3/2^+$ contributions]. The gray bars represent the corresponding systematic uncertainties of the data.}}
\label{fig:LM_C}      
\end{figure*}

In order to test the sensitivity of the Legendre moments to particular excited nucleon states, their $W$ dependencies were computed in the 0.4 to 0.6~GeV$^2$ $Q^2$ bin within the MAID2007 model \cite{refMAID2007} by switching the $\gamma_vnN^*$ electrocouplings of the $\Delta(1232)3/2^+$, $N(1440)1/2^+$, $N(1520)3/2^-$, and $N(1535)1/2^-$ resonances on and off. The results are shown in Figs.~\ref{fig:LM_A}-\ref{fig:LM_C}. The sensitivity is visualized by the difference between the full MAID2007 model prediction (blue solid curves) and the expectations when particular resonance contributions are turned off.

The results reveal a pronounced sensitivity of $A_0$-$A_3$, $B_0$-$B_2$, and $C_0$-$C_2$ Legendre moments to the contributions from the $\Delta(1232)3/2^+$ resonance. Switching off the $\Delta(1232)3/2^+$ affects the $W$ dependencies of the Legendre moments in the entire kinematic range covered by the measurements. The pronounced $\Delta(1232)3/2^+$ tail impacting the second and third resonance regions is related to the fact that the $\Delta(1232)3/2^+$ cross section is almost an order of magnitude bigger than the measured $\pi^-p$ cross sections in the second and the third resonance regions. Owing to isospin invariance, the $\Delta(1232)3/2^+$ electroexcitation amplitudes off the neutron and the proton should be the same. This prominent contribution of the $\Delta(1232)3/2^+$ resonance seen in our data is consistent with recent studies~\cite{Blin:2021twt} of the resonant contributions to the $F_2$ and $F_L$ inclusive structure functions.

The Legendre moments of the $\sigma_T + \epsilon \sigma_L$ and $\sigma_{TT}$ structure functions exhibit no significant sensitivity to the $N(1440)1/2^+$ resonance, while a moderate sensitivity to the contributions from this resonance can be observed in the $W$ dependence of the $\sigma_{LT}$ Legendre moments (see Fig.~\ref{fig:LM_C}). The $N(1440)1/2^+$ in $\pi^-p$ electroproduction can hence be best explored through the interference between longitudinal and transverse production amplitudes. The Breit-Wigner shape of the $N(1440)1/2^+$ resonance in the $W$ dependence of the $C_{1}$ to $C_{3}$ Legendre moments can only be produced in the interference with the imaginary part of the nonresonant amplitudes, which are small. This makes it difficult to observe the manifestation of the $N(1440)1/2^+$ resonance in the $W$ dependence of the $\sigma_{LT}$ structure function. Instead manifestations of this resonance can be seen in the interference between the real parts of the resonant and nonresonant contributions. Consequently any structure from $N(1440)1/2^+$ contributions to the $\sigma_{LT}$ moments would be expected to be shifted away from the Breit-Wigner mass of this resonance. 

Switching off the electrocouplings of the $N(1520)3/2^-$ and $N(1535)1/2^-$ resonances affects the $W$ dependencies of mostly all Legendre moments of all three structure functions in the second resonance region. However, since there are no available experimental results on the $N(1440)1/2^+$, $N(1520)3/2^-$, and $N(1535)1/2^-$  electrocouplings off bound neutrons, the observed sensitivity to the contributions from these states only indicates a good opportunity to determine their electrocouplings from the $\pi^-p$ differential cross sections presented here.

\section{\label{sec:sum} Conclusions and Outlook}
Exclusive differential cross sections of the electroproduction process ${\gamma}_vn(p) \rightarrow p{\pi}^{-}(p)$ off the bound neutron in deuterium have been extracted for the first time with an almost complete azimuthal pion-angle coverage in most of the $(W,Q^2)$ bins from the JLab CLAS ``e1e" dataset within the kinematic region of $W =$ 1.1$-$1.825~GeV and $Q^{2}=$ 0.4$-$1.0~GeV$^2$. The results on exclusive $\pi^-p$ electroproduction off bound neutrons with nearly complete coverage for the final hadron emission angles in the CM frame provide needed input to extend the models for description of FSI in  $\pi^-p$ photoproduction off bound neutrons \cite{Tarasov:2015sta,Tarasov:2011ec,Nakamura:2018cst,briscoe2021photoproduction,Anisovich:2017afs} towards electroproduction. Accounting for FSI at the amplitude level is of particular importance in paving a way to determine the $N^*$ electroexcitation amplitudes off neutrons. The quasifree cross sections have been evaluated within this analysis via a particular procedure to determine and separate the FSI contributions that are kinematically accessible through the measured observables. 
The FSI contributions in this kinematic region for ${\pi}^{-}p$ electroproduction are about $10\%-20\%$ on average. The azimuthal angular dependence of the extracted quasifree cross sections shows the typical photon-polarization-dependent behavior that is expected for any exclusive electroproduction process assuming the one-photon-exchange approximation, which is a general and otherwise model-independent constraint. Consistent results for the $\sigma_T + \epsilon \sigma_L$ structure function for $\pi^+n$ and $\pi^-p$ electroproduction off protons and neutrons, respectively, as observed in the first resonance region in the angular range where the contribution from the $\Delta(1232)3/2^+$ dominates, further support the reliable extraction of the quasifree cross sections.  
  
Additionally, all accessible associated structure functions, $\sigma_{T}+\epsilon\sigma_{L}$, $\sigma_{TT}$, and $\sigma_{LT}$, have been extracted based on the $\phi^{\text{c.m.\!}}_{\pi}$ dependence of the exclusive differential cross sections with statistical and appropriate systematic uncertainties. The extracted Legendre moments of these structure functions demonstrate the sensitivity to resonant contributions in the first and the second resonance regions. This observed sensitivity underlines the importance of the extracted quasifree cross section data for phenomenological extractions of the $n \to N^*$ electroexcitation amplitudes of various resonances, which will ultimately grant access to isospin-dependent structure effects in various nucleon excitations that emerge from the underlying strong interaction mechanisms.

 

Now, as we have established a method to extract fully exclusive quasifree differential cross sections off the bound neutron, it would be very valuable to extend the kinematic coverage for the $\pi^-p$ electroproduction data, particularly to very forward and very backward ${\pi}^{-}$ polar angles, $W>$ 1.6~GeV, and $Q^2>$ 1~GeV$^2$, by analyzing the data from those further fully exclusive deuterium target experiments with the new CLAS12 detector in Hall B at JLab. This would  allow us to expand the $W$ and $Q^{2}$ coverage and to obtain new information on the $Q^{2}$ evolution of the $n \to N^*$ electroexcitation amplitudes.

\begin{acknowledgments}
The authors thank the administrative and technical staff at Jefferson Laboratory and at all other participating institutions for their invaluable contributions to the success of the experiment. This work was supported in parts by the National Science Foundation (NSF) under Grant No. PHY 1812382, the U.S. Department of Energy (DOE) under Contract No. DE-AC05-06OR23177, the Physics Department of the University of South Carolina (USC), Jefferson Science Associates (JSA), the National Research Foundation of Korea, the Chilean National Agency of Research and Development ANID PIA/APOYO AFB180002, and the Skobeltsyn Nuclear Physics Institute and Physics Department at the Lomonosov Moscow State University.
\end{acknowledgments}

\appendix
\section{Cross section formalism}
The cross section for the exclusive ${\gamma}_vn\rightarrow p{\pi}^{-}$ reaction with an unpolarized electron beam and off unpolarized free neutrons is given by
\begin{eqnarray}
\frac{d^{4}\sigma}{dWdQ^{2}d\Omega^{\text{c.m.\!}}_{\pi}}
&&=\Gamma_{\upsilon}\left(W,Q^{2}\right)\frac{d\sigma}{d\Omega_{\pi}^{\text{c.m.\!}}}.
\label{eq:4foldcross}
\end{eqnarray}
The invariant mass $W$ and virtual photon momentum transfer $Q^{2}$ are calculated by 
\begin{equation}
W=\sqrt{Q^{2}+M_{n}^{2}+2M_{n}\left(E-E^{'}\right)}\;\;\text{and}
\label{eq:W}
\end{equation}
\begin{equation}
Q^{2} \simeq 4EE^{'}{\sin}^{2}\frac{\theta_{e}}{2}=2EE^{'}\left(1-\cos{\theta_{e}}\right)\;,
\label{eq:Q2}
\end{equation}
where $E$ is the electron beam energy, and $E^{'}$ and $\theta_{e}$ are the outgoing electron energy and scattering angle, respectively.  $\Omega_\pi^{\text{c.m.}}$ corresponds to the solid angle of the outgoing $\pi^-$, and ``c.m.''  denotes when variables are calculated in the CM frame. The virtual photon flux is defined as
\begin{eqnarray}
\Gamma_{\upsilon}\left(W,Q^{2}\right)&&= \frac{\Gamma_{\upsilon}\left(E^{'},\Omega_{e^{'}}\right) }{J\left(W,Q^{2}\right)} \nonumber \\
&&=\frac{\alpha}{4\pi}\frac{1}{E^{2}M_{n}^{2}}\frac{W\left(W^{2}-M_{n}^{2}\right)}{\left(1-\epsilon\right)Q^{2}}.
\label{eq:photonflux}
\end{eqnarray}

Since $ Q^{2}=-q^{\mu}q_{\mu}=\vec{q}\,^{2}-\nu^{2}$ and $Q^{2}\simeq 4EE^{'}{\sin}^{2}\frac{\theta_{e}}{2}$, the transverse polarization of the virtual photon $\epsilon$ also can be simplified as
\begin{eqnarray}
\epsilon&&=\left(1+2\left(\frac{\vec{q}\,^{2}}{Q^{2}}\right){\tan}^{2}\frac{\theta_{e}}{2}\right)^{-1}\\
&&=\left(1+2\left(1+\frac{\nu^{2}}{Q^{2}}\right){\tan}^{2}\frac{\theta_{e}}{2}\right)^{-1}\\
&&\simeq\left(1+2\frac{Q^{2}+\nu^{2}}{4EE^{'}-Q^{2}}\right)^{-1}.
\label{eq:episloMod}
\end{eqnarray}

The hadronic differential cross section is calculated from the fourfold differential cross section [Eq.~\eqref{eq:4foldcross}], which is finally extracted from the experimental yield,
\begin{equation}
\frac{d^2\sigma}{d\Omega_{\pi}^{\text{c.m.\!}}}=\frac{1}{\Gamma_{\upsilon}\left(W,Q^{2}\right)}\frac{d^{4}\sigma}{dWdQ^{2}d\Omega^{\text{c.m.\!}}_{\pi}}.
\label{eq:hadronic}
\end{equation}
For the exclusive ${\gamma}_vn(p)\rightarrow p{\pi}^{-}(p)$ reaction, we use the same equations to extract the hadronic differential cross section by ignoring the off-mass-shell effects when calculating the virtual photon flux. 
\label{appendix:a}

\nocite{*}
\newpage
\bibliography{bound_neutron_paper}
\end{document}